\documentclass[fleqn,usenatbib]{mnras}

\usepackage{newtxtext,newtxmath}

\usepackage[T1]{fontenc}

\DeclareRobustCommand{\VAN}[3]{#2}
\let\VANthebibliography\thebibliography
\def\thebibliography{\DeclareRobustCommand{\VAN}[3]{##3}\VANthebibliography}

\usepackage{graphicx}	
\usepackage{amsmath}	
\usepackage{xcolor}     
\usepackage{enumitem}

\newcommand{\colibre}{\textsc{colibre}}
\newcommand{\eagle}{\textsc{eagle}}
\newcommand{\tng}{\textsc{illustrisTNG}}
\newcommand{\illustris}{\textsc{illustris}}
\newcommand{\simba}{\textsc{simba}}
\newcommand{\horizon}{\textsc{horizon-AGN}}

\title[Interaction-driven SFR enhancement in COLIBRE]{The effect of galaxy interactions on star formation rates in the COLIBRE simulations}

\author[Q. W. E. van Zegveld et al.]{Quinten W. E. van Zegveld,$^{1}$\thanks{E-mail: zegveld@strw.leidenuniv.nl}
Evgenii Chaikin,$^{2,1}$
Joop Schaye,$^{1}$
David R. Patton,$^{3}$
Sara L. Ellison,$^{4}$ \newauthor
Alejandro Benítez-Llambay,$^5$
Filip Hu\v sko,$^1$
Robert J. McGibbon,$^1$
Sylvia Ploeckinger,$^6$ \newauthor
Alexander J. Richings,$^{7,8}$ and
Matthieu Schaller$^{9,1}$
\\
$^{1}$Leiden Observatory, Leiden University, PO Box 9513, 2300 RA Leiden, the Netherlands \\
$^{2}$Institute for Computational Cosmology, Department of Physics, University of Durham, South Road, Durham, DH1 3LE, UK\\
$^{3}$Department of Physics and Astronomy, Trent University, 1600 West Bank Drive, Peterborough, ON K9L 0G2, Canada\\
$^{4}$Department of Physics \& Astronomy, University of Victoria, Finnerty Road, Victoria, BC V8P 1A1, Canada\\
$^5$Dipartimento di Fisica "Giuseppe Occhialini", Università degli Studi di Milano-Bicocca\\
$^6$Department of Astrophysics, University of Vienna, Türkenschanzstrasse 17, A-1180 Vienna, Austria\\
$^{7}$Centre for Data Science, Artificial Intelligence and Modelling, University of Hull, Cottingham Road, Hull, HU6 7RX, UK\\
$^{8}$E. A. Milne Centre for Astrophysics, University of Hull, Cottingham Road, Hull, HU6 7RX, UK\\
$^9$Lorentz Institute for Theoretical Physics, Leiden University, PO Box 9506, 2300 RA Leiden, the Netherlands\\
}

\date{Accepted XXX. Received YYY; in original form ZZZ}

\pubyear{\the\year{}}

\begin{document}
\label{firstpage}
\pagerange{\pageref{firstpage}--\pageref{lastpage}}
\maketitle

\begin{abstract}
Observations and theory indicate that galaxy interactions enhance star formation rates (SFRs). However, the degree of enhancement and its dependence on the properties of the interacting galaxies vary across different studies. In this work, we use the \colibre{} simulations of galaxy formation to investigate the effect of interactions on the SFRs of star-forming galaxies at redshift $z\approx0$. The \colibre{} simulations capture the multiphase nature of the interstellar medium and have volumes up to $200^3$ and $400^3$~cMpc$^3$ at m6 (gas and dark-matter particle mass $\sim10^6~\mathrm{M_\odot}$) and m7 ($\sim10^7~\mathrm{M_\odot}$) resolutions, respectively. After constructing samples of interacting galaxies (with mass ratios $>0.1$) and isolated controls, matched in stellar mass, large- and small-scale environment, and redshift, we show that the average specific SFR (sSFR) of interacting galaxies is enhanced by up to a factor of $\approx2$ for separations of $\approx10$~kpc. The enhancement decreases with pair separation but remains significant out to $\approx200$~kpc. The enhancement increases with increasing numerical resolution, is more pronounced in the central regions of galaxies, and decreases with increasing stellar mass at fixed separation. Mergers with higher mass ratios induce stronger sSFR enhancement. We compare our results with observational data from the SDSS, finding good agreement in the dependence of the mean sSFR enhancement on separation, but underpredicting its normalisation by a factor of $\approx2$. Finally, we show that the pre-merger sSFR enhancement of resolved interactions accounts for $\approx2$ per cent of the $z\approx0$ cosmic SFR density.
\end{abstract}




\section{Introduction}
\label{sec:introduction}

Galaxy mergers are an indispensable part of galaxy evolution in a cosmological environment. They are considered the main mechanism for transforming disc galaxies into ellipticals \citep{1972ApJ...178..623T,1992ARA&A..30..705B}.
Both simulations and observations show that galaxy interactions can enhance star formation activity \citep[e.g.][]{Barnes1996,2000MNRAS.312..859S,2000ApJ...530..660B,2004MNRAS.355..874N,2007ApJ...660L..51L}. This enhancement is driven by gas inflows induced by gravitational torques between the interacting galaxies. In some cases, the inflowing gas is subsequently accreted onto the central black hole (BH), which can trigger active galactic nucleus (AGN) feedback \citep[e.g.][]{2005MNRAS.361..776S,DiMatteo2005,Mcalpine2018, RodriguezMontero2019, 2023ApJ...951...92B}.

One of the first indications of the importance of galaxy mergers came from \citet{Larson1978}. They showed that morphologically disturbed galaxies, caused by tidal interactions with another galaxy, are bluer than undisturbed galaxies as a result of elevated star formation and recent starbursts. Since then, many observational studies have investigated the effect of interactions on galaxy star formation rates (SFRs), typically finding increased star formation activity in interacting galaxies relative to their isolated counterparts \citep[e.g.][]{Ellison2008, Scudder2012, 2013MNRAS.435.3627E}. The star-formation enhancement, $Q$, is often quantified in terms of the ratio of SFRs or specific SFRs ($\text{sSFR} \equiv \text{SFR}/M_*$, where $M_*$ is the galaxy stellar mass) between interacting galaxies and their matched isolated control galaxies, with enhancements of $1 \lesssim Q \lesssim 3$ typically found \citep[e.g.][]{2010ApJ...713..330X, Patton2013,Cao2016}. Although most studies focus on low-redshift galaxies, merger-driven sSFR enhancement has also been observed at higher redshifts, including up to $z \approx 9$ \citep{Pusk2025,2026arXiv260628590O}.

While the galaxy merger rate (per galaxy) is known to increase with redshift (at least out to $z\approx 4$; e.g., \citealt{Zepf1989,2009MNRAS.394L..51B,2015MNRAS.449...49R,2022MNRAS.509.5918H}, with a possible plateau at higher redshifts; e.g., \citealt{2025MNRAS.540.2146P}), the interaction-driven sSFR enhancement has been predicted \citep[e.g.][]{Martin2017} and observationally measured \citep[e.g.][]{2022ApJ...940....4S} to decrease with redshift. In addition to triggering starbursts and enhancing sSFRs, interactions may also drive galaxy quenching, which typically occurs shortly after the merger is completed \citep[e.g.][]{Wilkinson2022, Ellison2024}. However, the rapid quenching of post-merger galaxies seen in observations is currently not well reproduced by cosmological simulations \citep[e.g.][]{quai2023}. 

Similar to most observational studies, numerical simulations of galaxies also find enhanced sSFRs and starbursts due to interactions and mergers. Idealised simulations of two merging galaxies in an isolated environment indicate that interactions enhance galaxy sSFRs, with the largest enhancement around the first pericentric passage as well as during the final coalescence \citep[e.g.][]{2000MNRAS.312..859S,Kim2009,Renaud2014}. These results are supported by cosmological hydrodynamical simulations of galaxy formation, which produce statistical samples of galaxies in different environments. Enhanced sSFRs in interacting galaxies are found in many such simulations, including in \horizon{} \citep{Dubois2014} by \citet{Martin2017}, in \eagle{} \citep{Schaye2015} by \citet{Patton2020}, in \tng{} \citep{Pillepich2018} by \citet{Schechter2025}, and in \simba{} \citep{Dave2019} by \citet{RodriguezMontero2019}.

Both simulations and observations show that interaction-induced sSFR enhancement can depend strongly on the properties of the interacting galaxies. \citet{Patton2013} found that the enhancement in the Sloan Digital Sky Survey (SDSS) generally becomes stronger at smaller pair separations. \citet{Schechter2025} showed in \tng{} that the enhancement is stronger for pair mass ratios closer to unity (i.e. major mergers). \citet{2007AJ....134..527W} demonstrated that the enhancement in SDSS is greater for bluer galaxies than for redder ones. \citet{Hani2020} found in \tng{} that the enhancement increases with gas fraction. \citet{2018ApJ...868...46S} used the Cosmic Assembly Near-infrared Deep Extragalactic Legacy Survey (CANDELS) catalogue to show that the enhancement is more significant in galaxies with stellar masses $M_* \lesssim 10^{10-11}~\mathrm{M}_\odot$ than in more massive objects.

The environment in which a galaxy resides can impact its star formation activity \citep[e.g.][]{2008MNRAS.390L...9C,2009MNRAS.393.1324B,2010MNRAS.407.1514E,2010ApJ...718.1158L}, and may therefore bias inferred sSFR enhancements if interacting and control galaxies reside in different environments. \citet{Patton2016} introduced a method for matching interacting and isolated galaxies based on their environment, in addition to matching internal properties such as stellar mass, and showed that ensuring consistent environments for interacting galaxies and their isolated controls is important for recovering sSFR enhancement at large separations. By applying this statistical method to both simulations and observations, \citet{Patton2020} detected a statistical sSFR enhancement in interacting galaxies out to $\approx 200$~kpc in pair separation.

While the general consensus from both simulations and observations is that galaxy interactions induce sSFR enhancement, the magnitude of this enhancement and its dependence on galaxy properties can differ significantly between studies, depending on the sample selection, the matching procedure of interacting and control galaxies, and the ingredients of the galaxy formation models. This motivates further research with larger observational samples, more accurate matching algorithms, and more detailed models of galaxy formation. 

In this work, we use the new cosmological hydrodynamical simulations of galaxy formation \colibre{} \citep{Schaye2025, Chaikin2025a} to investigate the impact of galaxy mergers on star formation activity in $z\approx 0$ star-forming galaxies. Unlike earlier galaxy simulations of representative volumes, such as \eagle{} and \tng, the \colibre{} simulations are available in eight times greater cosmological volumes, incorporate a detailed prescription for the formation and evolution of dust grains \citep{2026MNRAS.545f2040T}, and capture the multiphase nature of the interstellar medium \citep{2025arXiv250615773P}, allowing stars to form in cold, molecular gas. The presence of this cold gas phase is particularly important for studying interaction-induced SFR enhancement, as stars form inside cold molecular clouds in the real Universe.

\colibre{} has been shown to reproduce the observed galaxy stellar mass function \citep{Chaikin2026}, the luminosity functions from the UV to the submillimetre \citep{lu2026}, and the relations between galaxy stellar mass and SFR \citep{Chaikin2026}, size \citep{ludlow2026}, metallicity \citep{2026arXiv260625995S}, and the masses of black holes, atomic gas and molecular gas \citep{Schaye2025}. \colibre{} also reproduces the integrated and spatially-resolved atomic and molecular Kennicutt-Schmidt relations, as well as the dependence of their residuals on metallicity \citep{Lagos2025}. Here, we explore how the SFR enhancement in \colibre{} depends on the separation of interacting galaxies, their stellar mass, mass ratio, and the aperture in which sSFRs are measured, and compare the predicted enhancement with observational data on interacting galaxies from the SDSS compiled by \citet{Patton2013}. The layout of this paper is as follows. Section~\ref{sec:methods} describes the \colibre{} model and construction of our galaxy samples. Section~\ref{sec:results} shows our main results, Section~\ref{sec:discussion} provides the discussion, and Section~\ref{sec:conclusion} concludes this study.

\section{Methods}
\label{sec:methods}

\subsection{The COLIBRE simulations}
\label{sec:methods_colibre}

The \colibre{} galaxy formation model is described in detail in \citet{Schaye2025} and \citet{Chaikin2025a}. Here we provide only a brief summary.

The \colibre{} simulations were run using the astrophysical code \textsc{Swift} \citep{Schaller2024}, following the self-consistent evolution of gas, stars, dark matter, and BHs from an initial redshift of $z = 63$ to $z = 0$. \colibre{} assumes the $\Lambda$CDM `3x2pt + all external constraints' cosmology with $\Omega_{\rm m,0} = 0.306$, $\Omega_{\rm b, 0} = 0.0486$, $\sigma_8 = 0.807$, $h = 0.681$, $n_{s} = 0.967$ from \citet{2022PhRvD.105b3520A}. Gas hydrodynamics is solved using \textsc{Sphenix} \citep{2022MNRAS.511.2367B}, which is a density–energy smoothed particle hydrodynamics (SPH) scheme. The simulations are available at three resolutions, m7, m6, and m5, corresponding to baryonic and dark matter particle masses of, respectively, $m_{\rm gas} = 1.47 \times 10^7~\mathrm{M}_\odot$ and $m_{\rm dm} = 1.94 \times 10^7~\mathrm{M}_\odot$, $m_{\rm gas} = 1.8 \times 10^6~\mathrm{M}_\odot$ and $m_{\rm dm} = 2.4 \times 10^6~\mathrm{M}_\odot$, and $m_{\rm gas} = 2.3 \times 10^5~\mathrm{M}_\odot$ and $m_{\rm dm} = 3.0 \times 10^5~\mathrm{M}_\odot$.

In contrast to most previous simulations of galaxy formation, which typically used equal numbers of dark matter and baryonic particles, the initial conditions of \colibre{} contain four times as many dark matter particles as baryonic particles to reduce spurious energy transfer from dark matter to stars, which can have undesirable effects on galaxy stellar components \citep{2023MNRAS.525.5614L,2023MNRAS.519.5942W}. At m5 resolution, the Plummer-equivalent gravitational softening length for baryons and dark matter is set to the minimum of $0.35$~proper kpc and $0.9$~comoving kpc, while at m6 and m7 resolutions these values are increased by factors of 2 and 4, respectively. The lower-resolution simulations are available in larger cosmological volumes, with the maximum volumes for m5, m6, and m7 resolutions being $(100~\mathrm{cMpc})^3$, $(200~\mathrm{cMpc})^3$, and $(400~\mathrm{cMpc})^3$, respectively. All simulations used in this work are summarised in Table~\ref{table: simulations}, with the largest simulation at m6 resolution being the fiducial simulation for our analysis. The m5 model is not used in this work, as the largest m5 simulation that has already reached $z = 0$ has a volume of (25 cMpc)$^{3}$, which is insufficient for the construction of interacting and control galaxy samples.

\subsubsection{The COLIBRE subgrid model}
\label{subsubsection: colibre_model}

The radiative cooling and heating rates for hydrogen and helium, as well as their free electrons, are computed by the non-equilibrium thermochemistry solver \textsc{chimes} \citep{2014MNRAS.440.3349R,2014MNRAS.442.2780R}. The heating and cooling rates for metals are instead provided by \textsc{hybrid-chimes} \citep{2025arXiv250615773P}, based on tabulated species fractions computed by \textsc{chimes} under the assumption of ionization equilibrium and steady-state chemistry and corrected for the non-equilibrium electron density from the primordial species. We stress that \colibre{} does not impose a pressure or entropy floor, and that the formation of a cold, molecular phase is therefore explicitly simulated. The \colibre{} chemistry model tracks the abundances of 12 individual elements, which are diffused among SPH gas neighbouring particles using a velocity shear-based model for turbulent mixing \citep{2026MNRAS.tmp..607C}. \colibre{} models the formation and evolution of dust \citep{2026MNRAS.545f2040T}, which is coupled to the \textsc{chimes} solver.

Gas particles are labelled as star-forming if they satisfy a gravitational instability criterion, following \citet{2024MNRAS.532.3299N}. The instability condition requires that the (absolute) gravitational binding energy of a gas cloud (represented by the gas mass within the particle's SPH kernel) exceeds its kinetic energy due to thermal and turbulent motions. In \citet{2024MNRAS.532.3299N}, this condition is given by
\begin{equation}
\frac{\sigma_{\rm turb}^2 + \sigma_{\rm th}^2}{G \rho^{1/3} \langle m_{\rm ngb}\rangle^{2/3}} < 1 \, ,
\label{eq: instability_crit}
\end{equation}
where $\rho$ is the mass density of the gas particle, $\langle m_{\rm ngb}\rangle$ is the average mass within the SPH kernel, $\sigma_{\rm turb}$ is the 3D turbulent velocity dispersion, $\sigma_{\rm th}$ is the thermal velocity dispersion, and $G$ is the gravitational constant. The average kernel mass is calculated as $\langle m_{\rm ngb}\rangle = \langle N_{\rm ngb} \rangle \, m_{\rm gas}$, where $m_{\rm gas}$ is the mass of the gas particle for which the star-formation criterion is applied, and the effective number of SPH gas neighbours in the kernel is $\langle N_{\rm ngb} \rangle \approx 65$.

SFRs of star-forming gas particles are calculated using the \citet{1959ApJ...129..243S} law with a star formation efficiency per free-fall time of $\varepsilon = 0.01$. Importantly, \citet{Lagos2025} showed that \colibre{} reproduces the observed integrated and spatially resolved atomic and molecular Kennicutt--Schmidt relations at $z\approx0$, including their scatter, despite not having been explicitly tuned to match these observations.

Star-forming gas particles are stochastically converted into stellar particles, which represent simple stellar populations with a \citet{2003PASP..115..763C} initial stellar mass function. Once formed, stellar particles inject energy and momentum into their surrounding gas and enrich it with metals, following the subgrid model of \citet{2026MNRAS.tmp..607C}. Energy feedback from core-collapse SNe is implemented using a stochastic thermal-kinetic feedback model \citep{2012MNRAS.426..140D,2023MNRAS.523.3709C}, with modifications as described in \citet{Schaye2025}. In addition, \colibre{} includes stochastic thermal feedback from type-Ia SNe, as well as three early stellar feedback processes -- H~\textsc{ii} regions, stellar winds, and radiation pressure -- which are also implemented stochastically \citep{2025arXiv250925309B}.

BHs in \colibre{} are represented by collisionless BH particles, which are seeded inside Friends-of-Friends (FoF) haloes by converting the densest gas particle into a BH particle once the FoF halo mass reaches a predefined threshold, provided the halo does not already harbour a BH particle. Once seeded, BHs grow by accreting gas and by merging with other BH particles. The mass accretion rate is calculated using the turbulence- and vorticity-limited Bondi–Hoyle–Lyttleton formula from \citet{Krumholz_et_al_2006}, and is capped at 100 times the Eddington rate. Dynamical friction and mergers between BH particles are modelled following \citet{2022MNRAS.516..167B}. At each of the three resolutions, the \colibre{} simulations are available with purely thermal AGN feedback \citep{Booth2009} and with hybrid AGN feedback, which combines bipolar kinetic jets with thermal energy injection and accounts for the evolution of BH spin \citep{Husko2025}. This work focuses on the thermal AGN feedback model, as cosmological volumes that are $8\times$ larger are available for this model (see table~2 of \citealt{Schaye2025} for details). We verified that switching from the thermal to the hybrid AGN feedback model has a negligible impact on the results presented in this work.

The strengths of supernova and AGN feedback in the \colibre{} simulations with thermal AGN feedback were calibrated to reproduce the $z = 0$ observed galaxy stellar mass function from \citet{Driver2022} and the size–stellar mass relation from \citet{Hardwick2022}, as detailed in \citet{Chaikin2025a}. The calibration was performed within the stellar mass range $10^9 < M_*/\mathrm{M}_\odot < 10^{11.3}$, using Gaussian process emulators at m7 resolution and through manual adjustments of the model parameters at m6 and m5 resolutions. Additionally, and independently of the galaxy stellar mass function and size–stellar mass relation, the coupling efficiency of AGN feedback (i.e. the fraction of accreted mass-energy injected into the gas in AGN feedback) was adjusted to reproduce the observationally inferred BH masses in massive galaxies at $z = 0$. At each resolution, the hybrid AGN \colibre{} model was manually calibrated to the same data as the thermal model, with the addition of the $z = 0$ AGN luminosity function, as detailed in \citet{Husko2025}.

\subsubsection{Subhalo identification in COLIBRE}

\begin{table*}
\caption{The \colibre{} simulations used in this work. Column (1): simulation name; column (2): cosmological volume per dimension; column (3): mean initial gas particle mass; column (4): mean dark matter particle mass; column (5): number of gas particles in the initial conditions; column (6): number of dark matter particles in the initial conditions; column (7): AGN feedback model used; column (8): number of interacting galaxies at redshift $z < 0.2$ with stellar masses $10^8<M_*/\mathrm{M_\odot} < 10^{12}$; column (9): number of interacting galaxies at $z < 0.2$ with stellar masses $M_* > 10^{10}~\mathrm{M_\odot}$. The fiducial simulation (L200m6) is highlighted in bold. Note that we do not report the number of galaxies in the $10^8 < M_*/\mathrm{M_\odot} < 10^{12}$ mass range for the L400m7 and L100m6 simulations, as we study only galaxies with stellar masses $M_*>10^{10}~\mathrm{M_\odot}$ in these simulations.}
	\centering
	\begin{tabular}{lrrrrrlrr}
   \hline
   Name & $L_{\rm box}$ [cMpc] & $m_{\rm gas}$ [$\rm M_\odot$]  & $m_{\rm dm}$ [$\rm M_\odot$] & $N_{\rm gas}$ & $N_{\rm dm}$  & AGN model  & $N_{\rm int} \,( 10^8<M_*/\mathrm{M_\odot} < 10^{12})$ & $N_{\rm int} \,(M_* > 10^{10}~\mathrm{M_\odot})$ \\
	    \hline
    \multicolumn{9}{|c|}{\textit{m7 resolution}} \\
    L400m7 & $400$ & $1.47 \times 10^7$  & $1.94 \times 10^7$ & $3008^3$ & $4\times 3008^3$  & Thermal & -- & 1\,684~470 \\
    L200m7 & $200$ & $1.47 \times 10^7$  & $1.94 \times 10^7$ & $1504^3$ & $4\times 1504^3$ &  Thermal & 1\,676~413  & 234~296  \\
    \multicolumn{9}{|c|}{\textit{m6 resolution}} \\
    \textbf{L200m6} & $\mathbf{200}$ & $\mathbf{1.8 \times 10^6}$  & $\mathbf{2.4 \times 10^6}$ & $\mathbf{3008^3}$ & $\mathbf{4\times 3008^3}$    & \textbf{Thermal} & \textbf{1\,746~313} & \textbf{226~940} \\
    L100m6 & $100$ & $1.8 \times 10^6$  & $2.4 \times 10^6$ & $1504^3$ & $4\times 1504^3$   & Thermal & -- & 24~850 \\
    \hline
\end{tabular}
\label{table: simulations}
\end{table*}

During the simulation, an on-the-fly FoF group finder is run to identify groups of dark matter particles using a linking length of 0.2 times the mean inter-particle separation. Afterwards, baryonic particles are attached to the closest dark matter particle within the same linking length, if any.

Using the FoF information as input, the history-based halo finder HBT-HERONS \citep{Moreno2025} is then run to identify central and satellite subhaloes. HBT-HERONS is an improved version of HBT+ \citep{Han2018}, whose core assumption is that all substructures form hierarchically. Starting from the earliest available snapshot, an iterative unbinding algorithm is applied to identify self-bound structures within FoF groups. Once a given subhalo is found, its 10 most bound particles (DM or stars) are tracked across subsequent snapshots to follow the evolution of this object forward in time. By tracking subhaloes across redshifts, HBT-HERONS outperforms structure finders that rely only on information at the current redshift, which often face challenges in locating satellites near the centres of host haloes \citep{Moreno2025}. Furthermore, thanks to the inclusion of information from consecutive redshifts, the unphysically large fluctuations in the mass evolution of satellites seen for other structure finders are nearly absent in HBT-HERONS \citep{Chandro-gomez2025}.

HBT-HERONS considers two self-bound subhaloes to have merged once the distance in phase space (velocity and position) between the tracer particles of these objects falls below a certain threshold value \citep[see equation (6) in][]{Moreno2025}. Due to the hierarchical nature of HBT-HERONS, a satellite can only merge with its parent halo, which itself can be a central or a satellite of another halo, but not with another satellite that was independently accreted by the same host. This applies only to a small fraction of satellite–satellite galaxy mergers, which are themselves not very common \citep{Bahe2019}. For further details about HBT-HERONS, see \citet{Moreno2025}.

The subhalo centres used in this work are taken directly from the HBT-HERONS algorithm, which correspond to the position of the most bound particle within the subhalo. All other galaxy properties are calculated by the Spherical Overdensity and Aperture Processor \citep[SOAP,][]{mcgibbon2025}, which is run on the HBT-HERONS output. Unless stated otherwise, the sSFRs in this work are calculated using the stellar mass and instantaneous SFR within a 3D spherical aperture with a radius of 10 proper kpc, including only stellar and gas particles that are gravitationally bound to the subhalo. We compute the sSFR within a relatively small aperture in order to minimise contamination of the sSFR enhancement from aperture overlap between merging galaxies at close separations ($r_{\rm 3D} \lesssim 50~\mathrm{kpc}$), which can occur in merging galaxy pairs when gas particles from one galaxy are misassigned by the halo finder to the other galaxy. The dependence of the sSFR enhancement on the size of the aperture is discussed in \S\ref{sec:results_aperture}. Unless otherwise stated, all remaining properties (including galaxy stellar masses that are not used in the computation of the sSFR) are calculated within an aperture of 50 proper kpc, which is the fiducial aperture in \colibre.

\subsection{Construction of the galaxy samples}
\label{sec:methods_samples_3d}

To answer the question of whether \colibre{} galaxies experience enhanced star formation as a result of the presence of a close companion, we construct two samples: a sample of interacting galaxies and a control sample of matched isolated galaxies. In this section, we describe how these samples are constructed.

We first create a parent sample containing all interacting and isolated galaxy candidates ($\S$\ref{subsubsection:parent_sample_construction}). Within the parent sample, we then search for galaxies that are potentially undergoing interactions ($\S$\ref{subsubsection:interacting_sample_construction}; hereafter referred to as \textit{interacting galaxies}) and match each of them to a relatively isolated galaxy ($\S$\ref{subsubsection:control_sample_construction}; hereafter referred to as \textit{a control galaxy}) with similar stellar mass, redshift, and environment (see below). To this end, we follow the selection algorithm of \citet{Patton2016}. The steps in the entire selection procedure are as follows:

\begin{figure*}
    \centering
    \includegraphics[width=0.99\linewidth]{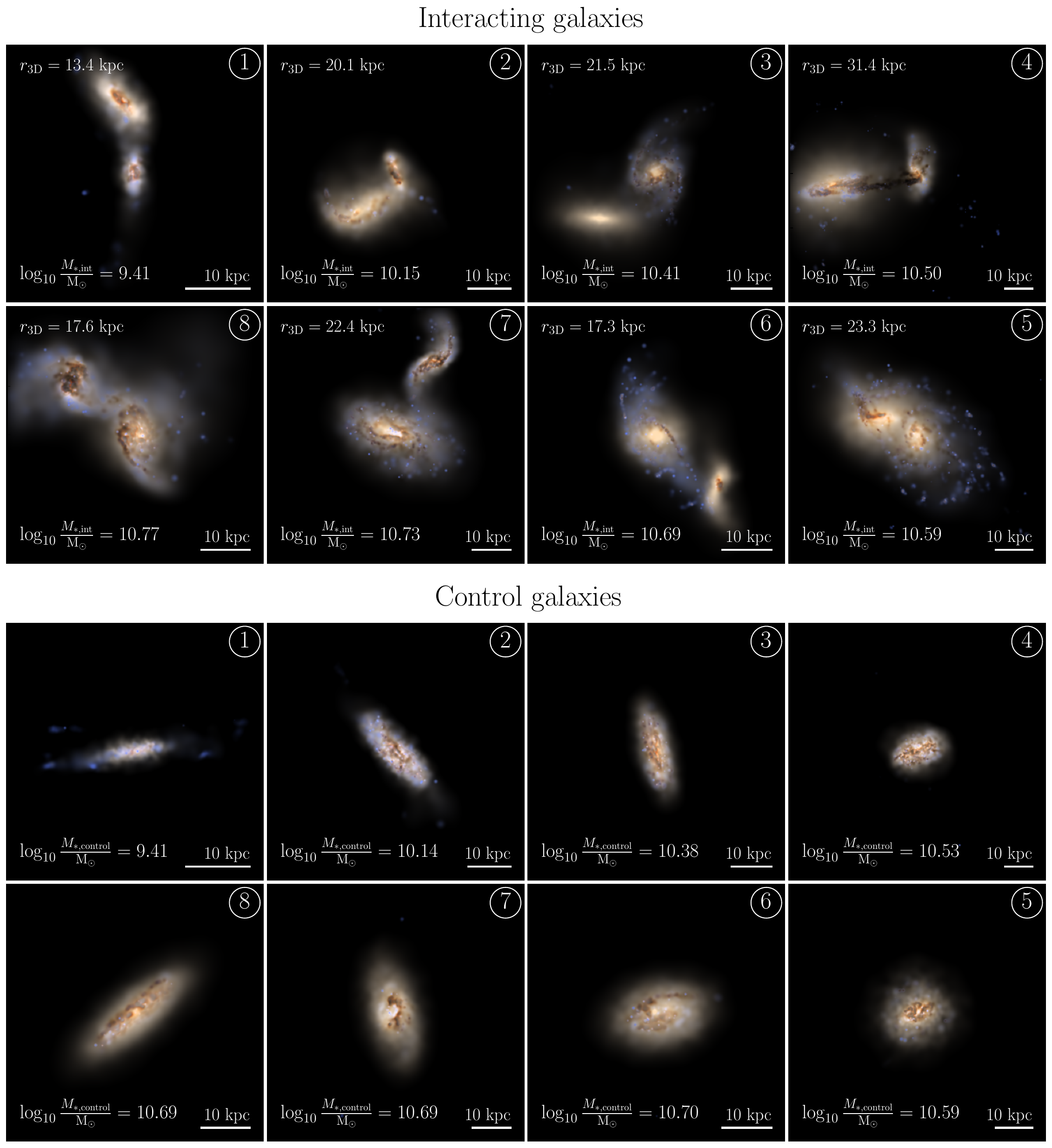}
    \caption{Visual impression of the \colibre{} interacting and control galaxies studied in this work. The definitions of interacting and control galaxies are given in $\S$\ref{sec:methods_samples_3d}. Each galaxy image shows the stellar light in \textit{HST} colours, including dust attenuation, computed using the 1D radiative transfer code \textsc{Partridge} (Hu\v sko et al., in preparation). The top two rows show eight interacting galaxies, selected from the \colibre{} L200m6 simulation at $z=0$. The stellar mass of each galaxy, indicated in the bottom left corner in solar mass units, increases in the clockwise direction starting from the top left panel. The bottom two rows show the eight control galaxies, matched to the eight interacting galaxies in the top two rows, as indicated by the number in the top right corner. All galaxies are viewed along the native $x$-axis of the simulation, which effectively corresponds to a random projection. The images are centred on the interacting (top two rows) and control (bottom two rows) galaxies. The size and depth of each image are equal to three times the 3D separation between the corresponding interacting galaxy and its closest companion, which is indicated in the top left corner of each panel in the top two rows.}
    \label{fig:visualisation}
\end{figure*}

\subsubsection{Construction of the parent sample}
\label{subsubsection:parent_sample_construction}

We start by selecting interacting and isolated galaxy candidates by applying the following constraints:
\begin{enumerate}

    \item We use all \colibre{} data outputs with redshifts less than or equal to $z = 0.2$. In total, this selection includes 14 outputs at redshifts $z = 0.00$, $0.01$, $0.02$, $0.03$, $0.04$, $0.05$, $0.06$, $0.08$, $0.10$, $0.12$, $0.14$, $0.16$, $0.18$, and $0.20$.

    \item At a given redshift, we select all galaxies within the stellar mass range $10^8 < M_*/\mathrm{M}_\odot < 10^{12}$. The lower bound is motivated by resolution requirements: at m6 \colibre{} resolution, all galaxies in the selected range contain at least $50$ stellar particles. Galaxies with stellar masses greater than $M_* = 10^{12}~\mathrm{M}_\odot$ are excluded due to their rarity, which prevents us from finding suitable matched isolated galaxies for them.
    
    \item From the selected stellar mass and redshift ranges, we retain only star-forming galaxies. Our star-formation cut ensures that galaxy SFRs are well sampled, which minimises the adverse effects of noise when comparing the SFRs of interacting and isolated galaxies. We define star-forming galaxies as those with an instantaneous sSFR greater than $10^{-11}~\mathrm{yr}^{-1}$, which is appropriate for the low redshifts considered in this work \citep[e.g.][]{2012MNRAS.424..232W}.

    \item Observational studies often remove AGN-dominated galaxies from their samples or treat them separately, as AGN emission can cause SFR measurements to become unreliable \citep[e.g.][]{Brinchmann2004}. Following this observational practice, we remove all simulated galaxies whose central black hole has a bolometric AGN luminosity greater than $10^{44}~\mathrm{erg~s}^{-1}$ (though we have checked that including these galaxies in our sample would have no significant impact on our results).
\end{enumerate}

\subsubsection{Construction of the interacting sample}
\label{subsubsection:interacting_sample_construction}

To construct the interacting sample, for a given galaxy within the parent sample, we locate the closest companion with a stellar mass greater than 10 per cent of the galaxy's stellar mass and calculate the separation between the centres of the two galaxies, where each centre corresponds to the most bound particle of the galaxy. We then impose the following additional cuts on the parent sample\footnote{Interacting galaxy candidates with stellar masses of $10^8~\mathrm{M}_\odot$ may have closest companions with stellar masses as low as $10^7~\mathrm{M}_\odot$. Although these low-mass companions are poorly resolved at our fiducial resolution, this should have only a marginal impact on our analysis, as we measure the sSFR enhancement only in the interacting galaxies and do not study the intrinsic properties of their closest companions.}:

\begin{enumerate}[start=5]
    \item To avoid galaxies whose closest companion (i.e. the closest galaxy with a stellar mass of more than 10~per cent of the galaxy's stellar mass) is much more massive than the galaxy itself, we remove galaxies whose closest companion has a stellar mass exceeding ten times the stellar mass of the interacting galaxy.
    
    \item In this work, we focus exclusively on the pre-merger phase of interacting galaxies. To avoid cases where two merging galaxies strongly overlap in their stellar mass distribution (and hence could in principle be considered merged) but HBT-HERONS still identifies them as separate objects (because the groups of the most bound particles do not yet overlap sufficiently in 6D phase space), we remove from the analysis all galaxy pairs with relatively close 3D separations. Specifically, following \citet{Patton2020}, a galaxy pair is removed if the 3D separation between the two galaxies is smaller than the sum of their 3D stellar half-mass radii. 
    
\end{enumerate}
We note that the galaxies that are the closest companions to interacting galaxy candidates are not required to be star-forming, and emphasize that the closest companions can be either more or less massive than their interacting counterparts. Furthermore, a given galaxy pair typically exists in multiple snapshots between $z=0$ and $0.2$. On average, a close interaction, i.e. a pair with a separation smaller than 50~kpc, contributes three times to the interacting galaxy sample. Lastly, we note that, although here we do not impose a strict limit on the maximum separation from the closest companion for a galaxy to be considered interacting, the maximum separation studied in this work is 300~kpc (except in $\S$\ref{sec:results_sfrd}, where we consider all possible separations to estimate the contribution of interaction-induced SFR enhancement to the cosmic SFR density at $z\approx0$).

\subsubsection{Construction of the control sample}
\label{subsubsection:control_sample_construction}

We create a control sample of galaxies that is very similar to the interacting galaxy sample, except for the presence of the closest companion. To build a control sample that is otherwise as statistically indistinguishable as possible from the interacting galaxy sample, starting from the parent galaxy sample from $\S$\ref{subsubsection:parent_sample_construction}, we follow an approach similar to \citet{Patton2016}, which describes how to match each interacting galaxy to a control galaxy with similar (i) redshift, (ii) stellar mass, (iii) local density, and (iv) isolation.

\begin{itemize}

\item We define the \textit{local density}, $N_2$, as the number of galaxies within 2~proper Mpc of the interacting galaxy with stellar masses more than 10 per cent of the interacting galaxy's stellar mass.

\item The \textit{isolation} of an interacting galaxy, $r_{\rm 3D,2}$, is defined as the 3D distance to its second closest companion, which is the second closest galaxy to the interacting galaxy with a stellar mass of more than 10 per cent of the interacting galaxy's stellar mass. The isolation of a control is then defined as the distance to its closest companion, $r_{\rm 3D}$.
\end{itemize}

Control galaxies can be matched to an interacting galaxy if their stellar masses differ by at most $0.05$~dex from that of the interacting galaxy, the local densities and isolation differ by at most 10 per cent, and the redshift is identical. The control galaxy with the most similar stellar mass, local density, and isolation is matched to each interacting galaxy\footnote{In principle, (i) an interacting galaxy can be matched to itself as a control galaxy (given the non-zero tolerance limits of the matching procedure), and (ii) the closest companion of an interacting galaxy can be selected as the control galaxy. We remove all such occurrences from our final sample.}. The weighting scheme used for this matching procedure is discussed in Appendix~\ref{app:matching}. 

To avoid biases near the edges of the ranges covered by the variables used for matching, we further constrain our interacting galaxy sample such that $10^{8.05} \leq M_*/\mathrm{M_\odot} \leq 10^{11.95}$, $N_2 \geq 2$, and $r_{\rm 3D,2} < 1.5$~Mpc. Appendix~\ref{app:matching} provides further details of this matching procedure, where we show that the resulting interacting and control samples have statistically indistinguishable stellar masses, local densities, and isolation properties. 

In the fiducial \colibre{} simulation, L200m6, our selection process yields 1\,746~313 interacting galaxies with stellar masses $10^8<M_*/\mathrm{M_\odot} < 10^{12}$, of which 118~248, 266~701, and 766~069 have separations smaller than 50, 100, and 300~kpc, respectively. Of the full sample, 226~940 galaxies have stellar masses $M_* > 10^{10}~\mathrm{M}_\odot$. The sample sizes for the other \colibre{} simulations analysed in this work are shown in Table~\ref{table: simulations}. An illustration of eight interacting galaxies and their matched controls selected from the \colibre{} L200m6 simulation is shown in Fig.~\ref{fig:visualisation}, in which each image shows stellar light in HST colours, including dust attenuation, computed by the 1D radiative transfer code \textsc{Partridge} (Hu\v sko et al., in preparation).

\subsection{Comparison to observations}
\label{sec:methods_samples_2d}

In order to test whether the sSFR enhancement in \colibre{} is consistent with that in the real Universe, in the second part of this work ($\S$\ref{sec:results_sdss}) we perform a comparison with the SDSS data from \citet{Patton2013}. Since observational studies rely on limited information, we construct a sample of interacting galaxies and controls that more closely follows the observational approach, with the following adjustments:
\begin{enumerate}

\item Instead of working with the three-dimensional separation $r_{\rm 3D}$, we use the projected separation $r_{\rm proj}$. For a given galaxy pair, the projected separation is computed in the $y$-$z$ plane of the simulated volume, which effectively corresponds to a random orientation of the pair.

\item When identifying close companions and calculating environmental properties, we include only galaxies with a line-of-sight (LOS) velocity difference $\Delta v < 1000$~km~s$^{-1}$. The velocity difference arises from both the Hubble expansion and galaxy peculiar velocities, and the LOS direction is taken to be the $x$-axis of the native coordinate system of the simulation. The peculiar velocity of each galaxy is computed using all particles within a 3D aperture of radius 50 kpc that are bound to the galaxy.
    
\item To reduce contamination from non-interacting pairs, galaxies with an LOS velocity difference of $\Delta v > 300$~km~s$^{-1}$ relative to their closest companion are excluded from the interacting sample (but may still be included in the control sample). 

\item Following \citet{Patton2020}, in $\S$\ref{subsubsection:control_sample_construction} we adopted a stellar mass tolerance of 0.05~dex for matching control galaxies to each interacting galaxy. For the comparison with SDSS, we instead use a tolerance of 0.1~dex, consistent with \citet{Patton2013}.

\item Finally, in the observational sample of \citet{Patton2013}, owing to the large size of the SDSS sample, interacting galaxies are matched to \textit{statistical} controls (rather than single controls), with the sSFR of each statistical control computed as the weighted average of at least 10 individual control galaxies. To maximise consistency between the \colibre{} and SDSS samples, we switch from our fiducial selection of single controls (described in $\S$\ref{subsubsection:control_sample_construction}) to statistical controls. The construction of the statistical control sample is largely the same as that of the individual control sample. Specifically, for each interacting galaxy, instead of selecting only the best-matched control galaxy, we retain all suitable control galaxies identified by the matching algorithm and construct a statistical control by computing their weighted average, following the weighting scheme described in Appendix~\ref{app:matching}. Furthermore, to maximise consistency with \citet{Patton2013}, interacting galaxies for which fewer than 10 suitable controls can be found are discarded. In the relevant stellar mass range ($10^9 < M_*/\mathrm{M}_\odot < 10^{12}$), at least 10 suitable control galaxies are found for more than 99~per cent of interacting galaxies. On average, we find 33 control galaxies per interacting galaxy, comparable to the SDSS average of 34.
\end{enumerate}
The sample construction for the comparison with the SDSS observations is redone using these adjustments. Note that the closest companion of a galaxy in the 2D case is not always the same galaxy as in the 3D analysis.

\section{Results}
\label{sec:results}

This section is organised as follows. In Section~\ref{sec:results_colibre}, we compare the SFRs in the interacting and control samples of \colibre{} galaxies, constructed as described in Section~\ref{sec:methods_samples_3d}. In Section~\ref{sec:results_sdss}, we compare the predictions of the \colibre{} simulations to observational data from SDSS. For this comparison, the interacting and control galaxy samples are built as described in Section~\ref{sec:methods_samples_2d}. Unless stated otherwise, all simulation results are shown for the L200m6 \colibre{} simulation. The interaction-induced sSFR enhancement in \colibre{} is converged with boxsize and increases with increasing resolution, which is shown in Section~\ref{sec:discussion_resolution}.

\subsection{Interaction-induced sSFR enhancement in \colibre{}}
\label{sec:results_colibre}

In this section, we address whether the sSFR of galaxies with a close companion is enhanced relative to control galaxies in \colibre, and how this enhancement depends on the separation between the interacting galaxy and its closest companion, the interacting galaxy stellar mass, and the mass ratio of the closest companion to the interacting galaxy. We also investigate whether interacting galaxies show enhanced atomic and molecular gas fractions relative to their control galaxies.

\subsubsection{The dependence on pair separation and galaxy stellar mass}

\begin{figure}
    \centering
    \includegraphics[width=0.5\textwidth]{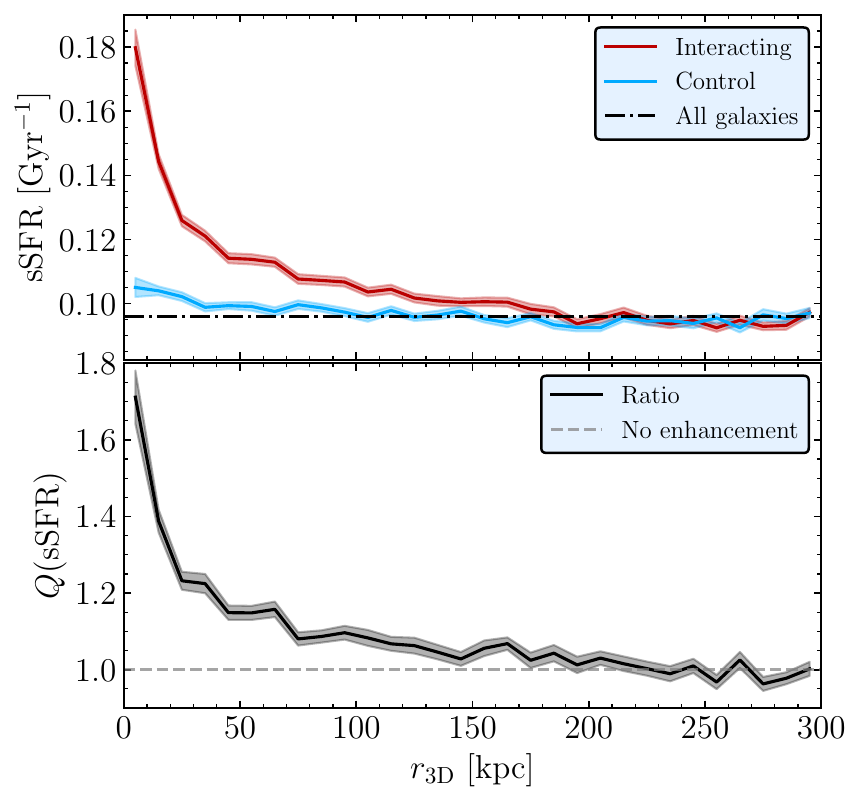}
    \caption{\textit{Top panel}: the mean sSFRs of interacting galaxies (red) and their matched controls (blue) as a function of the 3D separation between the interacting galaxies and their closest companions. The results are shown for galaxies from the \colibre{} L200m6 simulation at redshift $0 \leq z \leq 0.2$. All interacting galaxies have stellar masses $M_{\rm *,int} > 10^{10}~\mathrm{M_\odot}$ and a closest companion galaxy whose stellar mass is more than 10 per cent of theirs. For reference, the black dash-dotted line indicates the mean sSFR of all star-forming galaxies in the simulation within the same redshift and stellar mass range. \textit{Bottom panel}: the mean sSFR enhancement $Q(\mathrm{sSFR})$ of interacting galaxies relative to controls, plotted as a function of separation. The shaded regions indicate bootstrap errors on the mean value in each bin. The mean sSFRs of interacting galaxies are significantly larger than those of controls for separations $r_{\rm 3D} \lesssim 200~\mathrm{kpc}$, with the enhancement increasing rapidly at $r_{\rm 3D} \lesssim 50~\mathrm{kpc}$.}
    \label{fig:q_r}
\end{figure}

Using the interacting and control galaxy samples constructed in Section~\ref{sec:methods_samples_3d}, we calculate the mean sSFR in 10 proper kpc-wide bins of separation between interacting galaxies and their closest companions. We repeat this procedure for the control galaxies, assigning them to separation bins based on the separations of the interacting galaxies to which they were matched. Fig.~\ref{fig:q_r} shows the sSFR for both samples in the top panel and the mean sSFR enhancement of interacting galaxies relative to their controls in the bottom panel. Following \citet{Patton2020,Hani2020}, the mean sSFR enhancement, $Q(\mathrm{sSFR})$, is defined as

\begin{equation}
\label{eq:q_ssfr}
Q(\mathrm{sSFR}) \equiv \frac{\langle \mathrm{sSFR(interacting~galaxies)} \rangle}{\langle \mathrm{sSFR(control~galaxies)} \rangle} \, ,
\end{equation}
where the angular brackets in the numerator (denominator) indicate the average over sSFRs of interacting (control) galaxies in a given separation bin. In Fig.~\ref{fig:q_r}, we consider only interacting galaxies with stellar masses $M_{\rm *,int} > 10^{10}~\mathrm{M_\odot}$, which is the same mass threshold as that used in \citet{Patton2020}. For separations $r_{\rm 3D}<300~\mathrm{kpc}$, this results in a sample size of 109~985 interacting galaxies. 

\begin{figure}
    \centering
    \includegraphics[width=0.5\textwidth]{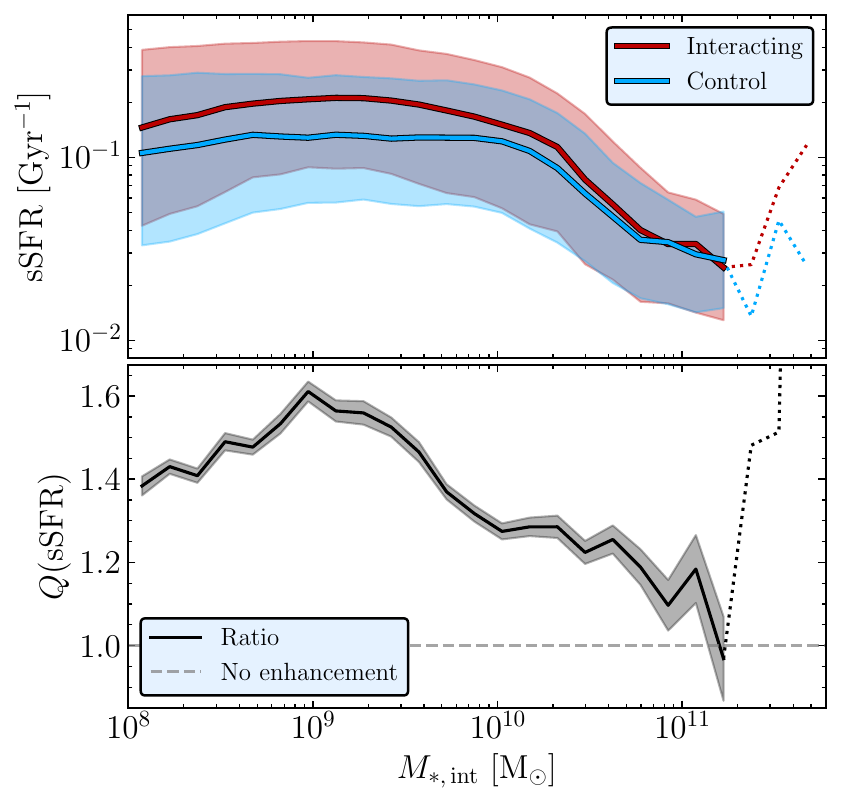}
    \caption{\textit{Top panel:} the median sSFR of interacting galaxies and their controls for interacting galaxies with pair separations $r_{\rm 3D} < 50~\mathrm{kpc}$ and stellar mass ratios $> 0.1$, shown as a function of interacting galaxy stellar mass over the range $10^8 < M_{*, \rm int}/\mathrm{M}_\odot < 10^{12}$. The interacting galaxies are indicated in red and the controls in blue. The shaded areas denote the 16$^{\rm th}$ to 84$^{\rm th}$ percentile scatter. The dotted lines indicate bins with fewer than 50 galaxies to highlight small-number statistics. \textit{Bottom panel:} the mean sSFR enhancement $Q(\mathrm{sSFR})$ as a function of interacting galaxy stellar mass. The shaded regions indicate bootstrap errors on the mean value in each bin. The sSFR enhancement is seen at all stellar masses, peaking at $M_{*, \rm int} \approx 10^{9}~\mathrm{M_\odot}$.}
    \label{fig:q_mstar}
\end{figure}

The mean sSFRs of interacting and control galaxies are shown by the red and blue curves, respectively, with shaded regions of the same colours indicating $1\sigma$ errors estimated via bootstrap sampling. For reference, we also show the mean sSFR of all star-forming galaxies in the simulation within the same stellar mass and redshift range, indicated by the black dash-dotted line. We find that the sSFR of interacting galaxies is enhanced relative to their controls for separations up to $\approx 200$~kpc. At $r_{\rm 3D} < 200$~kpc, the enhancement increases monotonically with decreasing separation, reaching $Q(\mathrm{sSFR}) \approx 1.1$ at $r_{\rm 3D} \approx 70$~kpc and further rising to $Q(\mathrm{sSFR}) \approx 1.7$ by $r_{\rm 3D} \approx 5$~kpc, which is the smallest separation bin we consider. At $r_{\rm 3D} > 200$~kpc, the sSFRs of the two samples converge, yielding $Q(\mathrm{sSFR}) \approx 1$, independent of separation.

The sSFRs of controls also increase at small separations ($r_{\mathrm{3D}}<30~\mathrm{kpc}$) relative to the mean sSFR of all star-forming galaxies within the same redshift and stellar mass range (the black dash-dotted line). As $r_{2,\mathrm{3D}}$ can take values only marginally larger than $r_{\mathrm{3D}}$ to satisfy our sample selection algorithm, and since we match interacting galaxies to controls based on isolation, interacting galaxies with two close companions at separations $r_{\mathrm{3D}}$ and $r_{2,\mathrm{3D}}$ are matched to controls with a single close companion at separation $r_{2,\mathrm{3D}}$. We find that $\approx 10$~per~cent of interacting galaxies in the lowest separation bin ($r_{\mathrm{3D}}<10~\mathrm{kpc}$) have a second closest companion within 50~kpc, resulting in controls that are also undergoing an interaction. As we aim to measure the sSFR enhancement induced by the closest companion of the interacting galaxies, rather than the contribution of other nearby galaxies, this slightly elevated sSFR of controls (which is also observed in similar studies; e.g. \citealt{Patton2013, Patton2020}) is expected and should account for the effect of additional galaxies near the interacting pairs. We have verified that the enhanced sSFRs of controls vanish for isolated pairs and controls.

While at small separations the enhancement is driven by gravitational tidal interactions of the merging galaxies, at separations approaching $200~\mathrm{kpc}$, gravity is unlikely to be strong enough to cause a significant enhancement, and is thus likely due to other effects. \citet{Patton2024} used the \tng{} simulations to investigate the probability that galaxy pairs observed at large separations have undergone an earlier pericentric passage. They found that 28 per cent of galaxies ($M_*>10^{10}~\mathrm{M}_\odot$) with separations of $200$--$500~\mathrm{kpc}$ from their closest companion have undergone such a passage within the past Gyr. Furthermore, \citet{2025MNRAS.537..915F} found in \tng{} that the interaction-induced sSFR enhancement peaks $\approx 0.1$~Gyr after the first pericentric passage, and that the sSFR can remain enhanced for $\approx 0.5$~Gyr after the pericentric passage. Such earlier encounters can therefore explain how galaxies with seemingly very large present-day separations can still show signs of enhanced sSFR.

Having established that the sSFR enhancement of interacting galaxies can be significant in \colibre, in Fig.~\ref{fig:q_mstar} we show it as a function of interacting galaxy stellar mass\footnote{By using the notation $M_{\rm *, int}$ (as opposed to $M_*$), we emphasise that the sSFR of control galaxies is binned according to the stellar masses of their corresponding interacting galaxies.}, $M_{*,\rm int}$. Here, we select only interacting galaxies with separations $r_{\rm 3D} < 50~\mathrm{kpc}$ (and their matched controls), i.e. where the sSFR enhancement is strongest according to Fig.~\ref{fig:q_r}, which results in a sample size of 118~248 interacting galaxies. The interacting and control galaxies, shown in red and blue, respectively, are binned in logarithmically spaced stellar mass bins with a width of $0.15$~dex. The median value in each bin is shown along with the $16^{\rm th}$ to $84^{\rm th}$ percentile scatter. The sSFR enhancement $Q(\mathrm{sSFR})$ is computed using equation~(\ref{eq:q_ssfr}), where the averages over sSFR are taken within interacting galaxy stellar mass bins. Bins with fewer than 50 galaxies are indicated by a dotted line to highlight small-number statistics.

The sSFR enhancement is visible across three orders of magnitude in interacting galaxy stellar mass ($10^{8} \lesssim M_{*,\rm int}/\mathrm{M}_\odot \lesssim 10^{11}$), and becomes substantial ($Q \gtrsim 1.2$) for $M_{*,\rm int} \lesssim 10^{10.5}~\mathrm{M}_\odot$,  peaking at $M_{*,\rm int} \approx 10^{9}~\mathrm{M}_\odot$ ($Q \approx 1.6$). At higher stellar masses ($M_{*,\rm int} \gtrsim 10^{10.5}~\mathrm{M}_\odot$), the enhancement weakens, approaching $Q \approx 1$ by $M_{*,\rm int} \sim 10^{11}~\mathrm{M}_\odot$. This trend is consistent with the findings from \citet{RodriguezMontero2019} and \citet{Hani2020} in the \simba{} and \tng{} simulations, respectively, and is likely due to the decrease in the fraction of gas that can fuel star formation in those massive objects, which we explore in \S\ref{appendix:f_gas}. 

At the low-mass end ($M_{*,\rm int} \lesssim 10^9~\mathrm{M}_\odot$), the decrease in $Q$ is likely a numerical effect, as the fraction of molecular gas (which fuels star formation) drops steeply with decreasing stellar mass below $M_{*,\rm int}  \sim 10^9~\mathrm{M_\odot}$, and this fall-off mass is not converged between the different \colibre{} resolutions \citep[see fig. 19 in][]{Schaye2025}. In \S\ref{appendix:f_gas}, we show that the enhancement in the molecular gas fraction has a shape very similar to that of the sSFR enhancement, confirming that the drop in the molecular gas fraction at $M_{*,\rm int}  \lesssim 10^9~\mathrm{M_\odot}$ is likely driving the drop in the sSFR enhancement at the low-mass end. Further evidence suggesting that this decrease in $Q$(sSFR) is a numerical effect is that (i) the peak in sSFR enhancement at $M_{*,\rm int} \sim 10^9~\mathrm{M}_\odot$ is nearly absent at m7 resolution (see Section~\ref{sec:discussion_resolution}), and that (ii) this peak disappears at m6 resolution if we require interacting and control galaxies to have (nearly) the same molecular gas mass (not shown). Another possible driver of the decrease in $Q$ at the low-mass end is a physical effect. Low-mass galaxies have shallower potential wells, which can prevent gas from becoming star-forming and in turn suppress the interaction-induced sSFR enhancement.

\begin{figure}
    \centering
    \includegraphics[width=0.5\textwidth]{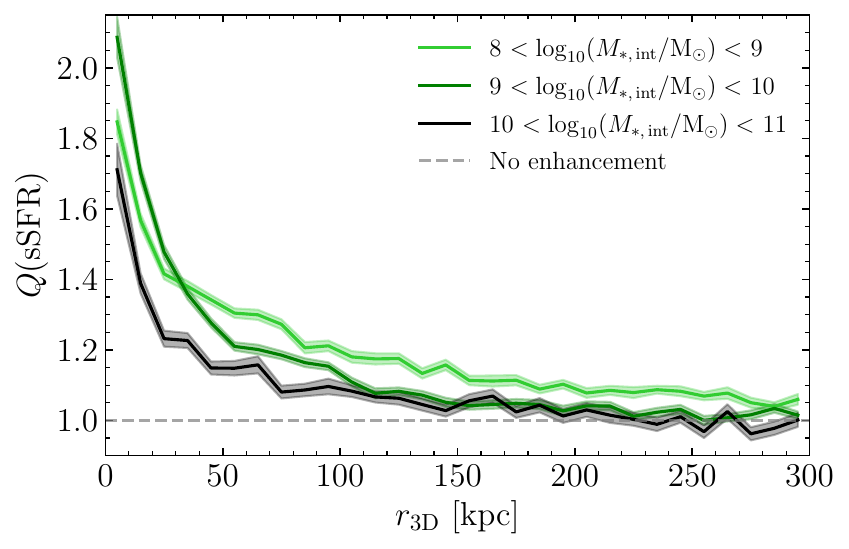}
    \caption{The sSFR enhancement $Q(\mathrm{sSFR})$ as a function of pair separation, shown in three 1-dex interacting galaxy stellar mass bins spanning the range $10^8 < M_{*, \rm int}/\mathrm{M}_\odot < 10^{11}$, indicated by colour. On average, $Q(\mathrm{sSFR})$ is larger for interacting galaxies with lower stellar mass. The lowest stellar mass bin does not fully converge to unity at large separations, which is discussed in Appendix~\ref{app:environment}.}
    \label{fig:q_r_mstar}
\end{figure}

Figs.~\ref{fig:q_r} and \ref{fig:q_mstar} showed the sSFR enhancement \textit{averaged} over stellar mass and separation, respectively. To better understand the dependence of sSFR on these properties, Fig.~\ref{fig:q_r_mstar} shows the sSFR enhancement as a function of separation for three interacting galaxy stellar mass bins -- from $10^8$ to $10^9~\mathrm{M}_\odot$, from $10^9$ to $10^{10}~\mathrm{M}_\odot$, and from $10^{10}$ to $10^{11}~\mathrm{M}_\odot$ -- with progressively darker colours indicating higher mass bins. All three bins exhibit significant sSFR enhancement at small separations ($r_{\rm 3D} \lesssim 50~\mathrm{kpc}$). At fixed separation, the enhancement is larger for lower-mass bins, except at $r_{\rm 3D} \lesssim 30~\mathrm{kpc}$, where the enhancement in the intermediate-mass bin exceeds that in the lowest-mass bin. In particular, as the separation decreases from $30$ to $5$~kpc, the enhancement in the lower-mass and intermediate-mass bins increases monotonically from $Q(\mathrm{sSFR}) \approx 1.4$ up to $\approx 1.85$ and $\approx 2.1$, respectively.

\begin{figure}
    \centering
    \includegraphics[width=0.5\textwidth]{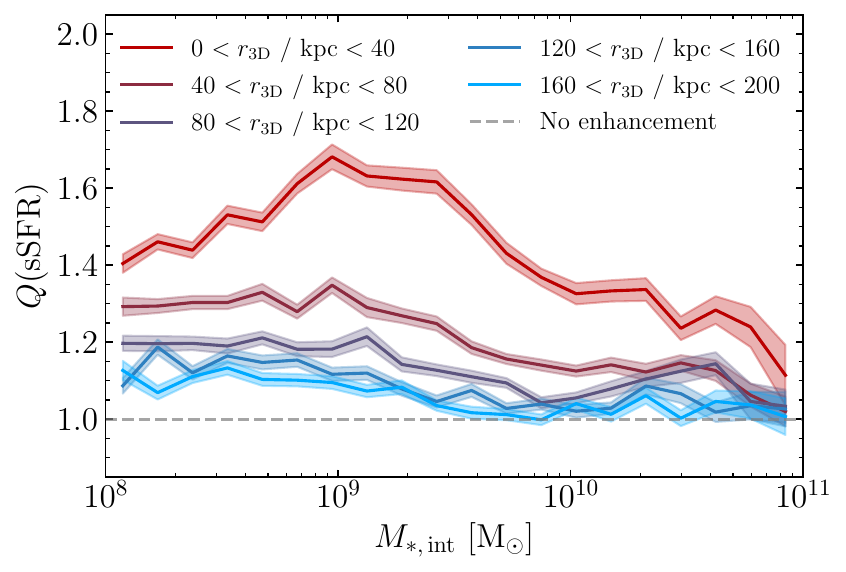}
    \caption{Mean sSFR enhancement of interacting galaxies relative to their controls, $Q(\mathrm{sSFR})$, in logarithmically spaced stellar mass bins, split into five separation bins indicated by colour. At fixed stellar mass, the sSFR enhancement decreases with increasing separation, with the largest values ($Q\approx 1.7$) found around $M_{*, \rm int} \sim 10^9~\mathrm{M_\odot}$ in the smallest separation bin.}
    \label{fig:q_mstar_r}
\end{figure}

Interestingly, the sSFR enhancement in the lowest stellar mass bin does not fully converge to unity at large separations, remaining at $Q(\mathrm{sSFR}) \approx 1.1$ even at $r_{\rm 3D} \gtrsim 250~\mathrm{kpc}$. This is likely due to minor imperfections in the environmental matching between interacting galaxies and their controls. We discuss this further in Appendix~\ref{app:environment}, where we show that matching on the local density within an aperture of 0.8~Mpc, rather than 2~Mpc, improves the environmental matching within the inner 1~Mpc, resulting in $Q(\mathrm{sSFR})$ converging to unity at large separations for the lowest stellar mass bin.

Fig.~\ref{fig:q_mstar_r} shows the mean sSFR enhancement in logarithmic stellar mass bins for five separation bins. The stellar masses span the range $10^8 < M_{*,\rm int}/\mathrm{M}_\odot < 10^{11}$. The five separation bins are equally spaced from 0 to 200~kpc and are indicated by colour, from blue to red for large and small separations, respectively. The shaded areas indicate the bootstrap error on each mean. We find that $Q(\mathrm{sSFR})$ decreases monotonically with increasing $r_{\rm 3D}$, which holds for all $M_{*,\rm int}$ values. The lowest separation bin ($0 < r_{\rm 3D} / \mathrm{kpc} < 40$) shows a clear peak in the sSFR enhancement for galaxies with stellar mass $M_* \sim 10^9~\mathrm{M}_\odot$, reaching $Q(\mathrm{sSFR}) \approx 1.7$. For the other separation bins, the sSFR enhancement depends only weakly on stellar mass, a peak is not visible, and $Q(\mathrm{sSFR})$ does not exceed $\approx 1.35$.

\subsubsection{The effect of galaxy interactions on gas fractions}
\label{appendix:f_gas}

\begin{figure}
    \centering
    \includegraphics[width=0.99\linewidth]{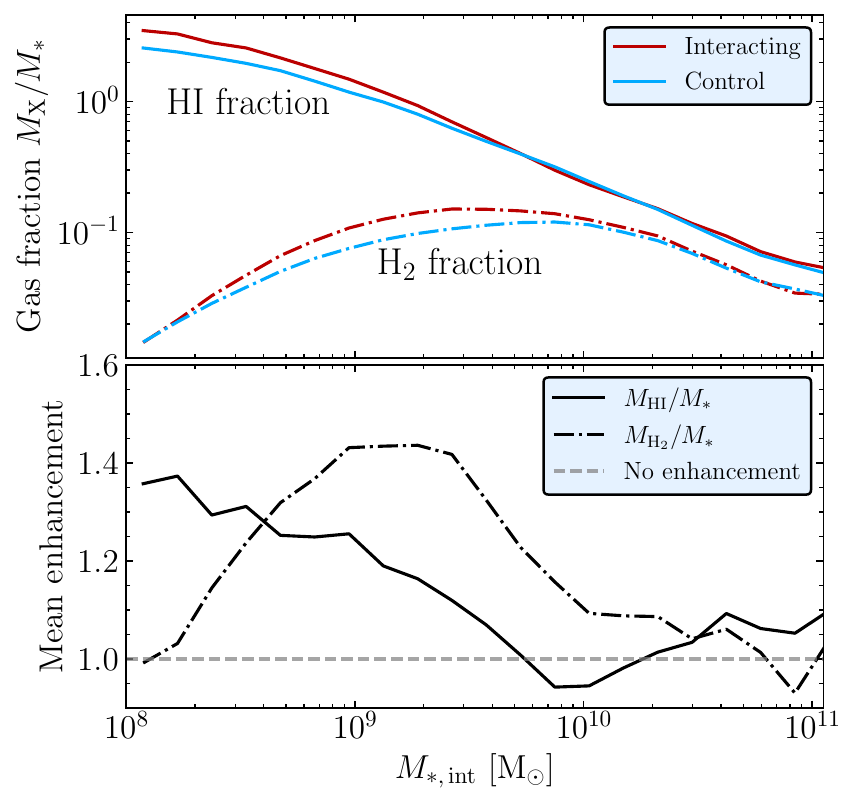}
    \caption{Atomic (solid) and molecular (dash-dotted) gas fractions computed within 10 kpc 3D apertures as a function of interacting galaxy stellar mass for the L200m6 simulation. The selection of interacting and control galaxies is the same as in Fig.~\ref{fig:q_mstar}. The top panel shows the mean gas fractions computed in 0.15~dex stellar mass bins for the interacting (red) and control (blue) galaxy samples. The bottom panel shows the ratio of the mean gas fractions of the interacting and control samples from the top panel. For clarity, we do not show uncertainties in the simulation predictions. The molecular gas fraction is maximally enhanced at $M_{\rm *,int}\sim 10^{9}~\mathrm{M_\odot}$, which coincides with the peak in the sSFR enhancement seen in Fig.~\ref{fig:q_mstar}. At higher $M_{\rm *,int}$, both H~\textsc{i} and H$_2$ gas fractions are relatively low, as expected for massive systems, whereas the drop in the molecular gas fraction below $M_{\rm *,int}\sim 10^{9}~\mathrm{M_\odot}$ is likely a resolution effect.}
    \label{fig:gas_fraction_vs_mstar}
\end{figure}

In this section, we examine the enhancement in atomic and molecular gas fractions of interacting galaxies and its correlation with the sSFR enhancement shown in previous figures. 

Fig.~\ref{fig:gas_fraction_vs_mstar} shows atomic and molecular gas fractions as a function of interacting galaxy stellar mass. We use the same sample of galaxies as in Fig.~\ref{fig:q_mstar}, which corresponds to star-forming interacting galaxies from the L200m6 simulation at $0<z<0.2$, with 3D separations less than 50 kpc and mass ratios greater than 0.1. The top panel shows the ratios of atomic mass (solid curves) and molecular mass (dash-dotted curves) to stellar mass, computed as means in 0.15-dex stellar mass bins. To estimate gas fractions of individual galaxies, similar to the sSFR in Fig.~\ref{fig:q_mstar}, we compute the H~\textsc{i} and H$_2$ masses within 10-kpc 3D apertures and divide them by the stellar mass within the same apertures. The bottom panel shows the mean enhancement in H~\textsc{i} and H$_2$ computed using equation~(\ref{eq:q_ssfr}), where we replaced sSFR with the gas fractions. 

We emphasize that the H~\textsc{i} and H$_2$ masses are direct predictions of the \colibre{} simulations and note that gas fractions were not among the properties used in the selection procedure described in $\S$\ref{sec:methods_samples_3d}. As a result, they can systematically differ between interacting and control samples, although they are indirectly constrained, as the selection requires both interacting and control galaxies to be star-forming and reside in similar environments.

We find that both the atomic and molecular gas fractions within the inner 10 kpc are relatively low at $M_{\rm *,int}\gtrsim 10^{10.5}~\mathrm{M_\odot}$, as expected for massive objects. The H~\textsc{i} gas fraction monotonically increases with decreasing stellar mass over the entire stellar mass range, while the H$_2$ fraction peaks around $M_{\rm *,int}\sim 10^{9.5}~\mathrm{M_\odot}$ and begins to drop at lower $M_{\rm *,int}$. These results are in line with fig.~19 from \citet{Schaye2025}, who showed that the drop in the molecular gas fraction at low stellar mass depends sensitively on numerical resolution.

The atomic gas fraction is enhanced for interacting galaxies with stellar masses $M_{\rm *,int}\lesssim10^{9.5}~\mathrm{M}_\odot$, with the mean enhancement increasing monotonically with decreasing stellar mass. The enhancement in the molecular gas fraction has a similar shape to that of the sSFR enhancement observed in Fig.~\ref{fig:q_mstar}, both peaking at $M_{\rm *,int}\sim 10^{9}~\mathrm{M_\odot}$, with the amplitude of the H$_2$ peak ($\approx 1.4$) being comparable to that of the sSFR peak ($\approx 1.6$). This can be expected, as the SFR and molecular gas content are tightly correlated \citep[e.g.][]{2008AJ....136.2846B,Lagos2025}, while the overall higher sSFR enhancement compared to the gas fraction enhancement can be attributed to the non-linearity of the star formation law. The H~\textsc{i} enhancement at $M_{\rm *,int}\lesssim10^{9.5}~\mathrm{M}_\odot$ can be attributed to the systematically higher total gas fractions of interacting galaxies at these stellar masses. The H$_2$ enhancement is likely driven by gravitational tidal interactions, which funnel gas towards the centres of interacting galaxies and compress it, thereby increasing the gas density and, consequently, the H$_2$ gas fraction.

Fig.~\ref{fig:gas_fraction_vs_mstar} confirms that the drop in the sSFR enhancement at $M_{\rm *,int}\lesssim 10^{9}~\mathrm{M_\odot}$ seen in Fig.~\ref{fig:q_mstar} is likely a resolution effect, as the sSFR enhancement follows that of the H$_2$ enhancement, and the latter depends strongly on resolution in this stellar mass range (see fig.~19 in \citealt{Schaye2025}).

\subsubsection{The effect of stellar mass ratio in galaxy pairs}

\begin{figure}
    \centering
    \includegraphics[width=0.5\textwidth]{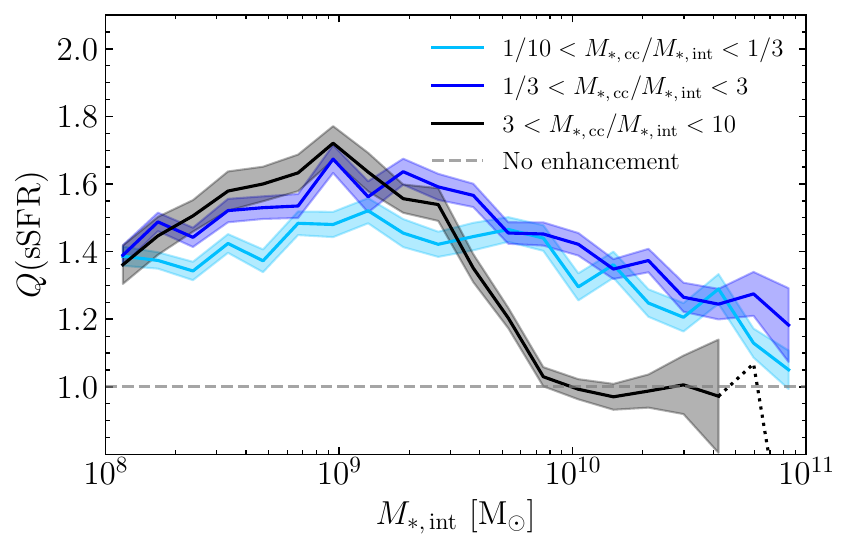}
    \caption{The sSFR enhancement $Q(\mathrm{sSFR})$ as a function of stellar mass for three stellar mass ratio bins in galaxy pairs: $M_{*,\mathrm{cc}}/M_{*,\mathrm{int}}$ from $1/10$ to $1/3$ (light-blue), from $1/3$ to $3$ (dark-blue), and from $3$ to $10$ (black), where $M_{*,\rm int}$ is the stellar mass of the interacting galaxy and $M_{*,\rm cc}$ is the stellar mass of its closest companion (mass ratios below $1/10$ are discussed in Appendix~\ref{app:mini_mergers}). Only pairs with separations $r_{\rm 3D} < 50~\mathrm{kpc}$ are considered. On average, higher mass ratios result in slightly stronger sSFR enhancement, except for the $3<M_{*,\mathrm{cc}}/M_{*,\mathrm{int}}<10$ mass ratio bin at $M_{*,\rm int}\gtrsim 10^{9.5}~\mathrm{M_\odot}$ where $Q(\mathrm{sSFR})$ decreases to unity.}
    \label{fig:mass_ratio}
\end{figure}

We next study the impact of stellar mass ratios on sSFR enhancement. Fig.~\ref{fig:mass_ratio} shows $Q(\mathrm{sSFR})$ as a function of interacting galaxy stellar mass from $10^8$ to $10^{11}~\mathrm{M}_\odot$, using only interacting galaxies with pair separations $r_{\rm 3D}<50~\mathrm{kpc}$. The results are shown for three mass ratio bins: $1/10<M_{*,\mathrm{cc}}/M_{*,\mathrm{int}}<1/3$ (light-blue), $1/3<M_{*,\mathrm{cc}}/M_{*,\mathrm{int}}<3$ (dark-blue), and $3<M_{*,\mathrm{cc}}/M_{*,\mathrm{int}}<10$ (black), where $M_{*,\rm int}$ is the stellar mass of the interacting galaxy and $M_{*,\rm cc}$ is the stellar mass of its closest companion.

For all three mass ratio bins, the sSFR is enhanced at $10^{8} < M_* < 10^{9.5}~\mathrm{M_\odot}$, with slightly stronger enhancement for larger $M_{*,\mathrm{cc}}/M_{*,\mathrm{int}}$ ratios, which we attribute to stronger effects of gravitational tides. In all three mass ratio bins, the sSFR enhancement reaches a maximum around $M_{*,\rm int}\sim10^{9}~\mathrm{M}_\odot$, with $Q(\mathrm{sSFR})\approx1.5$, $\approx1.6$, and $\approx1.7$ for the low, intermediate, and high mass ratio bins, respectively. 

At higher stellar masses ($10^{9.5} < M_* <10^{11}~\mathrm{M_\odot}$), the sSFR remains moderately enhanced for the low and intermediate mass ratio bins, but decreases to unity with increasing $M_*$ for the high mass ratio bin. This reduction is likely a consequence of the fact that, at high stellar masses and mass ratios $3<M_{*,\mathrm{cc}}/M_{*,\mathrm{int}}<10$, the interacting galaxy is more likely to be a satellite in a galaxy group or cluster environment. Such galaxies may have experienced strong environmental effects since their infall into the host halo (e.g. ram pressure stripping; \citealt{1999MNRAS.308..947A}), resulting in a reduced cold gas reservoir and consequently weaker sSFR enhancement.

In Appendix~\ref{app:mini_mergers}, we extend the results from Fig.~\ref{fig:mass_ratio} by considering two additional mass ratio bins, decreasing from $M_{*,\rm cc}/M_{*,\rm int}=1/10$ to $M_{*,\rm cc}/M_{*,\rm int}=1/100$, and find a monotonically decreasing enhancement with decreasing $M_{*,\rm cc}/M_{*,\rm int}$ in these lower mass ratio bins.

\subsubsection{The effect of the aperture size}
\label{sec:results_aperture}

\begin{figure}
    \centering
    \includegraphics[width=0.5\textwidth]{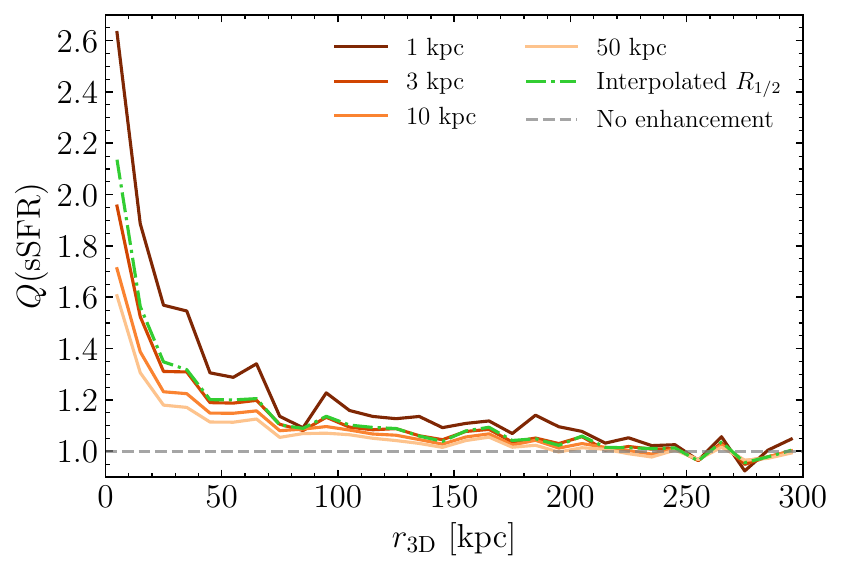}
    \caption{The mean sSFR enhancement of interacting galaxies relative to their controls as a function of 3D separation for different 3D aperture sizes within which the sSFRs are computed. The results are shown for interacting galaxies with stellar masses $M_*>10^{10}~\mathrm{M}_\odot$ and mass ratios $>0.1$. The sSFR enhancements measured within 3D apertures of radius 50, 10, 3, and 1~kpc are shown by solid curves in progressively darker colours. The dash-dotted curve indicates the sSFR enhancement in the interpolated $R_{1/2}$ aperture (see the main text for details). All apertures show a qualitatively similar dependence of sSFR enhancement on separation, with progressively stronger enhancement found for smaller apertures.  
    }
    \label{fig:aperture}
\end{figure}

The aperture within which the sSFR enhancement due to mergers is measured can strongly affect the inferred enhancement, as interaction-induced star formation is expected to peak most strongly in the central regions of galaxies \citep[e.g.][]{Yuan2012,2013MNRAS.435.3627E,2019MNRAS.482L..55T}. Fig.~\ref{fig:aperture} shows the sSFR enhancement as a function of separation for different aperture sizes, using interacting galaxies with $M_{*,\rm int}>10^{10}~\mathrm{M}_\odot$ and mass ratios $> 0.1$. The sSFR enhancements corresponding to fixed 3D aperture sizes of 50, 10, 3, and 1~proper kpc, within which stellar masses and SFRs are available in the \colibre{} output data produced by SOAP, are shown in progressively darker red colours, with 10~kpc being our fiducial choice. The sSFR enclosed within the stellar half-mass radius ($R_{1/2}$) is also shown (green dash-dotted line), and is obtained by interpolation between stellar masses and SFRs in the fixed 3D apertures\footnote{The sSFR enhancements within a 30~kpc 3D aperture, which are available in the \colibre{} SOAP catalogues and used to construct the interpolated quantities, are not shown in Fig.~\ref{fig:aperture}, as they are almost identical to those within the 50~kpc 3D aperture.} of 1, 3, 10, 30, and 50~kpc in logarithmic space.

Fig.~\ref{fig:aperture} shows that the sSFR enhancement is substantially stronger for smaller apertures. Namely, at small separations ($r_{\rm 3D} < 10$~kpc), it reaches $Q(\mathrm{sSFR}) \approx 1.6$ for our largest aperture with a size of 50~kpc, $Q(\mathrm{sSFR}) \approx 1.7$ for our fiducial 10~kpc aperture, $\approx 1.95$ for a 3~kpc aperture, $\approx 2.15$ within $R_{1/2}$, and up to $\approx 2.6$ in the 1~kpc aperture.

\subsubsection{The contribution of interactions to the global SFR density}
\label{sec:results_sfrd}

In the previous sections, we have 
found that galaxy interactions can significantly enhance sSFRs of galaxies in \colibre{}. In this section, we investigate the relative contribution of this enhancement to the cosmic SFR density at $z\approx 0$.

The interacting galaxies from our fiducial sample (Section~\ref{sec:methods_samples_3d}) across the full separation range ($0<r_{\rm 3D}/\mathrm{Mpc}<1.5$), stellar mass range ($10^8<M_{*,\rm int}/\mathrm{M}_\odot<10^{12}$), and mass ratio range ($0.1<M_{*,\rm cc}/M_{*,\rm int}<10$) have a mean sSFR of 0.153~Gyr$^{-1}$, and their controls 0.141~Gyr$^{-1}$. This results in a total sSFR enhancement averaged over all separations of 9.1~per cent. For massive galaxies ($M_*>10^{10}~\mathrm{M}_\odot$), the mean sSFRs of interacting and control galaxies are 0.101~Gyr$^{-1}$ and 0.097~Gyr$^{-1}$, respectively, resulting in an enhancement of 4.2~per cent. 

The contribution of galaxy interactions to the global SFR density, $f_{\rm SFRD}$, can be estimated as the ratio between the interaction-induced SFR excess from all interacting pairs and the total SFR of all galaxies within the cosmological volume,
\begin{equation}
   f_{\rm SFRD} =  \frac{\sum \mathrm{SFR(interacting~galaxies) - SFR(control~galaxies)}}{\sum \mathrm{SFR(all~galaxies~in~the~volume)}}\, .
\end{equation}
We find that for our fiducial sample at redshift $0 \leq z \leq 0.2$, across the whole range of separations ($0<r_{\rm 3D}/\mathrm{Mpc}<1.5$), stellar masses ($10^8<M_{*,\rm int}/\mathrm{M}_\odot<10^{12}$), and mass ratios ($0.1<M_{*,\rm cc}/M_{*,\rm int}<10$), the contribution of galaxy interactions to the total SFR is 2.1 per cent. In other words, the interaction-induced SFR excess accounts for only a small fraction of the cosmic SFR density at $z\approx 0$, which is $\approx 0.012{-}0.013\, \mathrm{M_\odot\,yr^{-1}\,cMpc^{-3}}$ \citep{Chaikin2026}.

In reality, however, the contribution of interactions to the cosmic SFR density is likely larger, as our estimate does not account for contributions from the post-merger phase following galaxy interactions or from mini mergers.

\subsection{Comparison with observations}
\label{sec:results_sdss}

To assess whether the sSFR enhancement in galaxy mergers predicted by the \colibre{} simulations is representative of that in the real Universe, in this section we compare \colibre{} to the sample of observed interacting and control galaxies from \citet{Patton2013}, which is based on measurements from the Sloan Digital Sky Survey \citep[SDSS;][]{York2000} Data Release 7 \citep[DR 7;][]{Abazajian2009}.

\subsubsection{Projected separation versus 3D separation}

\begin{figure}
    \centering
    \includegraphics[width=0.5\textwidth]{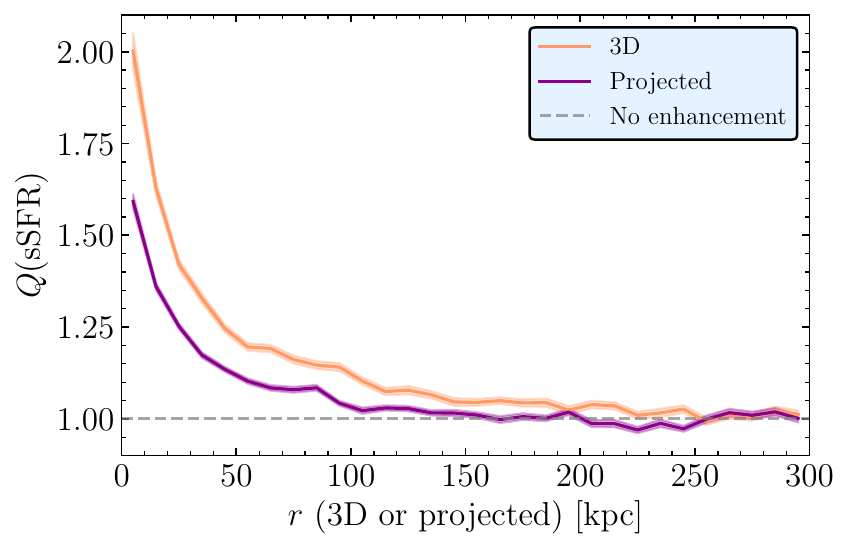}
    \caption{The mean sSFR enhancement of interacting galaxies with $M_{*, \rm int} > 10^{9}~\mathrm{M}_\odot$ relative to their controls, shown for both the fiducial sample (Section~\ref{sec:methods_samples_3d}; purple) and the sample using projected separation (Section~\ref{sec:methods_samples_2d}; orange) in the \colibre{} L200m6 simulation. The sSFR enhancement is binned in equally spaced pair-separation bins. The dashed grey line indicates no enhancement, and the shaded area shows the bootstrap error on each mean. While the sSFR of the fiducial sample is significantly enhanced for separations $r_{\rm 3D} \lesssim 200~\mathrm{kpc}$, the projected sample shows enhancement only for $r_{\rm proj} \lesssim 150~\mathrm{kpc}$. The sSFR enhancement in the projected sample is suppressed relative to the fiducial sample across the full range of pair separations.}
    \label{fig:2d3d}
\end{figure}

In this subsection, we investigate how the results from the previous sections are affected if, instead of using the true 3D separations of interacting galaxies, we adopt a method that mimics observational selection by using projected separations.

Fig.~\ref{fig:2d3d} shows the sSFR enhancement for our fiducial sample based on true 3D separations (purple; Section~\ref{sec:methods_samples_3d}) and for the projected separation sample, where close companions are identified using projected separations and the requirement that the LOS velocity difference is less than $1000~\mathrm{km~s^{-1}}$ (orange; Section~\ref{sec:methods_samples_2d}). The sSFR of the fiducial sample is significantly enhanced for separations $r_{\rm 3D} \lesssim 200~\mathrm{kpc}$, reaching a maximum enhancement of $Q(\rm{sSFR})\approx2$. In the projected sample, the enhancement is significant for $r_{\rm proj} \lesssim 150~\mathrm{kpc}$, with a maximum of $\approx 1.6$. Across the entire separation range, the sSFR enhancement of the projected sample is lower than that of the fiducial sample, until both converge to unity at large separations. \citet{Patton2020} found a similar difference between a sample constructed using 3D galaxy separations and one using projected distances in \tng.

This discrepancy is expected, as the projected sample can be viewed as a contaminated version of the 3D sample. Since, for a given galaxy pair, the projected separation is always smaller than the 3D separation, the sSFR enhancement for the same mergers is found at separations where $r_{\rm 3D} > r_{\rm proj}$. Additionally, contamination from pairs with small $r_{\rm proj}$ but large LOS distances reduces the measured enhancement, as these systems are not expected to be interacting at all.

\subsubsection{Comparison with SDSS}
\label{sec:results_sdss_sdss}

SDSS DR7 contains spectroscopic measurements of 929~555 galaxies spread over 9380 square degrees. \citet{Patton2013} selected galaxies from the main galaxy sample described by \citet{Strauss2002} with secure redshifts $0.02<z<0.2$ and extinction-corrected Petrosian apparent magnitudes of $14.0 \leq m_r \leq 17.77$. They used the total stellar mass estimates from \citet{Mendel2014} based on the photometry of \citet{Simard2011} and, to estimate the sSFRs, the fibre SFRs from \citet{Brinchmann2004} and fibre stellar masses from \citet{Mendel2014}. For each galaxy, they determined $r_{\rm proj}$, $r_{\rm proj,2}$, and $N_2$. They studied interacting and isolated galaxies that are star-forming (selected using the emission-line criterion from \citealt{2003MNRAS.346.1055K}), but did not restrict companions to be star-forming as well. Their approach yields $\approx211~000$ galaxies. For further details about their galaxy sample, see \citet{Patton2013}. 

We use the sample from \citet{Patton2013} for our comparison with predictions from the \colibre{} simulations. To maximize consistency with the observations, for the stellar mass estimates in \colibre{} we use a 50~kpc \textit{projected} aperture\footnote{The projected aperture of size 50~kpc includes all particles bound to the galaxy whose projected distance is within 50~kpc from the galaxy centre.} along the $x$-axis of the native Cartesian axes of the simulated volume, which effectively corresponds to a random orientation of galaxies. The sSFRs are defined within the aperture corresponding to the 3~arcsec fibre size of the SDSS observations at each redshift. This fibre size corresponds to $\approx 1$~kpc at $z = 0.02$ and $\approx 10$~kpc at $z = 0.2$. We calculate the sSFR by interpolating the stellar mass and SFR in logarithmic space between fixed projected apertures of 1, 3, 10, 30, and 50~kpc, selecting the appropriate bracketing apertures at each redshift. If, for a given galaxy, the SFR in at least one of the two apertures bracketing the fibre size is zero, we adopt the sSFR of the nearest larger aperture with a non-zero SFR. We select \colibre{} galaxy samples for this comparison following the algorithm described in Section~\ref{sec:methods_samples_2d}, using the \colibre{} outputs at the redshifts covered by the SDSS sample and within the same stellar mass range as the SDSS galaxies.

Compared to volume-limited samples, flux-limited observational surveys such as SDSS produce galaxy samples with an overabundance of galaxies with higher stellar masses and more massive closest companions, as these objects are easier to detect. To minimize biases in the comparison between \colibre{} and SDSS, galaxies from both samples are selected such that the stellar mass, mass ratio ($>1/10$), and redshift distributions are indistinguishable. As in the SDSS sample galaxies with stellar masses $<10^9~\mathrm{M_\odot}$ are relatively rare ($<4$~per cent), we impose a minimum stellar mass threshold of $10^9~\mathrm{M_\odot}$. The upper limit remains the same as in the rest of this work, $M_* = 10^{12}~\mathrm{M}_\odot$. The galaxies are divided into 30 stellar mass bins spanning $10^{9} \leq M_*/\mathrm{M_\odot} \leq 10^{12}$ with a width of 0.1~dex, 20 mass ratio bins from 0.1 to 10 with a width of 0.1~dex, and 12 redshift bins corresponding to the 12 \colibre{} outputs that satisfy $0.02 \leq z \leq 0.2$. The largest possible number of galaxies is selected from each three-dimensional bin such that the ratio of \colibre{} to SDSS galaxies is three to one. This sampling ratio preserves a relatively large fraction of SDSS galaxies (71 per cent) while retaining a large number of \colibre{} galaxies, resulting in small statistical uncertainties. In total, this approach yields 27~083 SDSS interacting galaxies and three times as many \colibre{} interacting galaxies.

\begin{figure}
    \centering
    \includegraphics[width=0.5\textwidth]{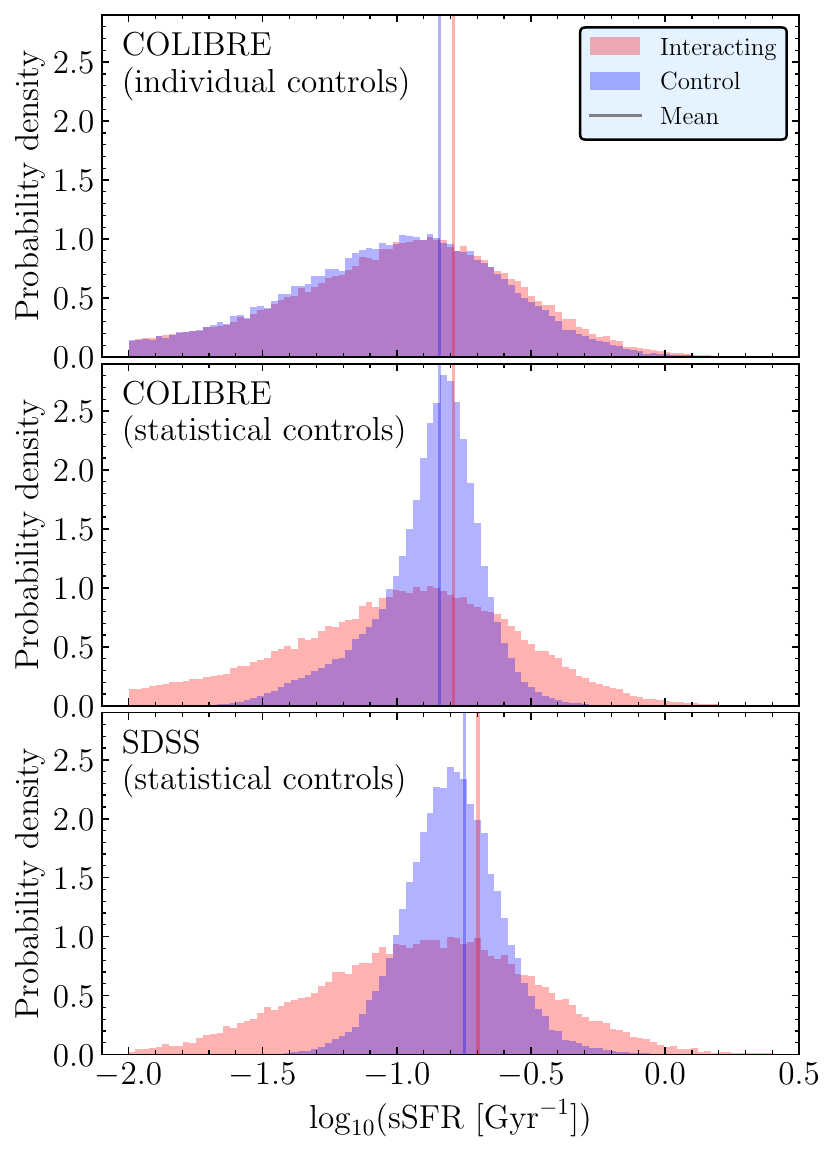}
    \caption{The distributions of sSFRs for interacting galaxies (red) and their controls (blue) in the \colibre{} L200m6 simulation and in the SDSS sample from \citet{Patton2013}. The interacting galaxies have stellar masses $10^9 < M_*/\mathrm{M_\odot} < 10^{12}$ and projected separations $r_{\rm proj}<300~\mathrm{kpc}$. The top and middle panels show the distributions in \colibre{} using individual and statistical controls, respectively (see text for details), while the bottom panel shows the SDSS distributions using statistical controls. The vertical lines indicate the mean sSFRs of the interacting and control galaxy distributions. \colibre{} and SDSS have very similar sSFR distributions for interacting galaxies and their statistical controls, and switching from individual to statistical controls in the simulation does not affect the mean sSFR enhancement (i.e. the separation between the red and blue vertical lines remains unchanged).}
    \label{fig:pdfs}
\end{figure}

In addition to matching the SDSS sample to the sample of \colibre{} interacting galaxies with statistical controls (constructed following the algorithm in Section~\ref{sec:methods_samples_2d}), we also match the SDSS sample to a sample of \colibre{} interacting galaxies with individual controls (constructed using the same algorithm, but without applying adjustments (iv) and (v) to the fiducial matching described in $\S$\ref{subsubsection:control_sample_construction}). This allows us to quantify the effect of switching from single to statistical controls in the simulation. Fig.~\ref{fig:pdfs} shows the probability density distributions of the sSFRs of interacting galaxies (red) and their controls (blue), together with their mean sSFRs (vertical lines). The top and middle panels show the distributions for \colibre{} using individual and statistical controls, respectively, while the bottom panel shows the SDSS interacting galaxies and their statistical controls.

The top panel shows that the sSFR distributions of \colibre{} interacting galaxies and their individual controls are very similar. The small excess of high-sSFR galaxies and deficit of low-sSFR galaxies in the interacting sample relative to the control sample produces a visible enhancement in the mean sSFR, while the overall shapes of the distributions remain nearly identical. Comparing the blue histograms in the top and middle panels reveals that switching from individual to statistical controls narrows the sSFR distribution of the control galaxies, as expected when averaging over multiple controls. At the same time, the virtually unchanged positions of the blue and red vertical lines in the top and middle panels confirm that switching from individual to statistical controls has a negligible impact on the mean sSFR enhancement.

Comparing the middle and bottom panels further reveals that the sSFR distributions of \colibre{} and SDSS are very similar, both for the interacting galaxies and their statistical controls. The slightly more extended tails of the sSFR distributions of both interacting and control galaxies in \colibre{} are most likely caused by the different criterion used to select star-forming galaxies, as we impose an sSFR threshold in \colibre{}, whereas the observations use emission-line diagnostics. Because the differences in the sSFR distributions between \colibre{} and SDSS shift the mean sSFRs of interacting galaxies and their statistical controls by a similar amount, their net effect on the mean sSFR enhancement, which is similar in \colibre{} and SDSS, is minor.

\begin{figure}
    \centering
    \includegraphics[width=0.5\textwidth]{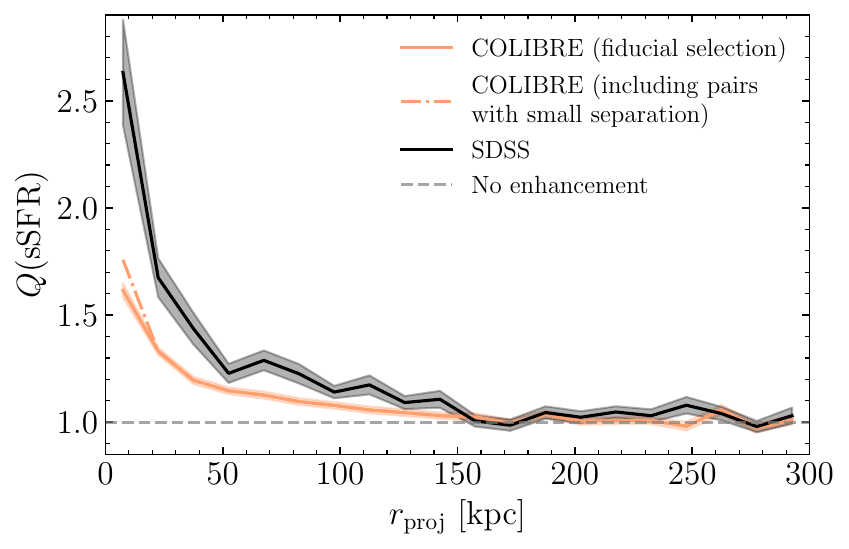}
    \caption{The mean sSFR enhancement of interacting galaxies relative to their statistical controls predicted by the \colibre{} L200m6 simulation (orange), compared with the observational data from SDSS data (black), shown in equally spaced projected separation bins. In both cases, only interacting galaxies with stellar mass $M_{*, \rm int} > 10^9~\mathrm{M}_\odot$ are considered. The enhancement is computed as the ratio of the mean sSFR of interacting galaxies to that of controls in each separation bin, and the shaded area indicates the bootstrap uncertainty on this ratio. The solid (dash-dotted) orange curve corresponds to the case in which interacting galaxies with very small distances to their closest companions are removed from (retained in) the \colibre{} sample (see main text for details). Both SDSS and \colibre{} galaxies show enhanced sSFR at separations $r_{\rm proj} \lesssim 150~\mathrm{kpc}$. The \colibre{} enhancement follows a similar dependence on separation as the observational data, albeit with a lower normalization.}
    \label{fig:sdss}
\end{figure}

The mean sSFR enhancement in \colibre{} (orange) and SDSS (black) galaxies, both using statistical controls, is shown in Fig.~\ref{fig:sdss} as a function of projected separation, with $1\sigma$ errors indicated by the shaded regions. As described in Section~\ref{sec:methods_samples_3d}, all interacting galaxies with a closest companion at a 3D separation smaller than the sum of their 3D stellar half-mass radii are removed from the analysis in this work, as galaxies in such interacting pairs can be too close to be visually distinguishable. Accordingly, we exclude these galaxies from our projected sample too (solid orange curve in Fig.~\ref{fig:sdss}), but also show the result when they are retained (i.e. when interacting galaxies are included regardless of how small the distance to their closest companion is; dash-dotted orange curve).

Both \colibre{} and SDSS galaxies show a significant sSFR enhancement for projected separations $r_{\rm proj} < 150$~kpc. The radial dependence of the enhancement is similar in both samples, but the normalization differs, with SDSS showing sSFR enhancements that are overall approximately twice as large as those predicted by \colibre{}. In particular, in the smallest considered separation bin, $r_{\rm proj} = 5~\mathrm{kpc}$, \colibre{} predicts $Q \approx 1.6$, while SDSS shows $Q \approx 2.65$. Comparing the solid and dash-dotted orange curves reveals that retaining pairs with very close separations in the \colibre{} sample increases the sSFR enhancement predicted by \colibre{} in the lowest separation bin from $Q \approx 1.6$ to $Q \approx 1.75$, which is still significantly below the enhancement seen in SDSS. However, we note that the difference between the mean enhancements is small compared to the range of sSFRs spanned by both the interacting and control samples (see Fig.~\ref{fig:pdfs}).

In Section~\ref{sec:discussion_resolution} we demonstrate that in \colibre{}, $Q(\text{sSFR})$ is an increasing function of resolution at fixed separation, so at least some of the discrepancy with the SDSS data could be reduced by switching from the m6 resolution to the higher resolution m5, when the m5 model is available in sufficiently large volumes (see table 2 in \citealt{Schaye2025}). We also note that the sSFR enhancement predicted by \colibre{} would increase by about 10~per~cent if we limited our sample of interacting pairs to only those in which both galaxies are star-forming (in our fiducial selection, the closest companions of the interacting galaxies are not required to be star-forming), making \colibre{} consistent with the SDSS data at $r_{\rm proj}\gtrsim 40~\mathrm{kpc}$ and reducing the discrepancy at smaller separations. This might be a reasonable adjustment, as observations are more likely to detect low-mass galaxies that are star-forming, since they are on average brighter than quiescent galaxies of the same mass. Lastly, we verified that the \colibre{} predictions presented here remain similar if we shift the sSFR threshold by up to 0.5~dex from the fiducial value of $10^{-11}~\mathrm{yr}^{-1}$.

\section{Discussion}
\label{sec:discussion}

\begin{figure}
    \centering
    \includegraphics[width=0.5\textwidth]{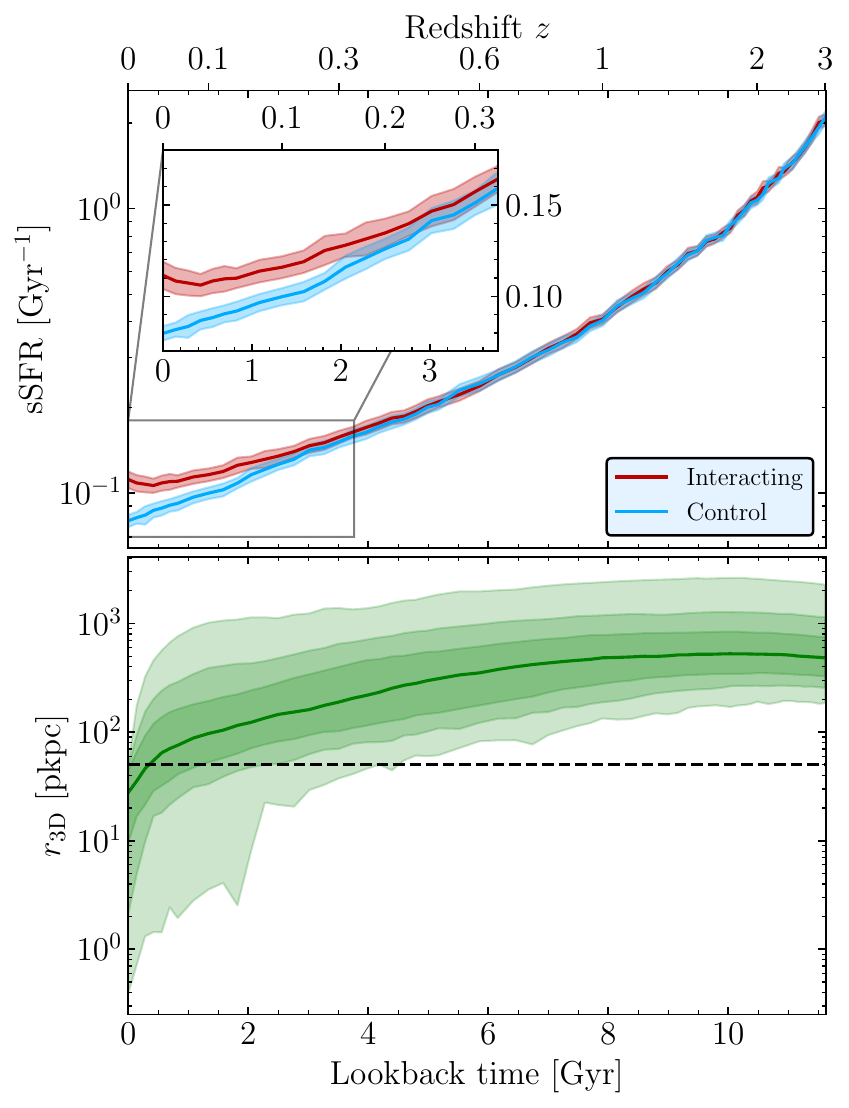}
    \caption{Evolution of interacting galaxies and their controls as a function of lookback time. The interacting galaxies are selected at redshift $z = 0$ to have stellar masses $M_{\rm *,int} > 10^{10}~\mathrm{M_\odot}$, separations $r_{\rm 3D} < 50$~kpc, and mass ratios $>0.1$. \textit{Top panel:} the mean sSFRs of interacting galaxies (red) and their controls (blue), along with a $2\sigma$ bootstrap standard error of the mean shown by the shaded regions. \textit{Bottom panel:} the median separation between interacting galaxies and their closest companions (green solid line), along with the $16^{\rm th}$ to $84^{\rm th}$, $5^{\rm th}$ to $95^{\rm th}$, and $1^{\rm st}$ to $99^{\rm th}$ percentile scatter (shaded regions). Close to the redshift of selection, where the median separation is by construction $\lesssim 50$~kpc, the sSFRs of interacting galaxies are greater than those of their matched controls. However, the differences between interacting and control galaxies become smaller at larger redshifts (corresponding to larger separations) and disappear completely above $z \approx 0.3$, at which point the median separation grows beyond $\approx 200~\mathrm{kpc}$.}
    \label{fig:time_evo}
\end{figure}

\subsection{Is sSFR enhancement due to interactions or differences in environment?}
In this work, we have studied how much the mean sSFR of interacting galaxies is enhanced relative to their isolated counterparts. In addition to stellar mass, we matched interacting galaxies to their isolated controls based on local density and isolation (see Section~\ref{sec:methods_samples_3d}), ensuring that interacting and control galaxies are drawn from similar environments so that the sSFR differences we find are due to the presence of the closest companions near the interacting galaxies. However, it is possible that subtle differences in the environments of interacting and control galaxies, particularly on smaller scales than the one on which the environment is defined, still remain and that they (partly) contribute to the sSFR enhancement found in the previous section.

To verify that the sSFR enhancement of interacting galaxies relative to their controls found in this work results from interactions and is not dominated by environmental differences between interacting and control galaxies, in Fig.~\ref{fig:time_evo} we show the time evolution of the properties of interacting galaxies and controls. We select interacting galaxies and their controls at redshift $z=0$ using our fiducial 3D selection (Section~\ref{sec:methods_samples_3d}) with interacting galaxy stellar masses $M_{\rm *,int} > 10^{10}~\mathrm{M_\odot}$, separations $r_\text{3D} < 50~\text{kpc}$, and mass ratios $>0.1$, and follow them back in time along the branch of the main progenitor, as provided by HBT-HERONS. The top panel shows the sSFRs of interacting and control galaxies, and the bottom panel shows the separations between interacting galaxies and the galaxies that are their closest companions at redshift $z=0$, both as functions of lookback time.

As expected, we find that the sample of interacting galaxies differs significantly from the sample of controls at redshift $z=0$, with a mean sSFR enhancement of $Q \approx 1.2$, consistent with the enhancement observed at separations $r < 50~\text{kpc}$ in Fig.~\ref{fig:q_mstar}. However, as the separation between galaxy pairs increases with increasing lookback time, the differences in sSFR between interacting and control galaxies become smaller, with the mean sSFRs of the interacting and control samples fully converging at $z \approx 0.3$, corresponding to a median separation of $r_{\rm 3D}\approx 200~\mathrm{kpc}$. The fact that the sSFRs of interacting galaxies, which are constructed to be interacting at redshift $z=0$, and their isolated controls converge as pair separation increases at higher redshift is a strong indication that the sSFR enhancement found in this work is caused by interactions and not by environmental differences between the samples of interacting and control galaxies.

\subsection{The effect of numerical resolution and cosmological volume}
\label{sec:discussion_resolution}

This section investigates the impact of numerical resolution and cosmological volume on the galaxy interaction-induced sSFR enhancement in \colibre{} found in this work. 

\begin{figure}
    \centering
    \includegraphics[width=0.5\textwidth]{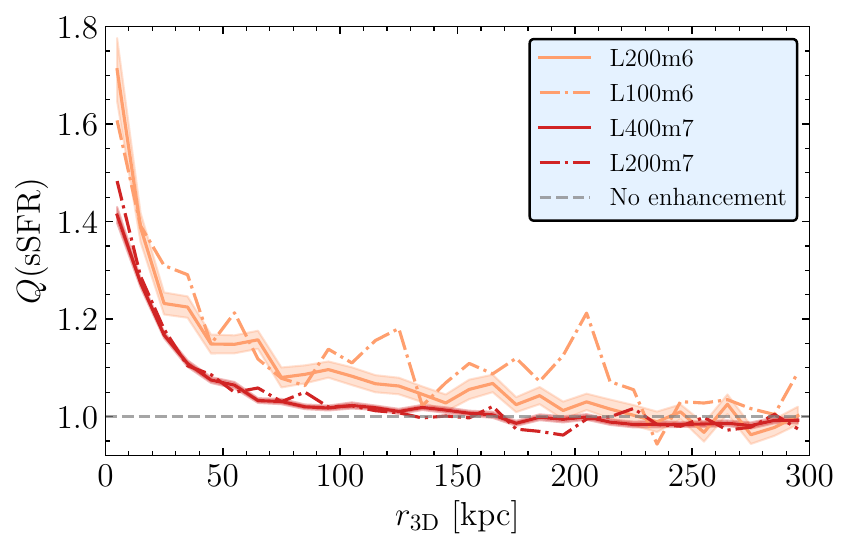}
    \caption{Interaction-induced sSFR enhancement in four \colibre{} simulations: L200m6 (orange, solid), L100m6 (orange, dash-dotted), L400m7 (red, solid), and L200m7 (red, dash-dotted). We include only interacting galaxies with stellar masses $M_{\rm *,int} > 10^{10}~\mathrm{M_\odot}$ and mass ratios $> 0.1$. At fixed resolution, increasing the box size results in no systematic differences in the sSFR enhancement, whereas increasing the resolution from m7 to m6 at fixed box size increases the enhancement by $\approx 10$–$15$ per cent at $r_{\rm 3D} < 100~\mathrm{kpc}$ and extends it out to $r_{\rm 3D} \approx 200~\mathrm{kpc}$.}
    \label{fig:resolution}
\end{figure}

In the majority of the analysis carried out in this work, we used the largest \colibre{} simulation at m6 resolution, which has a volume of $200^3~\text{cMpc}^3$. In Fig.~\ref{fig:resolution} we show the sSFR enhancement at m6 resolution (orange) in this fiducial simulation (solid) and in a cosmological volume of $100^3~\text{cMpc}^3$ (dash-dotted), as well as at m7 resolution (red) in cosmological volumes of $200^3$ (dash-dotted) and $400^3~\text{cMpc}^3$ (solid). For all four simulations, the construction of samples of interacting and control galaxies is done as described in Section~\ref{sec:methods_samples_3d}. We consider only interacting galaxies with stellar mass $M_* > 10^{10}~\mathrm{M_\odot}$.

The similarity between the curves for both simulations at m6 resolution and both simulations at m7 resolution indicates that the samples are converged with respect to box size. The sSFR enhancement at m7 resolution follows a similar trend as at m6, but with a lower normalisation ($Q\approx1.45$ for m7 versus \ $Q\approx1.7$ for m6 in the lowest-separation bin) and a smaller maximum distance over which the enhancement is observed ($r_{\rm 3D} < 130~\text{kpc}$ at m7 versus $r_{\rm 3D} < 200~\text{kpc}$ at m6). At fixed separation, the higher resolution results in a systematically larger enhancement, which is likely because at higher resolution, the central regions of galaxies where the enhancement is stronger (Fig.~\ref{fig:aperture}) are better resolved. This suggests that the discrepancy we find between \colibre{} and the SDSS observational data in Fig.~\ref{fig:sdss} can be alleviated by increasing the numerical resolution of the simulations. 

Our results are in qualitative agreement with \citet{Patton2020}, who performed convergence tests using the \tng{} simulations and found that enhancement is generally stronger at higher resolution. Lastly, \citet{Chaikin2026} showed that normalisation of the star-forming main sequence in \colibre{} is only broadly converged with resolution at low redshifts. Since we are interested in the relative differences between the sSFRs of interacting and control galaxies, these minor deviations are unlikely to affect our convergence test.

\begin{figure}
    \centering
    \includegraphics[width=0.5\textwidth]{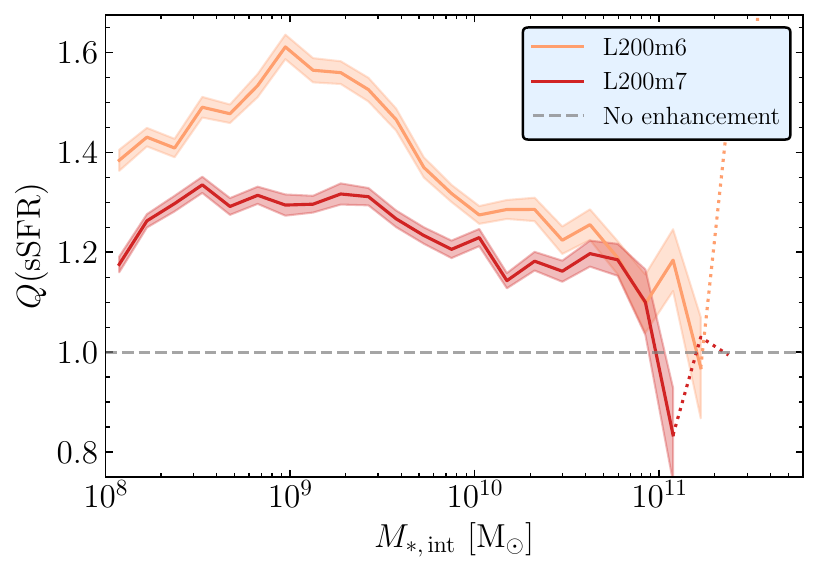}
    \caption{The mean sSFR enhancement of interacting galaxies relative to their controls in the L200m6 (orange) and L200m7 (red) \colibre{} simulations. The results are shown for interacting galaxies with separations $r_\text{3D}<50~\mathrm{kpc}$ and their controls. The sSFR enhancement at m7 resolution is present across the whole stellar mass range, but is significantly weaker than at m6 resolution, and shows no peak at $M_{*,\rm int}\sim10^9~\mathrm{M_\odot}$.}
    \label{fig:resolution2}
\end{figure}

Fig.~\ref{fig:resolution2} shows the difference between m6 and m7 resolution as a function of stellar mass, both for a box size of (200 cMpc)$^3$. The results are shown for interacting galaxies with a separation $r_\mathrm{3D}<50$~kpc in the fiducial galaxy samples. Similarly to Fig.~\ref{fig:resolution}, the sSFR enhancement at m7 resolution follows a similar trend as m6 but with a lower normalisation. The peak in $Q$ at m6 resolution for $M_{*,\text{int}}\approx10^9~\mathrm{M}_\odot$ is less pronounced at m7 resolution.

\subsection{Comparison with previous work}

Both this work and \citet{Patton2020} follow the procedure of \citet{Patton2016} that involves matching interacting galaxies to controls based on stellar mass, local density, and isolation. \citet{Patton2020} studied the sSFR enhancement in galaxy mergers in \eagle{} (the L0100N1504 simulation), \illustris{} (Illustris-1), and \tng{} (TNG100-1). They found that the sSFRs of interacting galaxies are enhanced relative to sSFRs of their controls at separations of up to $\approx150~\mathrm{kpc}$ in \eagle{} and \illustris, and up to $\approx200~\mathrm{kpc}$ in \tng. Similarly to \tng, we find that the sSFRs of galaxies in \colibre{} are enhanced out to $\approx200~\mathrm{kpc}$, albeit with a lower overall amplitude. At small separations ($r_{\rm 3D}<20~\mathrm{kpc}$), interacting galaxies and their controls with redshift $0 < z < 1$, stellar mass $10^{10} < M_*/\mathrm{M}_\odot < 10^{12}$, and mass ratio $\geq 1/10$ reach $Q \approx 2.0$ in \tng{}, $Q \approx 1.6$ in \eagle{}, and $Q \approx 1.7$ in \illustris{}. With a similar approach, except for our smaller redshift range of $0 < z < 0.2$, we find an sSFR enhancement in \colibre{} of up to $Q \approx 1.7$ (Fig.~\ref{fig:q_r}), which is similar to \eagle{} and \illustris, but clearly lower than in \tng. 

Despite the similar methodology, some important differences between our study and \citet{Patton2020} remain, which can account for, or further exacerbate, the differences in $Q$. First, \citet{Patton2020} measure sSFR within $R_{1/2}$ of each galaxy, while our work uses a fixed 10~kpc aperture (which is larger than $R_{1/2}$ for most \colibre{} galaxies at $M_* \lesssim 10^{11}~\mathrm{M_\odot}$). As sSFR enhancement due to mergers is expected to peak more strongly in the central parts of a galaxy \citep[e.g.][]{Yuan2012,2013MNRAS.435.3627E,2019MNRAS.482L..55T}, the aperture in which one defines the sSFR can have a significant impact on the measured enhancement. Indeed, Fig.~\ref{fig:aperture} shows that the measured enhancement in \colibre{} is markedly larger when using a smaller aperture, reaching $\approx 2.15$ for the sSFR defined within $R_{1/2}$. 

A second important difference is the inclusion of non-star-forming galaxies in the study of \citet{Patton2020}. Quenched galaxies possess less gas that can be available for star formation, so including quenched systems in our analysis would suppress the sSFR enhancement, particularly at the high-mass end ($M_* \gtrsim 10^{10}~\mathrm{M}_\odot$), where the quenched fraction in \colibre{} becomes significant (see fig.~18 in \citealt{Schaye2025}). Indeed, when we include quenched galaxies in our samples, the sSFR enhancement at low separations ($r_{\rm{3D}} < 10~\rm{kpc}$) drops from $Q\approx1.7$ to $\approx1.5$. We repeated our analysis following more closely the selection criteria of \citet{Patton2020}, where we (i) take redshifts $0\leq z\leq1$ instead of our fiducial choice of $0\leq z\leq0.2$, (ii) compute sSFRs within $R_{1/2}$ instead of 10-kpc apertures, and (iii) do not require interacting and control galaxies to be star-forming. Overall, the effects of these changes offset each other, and we find only minor differences compared to our fiducial selection. The sSFR enhancement reaches $Q\approx1.8$ at the lowest separations instead of $Q\approx1.7$, while becoming $5$--$10$ per cent less pronounced at $r_{\rm 3D}\gtrsim 30$~kpc (not shown).

Third, the \colibre, \eagle, \illustris, and \tng{} simulations have different implementations of galaxy formation processes, among which is the prescription for star formation. While all three other simulations require gas to be above a relatively low minimum density ($n_{\rm H}\sim0.1~\rm cm^{-3}$) and below some maximum temperature (or equivalent) to be star-forming, \colibre{} requires gas to be gravitationally unstable, which occurs when the gravitational binding energy of a gas cloud exceeds its kinetic energy due to thermal and turbulent motions (see Equation~\ref{eq: instability_crit}). Possibly, mergers in \colibre{} show somewhat smaller sSFR enhancement due to the explicit dependence of the star formation criterion on gas turbulence, which can be increased due to galaxy tidal interactions, and because higher densities than the thresholds used by the other simulations may be needed for instability.

We note that, although simulations such as \tng, \eagle{}, and \simba{} predict comparable galaxy stellar mass functions at $z \approx 0$, the haloes within which galaxies reside can possess significantly different gas fractions \citep{2024MNRAS.532.3417W}. This, in turn, can affect the amount of gas available for star formation and thus the amplitude of interaction-induced sSFR enhancement. We note, though, that the \colibre{} simulations reproduce the observed relations between stellar mass and H~\textsc{i} and H$_2$ gas fractions \citep{Schaye2025}.

Lastly, a recent study by \citet{Schechter2025} investigated the enhancement of sSFR (and black hole accretion rates) associated with galaxy mergers in \tng{} (TNG50-1) for redshifts $0.2 < z < 3$ and pair mass ratios $> 0.1$. Using simulation merger trees, they defined merging galaxies at a given redshift as those within 250~Myr of the merging event, and compared them to non-merging controls matched in stellar mass and redshift. At $z\approx0.2$, they found that the sSFR enhancement peaks at $10^{9} < M_*/\mathrm{M}_\odot < 10^{9.5}$ with $Q \approx 2$, decreasing to $\approx 1.8$ for lower stellar masses ($M_* \sim 10^{8}~\mathrm{M}_\odot$) and to $\approx 1.5$ for higher stellar masses ($M_* \sim 10^{10}~\mathrm{M}_\odot$), before approaching unity (albeit with large scatter) towards even higher masses ($M_* \sim 10^{11}~\mathrm{M}_\odot$). These findings are broadly consistent with the trends found in our Fig.~\ref{fig:q_mstar}, despite the methodological differences between our work and \citet{Schechter2025}.

\subsection{Caveats}

Following the matching procedure outlined in Section~\ref{sec:results_sdss}, we matched \colibre{} galaxies to SDSS galaxies based on stellar mass, mass ratio in galaxy pairs, and redshift in order to reduce biases between the SDSS and \colibre{} samples. However, because SDSS is flux-limited, the SDSS sample used in this work becomes incomplete at the low-mass end \citep[see fig.~10 in][]{Mendel2014}, especially for the closest companions of low-mass interacting galaxies with mass ratios less than $1$. Although we match \colibre{} and SDSS interacting galaxies based on pair mass ratios to alleviate the effect of incompleteness in SDSS, some residual systematics may still remain. 

Another possible shortcoming of using simulations to predict interaction-induced sSFR enhancement is a numerical effect in which a halo finder can incorrectly assign a fraction of the stellar mass and SFR of less massive galaxies in interacting pairs to more massive ones when their separation becomes very small (i.e. when the galaxies overlap). This is commonly referred to as numerical stripping. \citet{Patton2020}, whose results are based on the SUBFIND halo finder \citep{Springel2001}, attempted to correct for this effect, finding that the stellar masses of low-mass galaxies in interacting pairs can be reduced by up to $\approx 20$~per cent at separations $<10~\text{kpc}$ in some cases (note, however, that a significant fraction of this stellar mass reduction is due to physical stripping from tidal effects, which are well-resolved in the simulation).

In Appendix~\ref{app:numerical_stripping}, we repeat their test for \colibre{}, which uses the HBT-HERONS halo finder, finding a similar suppression of stellar masses at small separations. Since \citet{Patton2020} found that only a small fraction of galaxies ($\approx 1.2$~per cent in TNG100-1) are affected by numerical stripping, and because HBT-HERONS is expected to be less prone to this effect due to its history-based approach to halo identification, we do not apply any corrections to galaxy stellar masses in \colibre.

\section{Conclusion}

In this work, we have investigated the effect of galaxy interactions on star formation activity at redshift $z\approx 0$, using the \colibre{} simulations of galaxy formation \citep{Schaye2025, Chaikin2025a}. To construct the samples of interacting galaxies and isolated control galaxies, we used the largest \colibre{} simulation at m6 resolution (gas particle mass $m_{\rm gas}=1.8 \times 10^{6}~\mathrm{M_\odot}$), with a volume of $(200~\mathrm{cMpc})^3$. We restricted the samples to star-forming galaxies only ($\text{sSFR} > 10^{-11}~\mathrm{yr}^{-1}$) and matched interacting galaxies to isolated control galaxies by stellar mass, redshift, and environment, using the method from \citet{Patton2016}. Our main findings regarding the interaction-induced sSFR enhancement in \colibre{} (Section \ref{sec:results_colibre}) are:

\label{sec:conclusion}
\begin{itemize}
   \item The sSFR of interacting galaxies with stellar masses $10^{10}<M_{\rm *, int}/\mathrm{M}_\odot<10^{12}$ is enhanced with respect to their controls for pair separations $r_{\rm 3D}\lesssim 200$~kpc (Fig.~\ref{fig:q_r}). The mean sSFR enhancement, $Q$, increases monotonically with decreasing separation, reaching $Q\approx1.7$ in the smallest separation bin ($r_{\rm 3D}=5~\mathrm{kpc}$).
   \medskip
   \item The mean sSFR enhancement at separations $r_{\rm 3D} < 50~\mathrm{kpc}$ is $> 1.2$ for interacting galaxies with $M_{*, \rm int} \lesssim 10^{10.5}~\mathrm{M}_\odot$. It shows a non-monotonic dependence on $M_{*, \rm int}$, peaking at $Q \approx 1.6$ for $M_{*, \rm int} \approx  10^{9}~\mathrm{M}_\odot$. The sSFRs of more massive galaxies ($M_{*, \rm int} \gtrsim 10^{10.5}~\mathrm{M}_\odot$) are less enhanced, with the mean enhancement decreasing with increasing $M_{*, \rm int}$ and approaching $Q \approx 1$ by $M_{*, \rm int} \sim 10^{11}~\mathrm{M}_\odot$ (Fig.~\ref{fig:q_mstar}).
   \medskip
   \item At fixed pair separation, the enhancement generally becomes stronger for lower-mass interacting galaxies (Fig.~\ref{fig:q_r_mstar}), while at fixed interacting galaxy stellar mass, the enhancement increases monotonically with decreasing separation (Fig.~\ref{fig:q_mstar_r}).
   \medskip
   \item The mean enhancement of the molecular gas fraction peaks at $M_{*,\rm int}\sim10^{9}~\mathrm{M}_\odot$ (Fig. \ref{fig:gas_fraction_vs_mstar}), with the shape and amplitude of this peak being similar to those of the sSFR enhancement in Fig.~\ref{fig:q_mstar}. The mean atomic gas fraction is enhanced for galaxies with $M_{\rm *,int}\lesssim10^{9.5}~\mathrm{M}_\odot$, with the mean enhancement rising monotonically with decreasing stellar mass in this mass range.
   \medskip
   \item The mean sSFR enhancement of interacting galaxies in the $1/10 < M_{*,\mathrm{cc}}/M_{*,\mathrm{int}} < 1/3$, $1/3 < M_{*,\mathrm{cc}}/M_{*,\mathrm{int}} < 3$, and $3 < M_{*,\mathrm{cc}}/M_{*,\mathrm{int}} < 10$ mass ratio bins is present at $10^8 < M_{*,\rm int}/\mathrm{M}_\odot < 10^{9.5}$, with slightly stronger enhancement in the higher mass ratio bins (Fig.~\ref{fig:mass_ratio}). At higher masses, $10^{9.5}<M_{*, \rm int}/\mathrm{M}_\odot < 10^{11}$, the enhancement remains visible in the low and intermediate mass ratio bins, but decreases to unity in the high mass ratio bin.
   \medskip
   \item Interaction-induced sSFR enhancement is strongly dependent on the radius of the aperture within which the sSFR is measured. Our fiducial 3D aperture of 10~kpc results in $Q\approx 1.7$ in the smallest separation bin ($r_{\rm 3D} < 10~\mathrm{kpc}$), which increases to $Q\approx1.95$ for the $3$~kpc aperture, and further to $Q\approx2.15$ and $Q\approx2.6$ for sSFRs measured within $R_{1/2}$ and 1~kpc 3D apertures, respectively (Fig.~\ref{fig:aperture}).
   \medskip
   \item The contribution of the SFR excess induced by pre-merger galaxy interactions (for interacting galaxies with stellar masses $10^8 < M_{*,\rm int}/\mathrm{M}_\odot < 10^{12}$, mass ratios $0.1 < M_{*,\rm cc}/M_{*,\rm int} < 10$, and across all separations) to the cosmic SFR density at $z\approx0$ is $\approx2.1$~per~cent (\S\ref{sec:results_sfrd}).

\end{itemize}

Having studied how the sSFR enhancement in \colibre{} depends on pair separation, galaxy stellar mass, pair mass ratio, and the radius within which the sSFR is measured, we proceeded to compare \colibre{} predictions with observational data from SDSS DR7 (Section \ref{sec:results_sdss}). We adopted the methodology of \citet{Patton2013}, who constructed the SDSS galaxy sample in which interacting and control galaxies are matched based on stellar mass, redshift, and environment. To maximise consistency in the comparison with their data, we measured sSFR in \colibre{} within projected (i.e. 2D) apertures with a size corresponding to the SDSS fibre (a few kpc at $z\approx 0$), built a \colibre{} sample of interacting galaxies based on projected separation, $r_{\rm proj}$, distance in velocity space along the LOS, and switched to using statistical controls as opposed to individual control galaxies. 

We found that \colibre{} produces sSFR distributions of interacting and statistical control galaxies that are very similar to those in SDSS (Fig.~\ref{fig:pdfs}). The simulation also reproduces the shape of the mean sSFR enhancement as a function of projected separation seen in SDSS, with both \colibre{} and SDSS showing an enhancement out to $r_{\rm proj}\approx 150~\mathrm{kpc}$ (Fig.~\ref{fig:sdss}). However, \colibre{} predicts a lower normalisation than SDSS. The differences are largest in the lowest separation bin, where \colibre{} predicts $Q \approx 1.6$ compared to $Q \approx 2.65$ in SDSS. While this difference in the means is small compared to the distributions of the sSFRs for both interacting and control galaxies (Fig.~\ref{fig:pdfs}), it is statistically significant. In Section~\ref{sec:discussion_resolution}, we showed that the sSFR enhancement is converged with simulation volume but becomes stronger at higher numerical resolution. Therefore, increasing the \colibre{} resolution from our fiducial m6 resolution to m5 could potentially reduce some of the discrepancy with SDSS.

Overall, this work highlights that galaxy interactions tend to enhance galaxy SFRs, with the stellar masses of galaxies for which the mean interaction-driven SFR enhancement is significant ranging from $10^{8}$ to $10^{11}~\mathrm{M_\odot}$ and having pair separations out to $\approx 200$~kpc. Beyond SFR enhancements, interactions can also affect the gas distribution and metallicity, which will be analysed in Serrano Rodriguez et al. (in preparation). Future work could investigate post-mergers and galaxy interactions at high redshift in \colibre{}.

\section*{Acknowledgments}

We thank the SDSS collaborations for making their data available. This work used the DiRAC@Durham facility managed by the Institute for Computational Cosmology on behalf of the STFC DiRAC HPC Facility (www.dirac.ac.uk). The equipment was funded by BEIS capital funding via STFC capital grants ST/K00042X/1, ST/P002293/1, ST/R002371/1 and ST/S002502/1, Durham University and STFC operations grant ST/R000832/1. DiRAC is part of the National e-Infrastructure. This project has received funding from the Netherlands Organization for Scientific Research (NWO) through research programme Athena 184.034.002. ABL acknowledges support by the Italian Ministry for Universities (MUR) program `Dipartimenti di Eccellenza 2023-2027' within the Centro Bicocca di Cosmologia Quantitativa (BiCoQ), and support by UNIMIB's Fondo Di Ateneo Quota Competitiva (project 2024-ATEQC-0050). EC acknowledges support from STFC consolidated grant ST/X001075/1. 

\section*{Data Availability}

The data underlying this article will be shared on reasonable request to the corresponding author. The public version of the \textsc{Swift} simulation code is available at \href{http://www.swiftsim.com}{www.swiftsim.com}. The \textsc{Swift} modules related to the \colibre{} galaxy formation model will be integrated into the public version after the public release of \colibre. The \textsc{chimes} astrochemistry code is publicly available at \href{https://richings.bitbucket.io/chimes/home.html}{https://richings.bitbucket.io/chimes/home.html}. The HBT-HERONS  halo finder is available at \url{https://hbt-herons.strw.leidenuniv.nl/}.

\bibliographystyle{mnras}
\bibliography{bibliography}

\begin{thebibliography}{}
\makeatletter
\relax
\def\mn@urlcharsother{\let\do\@makeother \do\$\do\&\do\#\do\^\do\_\do\%\do\~}
\def\mn@doi{\begingroup\mn@urlcharsother \@ifnextchar [ {\mn@doi@} {\mn@doi@[]}}
\def\mn@doi@[#1]#2{\def\@tempa{#1}\ifx\@tempa\@empty \href {http://dx.doi.org/#2} {doi:#2}\else \href {http://dx.doi.org/#2} {#1}\fi \endgroup}
\def\mn@eprint#1#2{\mn@eprint@#1:#2::\@nil}
\def\mn@eprint@arXiv#1{\href {http://arxiv.org/abs/#1} {{\tt arXiv:#1}}}
\def\mn@eprint@dblp#1{\href {http://dblp.uni-trier.de/rec/bibtex/#1.xml} {dblp:#1}}
\def\mn@eprint@#1:#2:#3:#4\@nil{\def\@tempa {#1}\def\@tempb {#2}\def\@tempc {#3}\ifx \@tempc \@empty \let \@tempc \@tempb \let \@tempb \@tempa \fi \ifx \@tempb \@empty \def\@tempb {arXiv}\fi \@ifundefined {mn@eprint@\@tempb}{\@tempb:\@tempc}{\expandafter \expandafter \csname mn@eprint@\@tempb\endcsname \expandafter{\@tempc}}}

\bibitem[\protect\citeauthoryear{{Abadi}, {Moore}  \& {Bower}}{{Abadi} et~al.}{1999}]{1999MNRAS.308..947A}
{Abadi} M.~G.,  {Moore} B.,   {Bower} R.~G.,  1999, \mn@doi [\mnras] {10.1046/j.1365-8711.1999.02715.x}, \href {https://ui.adsabs.harvard.edu/abs/1999MNRAS.308..947A} {308, 947}

\bibitem[\protect\citeauthoryear{{Abazajian} et~al.,}{{Abazajian} et~al.}{2009}]{Abazajian2009}
{Abazajian} K.~N.,  et~al., 2009, \mn@doi [\apjs] {10.1088/0067-0049/182/2/543}, \href {https://ui.adsabs.harvard.edu/abs/2009ApJS..182..543A} {182, 543}

\bibitem[\protect\citeauthoryear{{Abbott} et~al.,}{{Abbott} et~al.}{2022}]{2022PhRvD.105b3520A}
{Abbott} T.~M.~C.,  et~al., 2022, \mn@doi [\prd] {10.1103/PhysRevD.105.023520}, \href {https://ui.adsabs.harvard.edu/abs/2022PhRvD.105b3520A} {105, 023520}

\bibitem[\protect\citeauthoryear{{Bah{\'e}} et~al.,}{{Bah{\'e}} et~al.}{2019}]{Bahe2019}
{Bah{\'e}} Y.~M.,  et~al., 2019, \mn@doi [\mnras] {10.1093/mnras/stz361}, \href {https://ui.adsabs.harvard.edu/abs/2019MNRAS.485.2287B} {485, 2287}

\bibitem[\protect\citeauthoryear{{Bah{\'e}} et~al.,}{{Bah{\'e}} et~al.}{2022}]{2022MNRAS.516..167B}
{Bah{\'e}} Y.~M.,  et~al., 2022, \mn@doi [\mnras] {10.1093/mnras/stac1339}, \href {https://ui.adsabs.harvard.edu/abs/2022MNRAS.516..167B} {516, 167}

\bibitem[\protect\citeauthoryear{{Bamford} et~al.,}{{Bamford} et~al.}{2009}]{2009MNRAS.393.1324B}
{Bamford} S.~P.,  et~al., 2009, \mn@doi [\mnras] {10.1111/j.1365-2966.2008.14252.x}, \href {https://ui.adsabs.harvard.edu/abs/2009MNRAS.393.1324B} {393, 1324}

\bibitem[\protect\citeauthoryear{{Barnes} \& {Hernquist}}{{Barnes} \& {Hernquist}}{1992}]{1992ARA&A..30..705B}
{Barnes} J.~E.,  {Hernquist} L.,  1992, \mn@doi [\araa] {10.1146/annurev.aa.30.090192.003421}, \href {https://ui.adsabs.harvard.edu/abs/1992ARA&A..30..705B} {30, 705}

\bibitem[\protect\citeauthoryear{{Barnes} \& {Hernquist}}{{Barnes} \& {Hernquist}}{1996}]{Barnes1996}
{Barnes} J.~E.,  {Hernquist} L.,  1996, \mn@doi [\apj] {10.1086/177957}, \href {https://ui.adsabs.harvard.edu/abs/1996ApJ...471..115B} {471, 115}

\bibitem[\protect\citeauthoryear{{Barrows}, {Comerford}, {Stern}  \& {Assef}}{{Barrows} et~al.}{2023}]{2023ApJ...951...92B}
{Barrows} R.~S.,  {Comerford} J.~M.,  {Stern} D.,   {Assef} R.~J.,  2023, \mn@doi [\apj] {10.3847/1538-4357/acd2d3}, \href {https://ui.adsabs.harvard.edu/abs/2023ApJ...951...92B} {951, 92}

\bibitem[\protect\citeauthoryear{{Barton}, {Geller}  \& {Kenyon}}{{Barton} et~al.}{2000}]{2000ApJ...530..660B}
{Barton} E.~J.,  {Geller} M.~J.,   {Kenyon} S.~J.,  2000, \mn@doi [\apj] {10.1086/308392}, \href {https://ui.adsabs.harvard.edu/abs/2000ApJ...530..660B} {530, 660}

\bibitem[\protect\citeauthoryear{{Ben{\'\i}tez-Llambay} et~al.,}{{Ben{\'\i}tez-Llambay} et~al.}{2026}]{2025arXiv250925309B}
{Ben{\'\i}tez-Llambay} A.,  et~al., 2026, \mn@doi [\mnras] {10.1093/mnras/stag268}, \href {https://ui.adsabs.harvard.edu/abs/2026MNRAS.546ag268B} {546, stag268}

\bibitem[\protect\citeauthoryear{{Bigiel}, {Leroy}, {Walter}, {Brinks}, {de Blok}, {Madore}  \& {Thornley}}{{Bigiel} et~al.}{2008}]{2008AJ....136.2846B}
{Bigiel} F.,  {Leroy} A.,  {Walter} F.,  {Brinks} E.,  {de Blok} W.~J.~G.,  {Madore} B.,   {Thornley} M.~D.,  2008, \mn@doi [\aj] {10.1088/0004-6256/136/6/2846}, \href {https://ui.adsabs.harvard.edu/abs/2008AJ....136.2846B} {136, 2846}

\bibitem[\protect\citeauthoryear{{Bluck}, {Conselice}, {Bouwens}, {Daddi}, {Dickinson}, {Papovich}  \& {Yan}}{{Bluck} et~al.}{2009}]{2009MNRAS.394L..51B}
{Bluck} A. F.~L.,  {Conselice} C.~J.,  {Bouwens} R.~J.,  {Daddi} E.,  {Dickinson} M.,  {Papovich} C.,   {Yan} H.,  2009, \mn@doi [\mnras] {10.1111/j.1745-3933.2008.00608.x}, \href {https://ui.adsabs.harvard.edu/abs/2009MNRAS.394L..51B} {394, L51}

\bibitem[\protect\citeauthoryear{{Booth} \& {Schaye}}{{Booth} \& {Schaye}}{2009}]{Booth2009}
{Booth} C.~M.,  {Schaye} J.,  2009, \mn@doi [\mnras] {10.1111/j.1365-2966.2009.15043.x}, \href {https://ui.adsabs.harvard.edu/abs/2009MNRAS.398...53B} {398, 53}

\bibitem[\protect\citeauthoryear{{Borrow}, {Schaller}, {Bower}  \& {Schaye}}{{Borrow} et~al.}{2022}]{2022MNRAS.511.2367B}
{Borrow} J.,  {Schaller} M.,  {Bower} R.~G.,   {Schaye} J.,  2022, \mn@doi [\mnras] {10.1093/mnras/stab3166}, \href {https://ui.adsabs.harvard.edu/abs/2022MNRAS.511.2367B} {511, 2367}

\bibitem[\protect\citeauthoryear{{Brinchmann}, {Charlot}, {White}, {Tremonti}, {Kauffmann}, {Heckman}  \& {Brinkmann}}{{Brinchmann} et~al.}{2004}]{Brinchmann2004}
{Brinchmann} J.,  {Charlot} S.,  {White} S.~D.~M.,  {Tremonti} C.,  {Kauffmann} G.,  {Heckman} T.,   {Brinkmann} J.,  2004, \mn@doi [\mnras] {10.1111/j.1365-2966.2004.07881.x}, \href {https://ui.adsabs.harvard.edu/abs/2004MNRAS.351.1151B} {351, 1151}

\bibitem[\protect\citeauthoryear{{Cao} et~al.,}{{Cao} et~al.}{2016}]{Cao2016}
{Cao} C.,  et~al., 2016, \mn@doi [\apjs] {10.3847/0067-0049/222/2/16}, \href {https://ui.adsabs.harvard.edu/abs/2016ApJS..222...16C} {222, 16}

\bibitem[\protect\citeauthoryear{{Ceccarelli}, {Padilla}  \& {Lambas}}{{Ceccarelli} et~al.}{2008}]{2008MNRAS.390L...9C}
{Ceccarelli} L.,  {Padilla} N.,   {Lambas} D.~G.,  2008, \mn@doi [\mnras] {10.1111/j.1745-3933.2008.00520.x}, \href {https://ui.adsabs.harvard.edu/abs/2008MNRAS.390L...9C} {390, L9}

\bibitem[\protect\citeauthoryear{{Chabrier}}{{Chabrier}}{2003}]{2003PASP..115..763C}
{Chabrier} G.,  2003, \mn@doi [\pasp] {10.1086/376392}, \href {https://ui.adsabs.harvard.edu/abs/2003PASP..115..763C} {115, 763}

\bibitem[\protect\citeauthoryear{{Chaikin}, {Schaye}, {Schaller}, {Ben{\'\i}tez-Llambay}, {Nobels}  \& {Ploeckinger}}{{Chaikin} et~al.}{2023}]{2023MNRAS.523.3709C}
{Chaikin} E.,  {Schaye} J.,  {Schaller} M.,  {Ben{\'\i}tez-Llambay} A.,  {Nobels} F. S.~J.,   {Ploeckinger} S.,  2023, \mn@doi [\mnras] {10.1093/mnras/stad1626}, \href {https://ui.adsabs.harvard.edu/abs/2023MNRAS.523.3709C} {523, 3709}

\bibitem[\protect\citeauthoryear{{Chaikin} et~al.,}{{Chaikin} et~al.}{2026a}]{Chaikin2025a}
{Chaikin} E.,  et~al., 2026a, \mn@doi [\mnras] {10.1093/mnras/stag300}, \href {https://ui.adsabs.harvard.edu/abs/2026MNRAS.548ag300C} {548, stag300}

\bibitem[\protect\citeauthoryear{{Chaikin} et~al.,}{{Chaikin} et~al.}{2026b}]{Chaikin2026}
{Chaikin} E.,  et~al., 2026b, \mn@doi [\mnras] {10.1093/mnras/stag740}, \href {https://ui.adsabs.harvard.edu/abs/2026MNRAS.548ag740C} {548, stag740}

\bibitem[\protect\citeauthoryear{{Chandro-G{\'o}mez} et~al.,}{{Chandro-G{\'o}mez} et~al.}{2025}]{Chandro-gomez2025}
{Chandro-G{\'o}mez} {\'A}.,  et~al., 2025, \mn@doi [\mnras] {10.1093/mnras/staf519}, \href {https://ui.adsabs.harvard.edu/abs/2025MNRAS.539..776C} {539, 776}

\bibitem[\protect\citeauthoryear{{Correa} et~al.,}{{Correa} et~al.}{2026}]{2026MNRAS.tmp..607C}
{Correa} C.~A.,  et~al., 2026, \mn@doi [\mnras] {10.1093/mnras/stag645}, \href {https://ui.adsabs.harvard.edu/abs/2026MNRAS.tmp..607C} {}

\bibitem[\protect\citeauthoryear{{Dalla Vecchia} \& {Schaye}}{{Dalla Vecchia} \& {Schaye}}{2012}]{2012MNRAS.426..140D}
{Dalla Vecchia} C.,  {Schaye} J.,  2012, \mn@doi [\mnras] {10.1111/j.1365-2966.2012.21704.x}, \href {https://ui.adsabs.harvard.edu/abs/2012MNRAS.426..140D} {426, 140}

\bibitem[\protect\citeauthoryear{{Dav{\'e}}, {Angl{\'e}s-Alc{\'a}zar}, {Narayanan}, {Li}, {Rafieferantsoa}  \& {Appleby}}{{Dav{\'e}} et~al.}{2019}]{Dave2019}
{Dav{\'e}} R.,  {Angl{\'e}s-Alc{\'a}zar} D.,  {Narayanan} D.,  {Li} Q.,  {Rafieferantsoa} M.~H.,   {Appleby} S.,  2019, \mn@doi [\mnras] {10.1093/mnras/stz937}, \href {https://ui.adsabs.harvard.edu/abs/2019MNRAS.486.2827D} {486, 2827}

\bibitem[\protect\citeauthoryear{{Di Matteo}, {Springel}  \& {Hernquist}}{{Di Matteo} et~al.}{2005}]{DiMatteo2005}
{Di Matteo} T.,  {Springel} V.,   {Hernquist} L.,  2005, \mn@doi [\nat] {10.1038/nature03335}, \href {https://ui.adsabs.harvard.edu/abs/2005Natur.433..604D} {433, 604}

\bibitem[\protect\citeauthoryear{{Driver} et~al.,}{{Driver} et~al.}{2022}]{Driver2022}
{Driver} S.~P.,  et~al., 2022, \mn@doi [\mnras] {10.1093/mnras/stac472}, \href {https://ui.adsabs.harvard.edu/abs/2022MNRAS.513..439D} {513, 439}

\bibitem[\protect\citeauthoryear{{Dubois} et~al.,}{{Dubois} et~al.}{2014}]{Dubois2014}
{Dubois} Y.,  et~al., 2014, \mn@doi [\mnras] {10.1093/mnras/stu1227}, \href {https://ui.adsabs.harvard.edu/abs/2014MNRAS.444.1453D} {444, 1453}

\bibitem[\protect\citeauthoryear{{Ellison}, {Patton}, {Simard}  \& {McConnachie}}{{Ellison} et~al.}{2008}]{Ellison2008}
{Ellison} S.~L.,  {Patton} D.~R.,  {Simard} L.,   {McConnachie} A.~W.,  2008, \mn@doi [\aj] {10.1088/0004-6256/135/5/1877}, \href {https://ui.adsabs.harvard.edu/abs/2008AJ....135.1877E} {135, 1877}

\bibitem[\protect\citeauthoryear{{Ellison}, {Patton}, {Simard}, {McConnachie}, {Baldry}  \& {Mendel}}{{Ellison} et~al.}{2010}]{2010MNRAS.407.1514E}
{Ellison} S.~L.,  {Patton} D.~R.,  {Simard} L.,  {McConnachie} A.~W.,  {Baldry} I.~K.,   {Mendel} J.~T.,  2010, \mn@doi [\mnras] {10.1111/j.1365-2966.2010.17076.x}, \href {https://ui.adsabs.harvard.edu/abs/2010MNRAS.407.1514E} {407, 1514}

\bibitem[\protect\citeauthoryear{{Ellison}, {Mendel}, {Patton}  \& {Scudder}}{{Ellison} et~al.}{2013}]{2013MNRAS.435.3627E}
{Ellison} S.~L.,  {Mendel} J.~T.,  {Patton} D.~R.,   {Scudder} J.~M.,  2013, \mn@doi [\mnras] {10.1093/mnras/stt1562}, \href {https://ui.adsabs.harvard.edu/abs/2013MNRAS.435.3627E} {435, 3627}

\bibitem[\protect\citeauthoryear{{Ellison}, {Ferreira}, {Wild}, {Wilkinson}, {Rowlands}  \& {Patton}}{{Ellison} et~al.}{2024}]{Ellison2024}
{Ellison} S.,  {Ferreira} L.,  {Wild} V.,  {Wilkinson} S.,  {Rowlands} K.,   {Patton} D.~R.,  2024, \mn@doi [The Open Journal of Astrophysics] {10.33232/001c.127779}, \href {https://ui.adsabs.harvard.edu/abs/2024OJAp....7E.121E} {7, 121}

\bibitem[\protect\citeauthoryear{{Faria}, {Patton}, {Courteau}, {Ellison}  \& {Brown}}{{Faria} et~al.}{2025}]{2025MNRAS.537..915F}
{Faria} L.,  {Patton} D.~R.,  {Courteau} S.,  {Ellison} S.,   {Brown} W.,  2025, \mn@doi [\mnras] {10.1093/mnras/staf124}, \href {https://ui.adsabs.harvard.edu/abs/2025MNRAS.537..915F} {537, 915}

\bibitem[\protect\citeauthoryear{{Forouhar Moreno}, {Helly}, {McGibbon}, {Schaye}, {Schaller}, {Han}, {Kugel}  \& {Bah{\'e}}}{{Forouhar Moreno} et~al.}{2025}]{Moreno2025}
{Forouhar Moreno} V.~J.,  {Helly} J.,  {McGibbon} R.,  {Schaye} J.,  {Schaller} M.,  {Han} J.,  {Kugel} R.,   {Bah{\'e}} Y.~M.,  2025, \mn@doi [\mnras] {10.1093/mnras/staf1478}, \href {https://ui.adsabs.harvard.edu/abs/2025MNRAS.tmp.1440M} {}

\bibitem[\protect\citeauthoryear{{Han}, {Cole}, {Frenk}, {Benitez-Llambay}  \& {Helly}}{{Han} et~al.}{2018}]{Han2018}
{Han} J.,  {Cole} S.,  {Frenk} C.~S.,  {Benitez-Llambay} A.,   {Helly} J.,  2018, \mn@doi [\mnras] {10.1093/mnras/stx2792}, \href {https://ui.adsabs.harvard.edu/abs/2018MNRAS.474..604H} {474, 604}

\bibitem[\protect\citeauthoryear{{Hani}, {Gosain}, {Ellison}, {Patton}  \& {Torrey}}{{Hani} et~al.}{2020}]{Hani2020}
{Hani} M.~H.,  {Gosain} H.,  {Ellison} S.~L.,  {Patton} D.~R.,   {Torrey} P.,  2020, \mn@doi [\mnras] {10.1093/mnras/staa459}, \href {https://ui.adsabs.harvard.edu/abs/2020MNRAS.493.3716H} {493, 3716}

\bibitem[\protect\citeauthoryear{{Hardwick}, {Cortese}, {Obreschkow}, {Catinella}  \& {Cook}}{{Hardwick} et~al.}{2022}]{Hardwick2022}
{Hardwick} J.~A.,  {Cortese} L.,  {Obreschkow} D.,  {Catinella} B.,   {Cook} R. H.~W.,  2022, \mn@doi [\mnras] {10.1093/mnras/stab3261}, \href {https://ui.adsabs.harvard.edu/abs/2022MNRAS.509.3751H} {509, 3751}

\bibitem[\protect\citeauthoryear{{He} et~al.,}{{He} et~al.}{2026}]{2026arXiv260403105H}
{He} F.,  et~al., 2026, \mn@doi [arXiv e-prints] {10.48550/arXiv.2604.03105}, \href {https://ui.adsabs.harvard.edu/abs/2026arXiv260403105H} {p. arXiv:2604.03105}

\bibitem[\protect\citeauthoryear{{Hu{\v{s}}ko}, {Lacey}  \& {Baugh}}{{Hu{\v{s}}ko} et~al.}{2022}]{2022MNRAS.509.5918H}
{Hu{\v{s}}ko} F.,  {Lacey} C.~G.,   {Baugh} C.~M.,  2022, \mn@doi [\mnras] {10.1093/mnras/stab3324}, \href {https://ui.adsabs.harvard.edu/abs/2022MNRAS.509.5918H} {509, 5918}

\bibitem[\protect\citeauthoryear{{Hu{\v{s}}ko} et~al.,}{{Hu{\v{s}}ko} et~al.}{2025}]{Husko2025}
{Hu{\v{s}}ko} F.,  et~al., 2025, \mn@doi [arXiv e-prints] {10.48550/arXiv.2509.05179}, \href {https://ui.adsabs.harvard.edu/abs/2025arXiv250905179H} {p. arXiv:2509.05179}

\bibitem[\protect\citeauthoryear{{Kauffmann} et~al.,}{{Kauffmann} et~al.}{2003}]{2003MNRAS.346.1055K}
{Kauffmann} G.,  et~al., 2003, \mn@doi [\mnras] {10.1111/j.1365-2966.2003.07154.x}, \href {https://ui.adsabs.harvard.edu/abs/2003MNRAS.346.1055K} {346, 1055}

\bibitem[\protect\citeauthoryear{{Kim}, {Wise}  \& {Abel}}{{Kim} et~al.}{2009}]{Kim2009}
{Kim} J.-h.,  {Wise} J.~H.,   {Abel} T.,  2009, \mn@doi [\apjl] {10.1088/0004-637X/694/2/L123}, \href {https://ui.adsabs.harvard.edu/abs/2009ApJ...694L.123K} {694, L123}

\bibitem[\protect\citeauthoryear{{Krumholz}, {McKee}  \& {Klein}}{{Krumholz} et~al.}{2006}]{Krumholz_et_al_2006}
{Krumholz} M.~R.,  {McKee} C.~F.,   {Klein} R.~I.,  2006, \mn@doi [\apj] {10.1086/498844}, \href {https://ui.adsabs.harvard.edu/abs/2006ApJ...638..369K} {638, 369}

\bibitem[\protect\citeauthoryear{{Lagos} et~al.,}{{Lagos} et~al.}{2025}]{Lagos2025}
{Lagos} C. d.~P.,  et~al., 2025, \mn@doi [arXiv e-prints] {10.48550/arXiv.2512.11309}, \href {https://ui.adsabs.harvard.edu/abs/2025arXiv251211309L} {p. arXiv:2512.11309}

\bibitem[\protect\citeauthoryear{{Larson} \& {Tinsley}}{{Larson} \& {Tinsley}}{1978}]{Larson1978}
{Larson} R.~B.,  {Tinsley} B.~M.,  1978, \mn@doi [\apj] {10.1086/155753}, \href {https://ui.adsabs.harvard.edu/abs/1978ApJ...219...46L} {219, 46}

\bibitem[\protect\citeauthoryear{{Lin} et~al.,}{{Lin} et~al.}{2007}]{2007ApJ...660L..51L}
{Lin} L.,  et~al., 2007, \mn@doi [\apjl] {10.1086/517919}, \href {https://ui.adsabs.harvard.edu/abs/2007ApJ...660L..51L} {660, L51}

\bibitem[\protect\citeauthoryear{{Lin} et~al.,}{{Lin} et~al.}{2010}]{2010ApJ...718.1158L}
{Lin} L.,  et~al., 2010, \mn@doi [\apj] {10.1088/0004-637X/718/2/1158}, \href {https://ui.adsabs.harvard.edu/abs/2010ApJ...718.1158L} {718, 1158}

\bibitem[\protect\citeauthoryear{{Lu} et~al.,}{{Lu} et~al.}{2026}]{lu2026}
{Lu} S.,  et~al., 2026, arXiv e-prints, \href {https://ui.adsabs.harvard.edu/abs/2026arXiv260506782L} {p. arXiv:2605.06782}

\bibitem[\protect\citeauthoryear{{Ludlow}, {Fall}, {Wilkinson}, {Schaye}  \& {Obreschkow}}{{Ludlow} et~al.}{2023}]{2023MNRAS.525.5614L}
{Ludlow} A.~D.,  {Fall} S.~M.,  {Wilkinson} M.~J.,  {Schaye} J.,   {Obreschkow} D.,  2023, \mn@doi [\mnras] {10.1093/mnras/stad2615}, \href {https://ui.adsabs.harvard.edu/abs/2023MNRAS.525.5614L} {525, 5614}

\bibitem[\protect\citeauthoryear{{Ludlow} et~al.,}{{Ludlow} et~al.}{2026}]{ludlow2026}
{Ludlow} A.~D.,  et~al., 2026, \mn@doi [arXiv e-prints] {10.48550/arXiv.2603.26200}, \href {https://ui.adsabs.harvard.edu/abs/2026arXiv260326200L} {p. arXiv:2603.26200}

\bibitem[\protect\citeauthoryear{{Martin}, {Kaviraj}, {Devriendt}, {Dubois}, {Laigle}  \& {Pichon}}{{Martin} et~al.}{2017}]{Martin2017}
{Martin} G.,  {Kaviraj} S.,  {Devriendt} J.~E.~G.,  {Dubois} Y.,  {Laigle} C.,   {Pichon} C.,  2017, \mn@doi [\mnras] {10.1093/mnrasl/slx136}, \href {https://ui.adsabs.harvard.edu/abs/2017MNRAS.472L..50M} {472, L50}

\bibitem[\protect\citeauthoryear{{McAlpine}, {Bower}, {Rosario}, {Crain}, {Schaye}  \& {Theuns}}{{McAlpine} et~al.}{2018}]{Mcalpine2018}
{McAlpine} S.,  {Bower} R.~G.,  {Rosario} D.~J.,  {Crain} R.~A.,  {Schaye} J.,   {Theuns} T.,  2018, \mn@doi [\mnras] {10.1093/mnras/sty2489}, \href {https://ui.adsabs.harvard.edu/abs/2018MNRAS.481.3118M} {481, 3118}

\bibitem[\protect\citeauthoryear{{McGibbon}, {Helly}, {Schaye}, {Schaller}  \& {Vandenbroucke}}{{McGibbon} et~al.}{2025}]{mcgibbon2025}
{McGibbon} R.,  {Helly} J.,  {Schaye} J.,  {Schaller} M.,   {Vandenbroucke} B.,  2025, \mn@doi [The Journal of Open Source Software] {10.21105/joss.08252}, \href {https://ui.adsabs.harvard.edu/abs/2025JOSS...10.8252M} {10, 8252}

\bibitem[\protect\citeauthoryear{{Mendel}, {Simard}, {Palmer}, {Ellison}  \& {Patton}}{{Mendel} et~al.}{2014}]{Mendel2014}
{Mendel} J.~T.,  {Simard} L.,  {Palmer} M.,  {Ellison} S.~L.,   {Patton} D.~R.,  2014, \mn@doi [\apjs] {10.1088/0067-0049/210/1/3}, \href {https://ui.adsabs.harvard.edu/abs/2014ApJS..210....3M} {210, 3}

\bibitem[\protect\citeauthoryear{{Nikolic}, {Cullen}  \& {Alexander}}{{Nikolic} et~al.}{2004}]{2004MNRAS.355..874N}
{Nikolic} B.,  {Cullen} H.,   {Alexander} P.,  2004, \mn@doi [\mnras] {10.1111/j.1365-2966.2004.08366.x}, \href {https://ui.adsabs.harvard.edu/abs/2004MNRAS.355..874N} {355, 874}

\bibitem[\protect\citeauthoryear{{Nobels}, {Schaye}, {Schaller}, {Ploeckinger}, {Chaikin}  \& {Richings}}{{Nobels} et~al.}{2024}]{2024MNRAS.532.3299N}
{Nobels} F. S.~J.,  {Schaye} J.,  {Schaller} M.,  {Ploeckinger} S.,  {Chaikin} E.,   {Richings} A.~J.,  2024, \mn@doi [\mnras] {10.1093/mnras/stae1390}, \href {https://ui.adsabs.harvard.edu/abs/2024MNRAS.532.3299N} {532, 3299}

\bibitem[\protect\citeauthoryear{{Omori} et~al.,}{{Omori} et~al.}{2026}]{2026arXiv260628590O}
{Omori} K.~C.,  et~al., 2026, \mn@doi [arXiv e-prints] {10.48550/arXiv.2606.28590}, \href {https://ui.adsabs.harvard.edu/abs/2026arXiv260628590O} {p. arXiv:2606.28590}

\bibitem[\protect\citeauthoryear{{Patton}, {Torrey}, {Ellison}, {Mendel}  \& {Scudder}}{{Patton} et~al.}{2013}]{Patton2013}
{Patton} D.~R.,  {Torrey} P.,  {Ellison} S.~L.,  {Mendel} J.~T.,   {Scudder} J.~M.,  2013, \mn@doi [\mnras] {10.1093/mnrasl/slt058}, \href {https://ui.adsabs.harvard.edu/abs/2013MNRAS.433L..59P} {433, L59}

\bibitem[\protect\citeauthoryear{{Patton}, {Qamar}, {Ellison}, {Bluck}, {Simard}, {Mendel}, {Moreno}  \& {Torrey}}{{Patton} et~al.}{2016}]{Patton2016}
{Patton} D.~R.,  {Qamar} F.~D.,  {Ellison} S.~L.,  {Bluck} A. F.~L.,  {Simard} L.,  {Mendel} J.~T.,  {Moreno} J.,   {Torrey} P.,  2016, \mn@doi [\mnras] {10.1093/mnras/stw1494}, \href {https://ui.adsabs.harvard.edu/abs/2016MNRAS.461.2589P} {461, 2589}

\bibitem[\protect\citeauthoryear{{Patton} et~al.,}{{Patton} et~al.}{2020}]{Patton2020}
{Patton} D.~R.,  et~al., 2020, \mn@doi [\mnras] {10.1093/mnras/staa913}, \href {https://ui.adsabs.harvard.edu/abs/2020MNRAS.494.4969P} {494, 4969}

\bibitem[\protect\citeauthoryear{{Patton}, {Faria}, {Hani}, {Torrey}, {Ellison}, {Thakur}  \& {Westlake}}{{Patton} et~al.}{2024}]{Patton2024}
{Patton} D.~R.,  {Faria} L.,  {Hani} M.~H.,  {Torrey} P.,  {Ellison} S.~L.,  {Thakur} S.~D.,   {Westlake} R.~I.,  2024, \mn@doi [\mnras] {10.1093/mnras/stae608}, \href {https://ui.adsabs.harvard.edu/abs/2024MNRAS.529.1493P} {529, 1493}

\bibitem[\protect\citeauthoryear{{Pillepich} et~al.,}{{Pillepich} et~al.}{2018}]{Pillepich2018}
{Pillepich} A.,  et~al., 2018, \mn@doi [\mnras] {10.1093/mnras/stx2656}, \href {https://ui.adsabs.harvard.edu/abs/2018MNRAS.473.4077P} {473, 4077}

\bibitem[\protect\citeauthoryear{{Ploeckinger}, {Richings}, {Schaye}, {Trayford}, {Schaller}  \& {Chaikin}}{{Ploeckinger} et~al.}{2025}]{2025arXiv250615773P}
{Ploeckinger} S.,  {Richings} A.~J.,  {Schaye} J.,  {Trayford} J.~W.,  {Schaller} M.,   {Chaikin} E.,  2025, \mn@doi [arXiv e-prints] {10.48550/arXiv.2506.15773}, \href {https://ui.adsabs.harvard.edu/abs/2025arXiv250615773P} {p. arXiv:2506.15773}

\bibitem[\protect\citeauthoryear{{Pusk{\'a}s} et~al.,}{{Pusk{\'a}s} et~al.}{2025a}]{Pusk2025}
{Pusk{\'a}s} D.,  et~al., 2025a, \mn@doi [arXiv e-prints] {10.48550/arXiv.2510.14743}, \href {https://ui.adsabs.harvard.edu/abs/2025arXiv251014743P} {p. arXiv:2510.14743}

\bibitem[\protect\citeauthoryear{{Pusk{\'a}s} et~al.,}{{Pusk{\'a}s} et~al.}{2025b}]{2025MNRAS.540.2146P}
{Pusk{\'a}s} D.,  et~al., 2025b, \mn@doi [\mnras] {10.1093/mnras/staf813}, \href {https://ui.adsabs.harvard.edu/abs/2025MNRAS.540.2146P} {540, 2146}

\bibitem[\protect\citeauthoryear{{Quai}, {Byrne-Mamahit}, {Ellison}, {Patton}  \& {Hani}}{{Quai} et~al.}{2023}]{quai2023}
{Quai} S.,  {Byrne-Mamahit} S.,  {Ellison} S.~L.,  {Patton} D.~R.,   {Hani} M.~H.,  2023, \mn@doi [\mnras] {10.1093/mnras/stac3713}, \href {https://ui.adsabs.harvard.edu/abs/2023MNRAS.519.2119Q} {519, 2119}

\bibitem[\protect\citeauthoryear{{Renaud}, {Bournaud}, {Kraljic}  \& {Duc}}{{Renaud} et~al.}{2014}]{Renaud2014}
{Renaud} F.,  {Bournaud} F.,  {Kraljic} K.,   {Duc} P.-A.,  2014, \mn@doi [\mnras] {10.1093/mnrasl/slu050}, \href {https://ui.adsabs.harvard.edu/abs/2014MNRAS.442L..33R} {442, L33}

\bibitem[\protect\citeauthoryear{{Richings}, {Schaye}  \& {Oppenheimer}}{{Richings} et~al.}{2014a}]{2014MNRAS.440.3349R}
{Richings} A.~J.,  {Schaye} J.,   {Oppenheimer} B.~D.,  2014a, \mn@doi [\mnras] {10.1093/mnras/stu525}, \href {https://ui.adsabs.harvard.edu/abs/2014MNRAS.440.3349R} {440, 3349}

\bibitem[\protect\citeauthoryear{{Richings}, {Schaye}  \& {Oppenheimer}}{{Richings} et~al.}{2014b}]{2014MNRAS.442.2780R}
{Richings} A.~J.,  {Schaye} J.,   {Oppenheimer} B.~D.,  2014b, \mn@doi [\mnras] {10.1093/mnras/stu1046}, \href {https://ui.adsabs.harvard.edu/abs/2014MNRAS.442.2780R} {442, 2780}

\bibitem[\protect\citeauthoryear{{Rodriguez-Gomez} et~al.,}{{Rodriguez-Gomez} et~al.}{2015}]{2015MNRAS.449...49R}
{Rodriguez-Gomez} V.,  et~al., 2015, \mn@doi [\mnras] {10.1093/mnras/stv264}, \href {https://ui.adsabs.harvard.edu/abs/2015MNRAS.449...49R} {449, 49}

\bibitem[\protect\citeauthoryear{{Rodr{\'\i}guez Montero}, {Dav{\'e}}, {Wild}, {Angl{\'e}s-Alc{\'a}zar}  \& {Narayanan}}{{Rodr{\'\i}guez Montero} et~al.}{2019}]{RodriguezMontero2019}
{Rodr{\'\i}guez Montero} F.,  {Dav{\'e}} R.,  {Wild} V.,  {Angl{\'e}s-Alc{\'a}zar} D.,   {Narayanan} D.,  2019, \mn@doi [\mnras] {10.1093/mnras/stz2580}, \href {https://ui.adsabs.harvard.edu/abs/2019MNRAS.490.2139R} {490, 2139}

\bibitem[\protect\citeauthoryear{{Schaller} et~al.,}{{Schaller} et~al.}{2024}]{Schaller2024}
{Schaller} M.,  et~al., 2024, \mn@doi [\mnras] {10.1093/mnras/stae922}, \href {https://ui.adsabs.harvard.edu/abs/2024MNRAS.530.2378S} {530, 2378}

\bibitem[\protect\citeauthoryear{{Schaye} et~al.,}{{Schaye} et~al.}{2015}]{Schaye2015}
{Schaye} J.,  et~al., 2015, \mn@doi [\mnras] {10.1093/mnras/stu2058}, \href {https://ui.adsabs.harvard.edu/abs/2015MNRAS.446..521S} {446, 521}

\bibitem[\protect\citeauthoryear{{Schaye} et~al.,}{{Schaye} et~al.}{2026}]{Schaye2025}
{Schaye} J.,  et~al., 2026, \mn@doi [\mnras] {10.1093/mnras/stag375}, \href {https://ui.adsabs.harvard.edu/abs/2026MNRAS.548ag375S} {548, stag375}

\bibitem[\protect\citeauthoryear{{Schechter} et~al.,}{{Schechter} et~al.}{2025}]{Schechter2025}
{Schechter} A.~L.,  et~al., 2025, \mn@doi [\apj] {10.3847/1538-4357/ade791}, \href {https://ui.adsabs.harvard.edu/abs/2025ApJ...989..149S} {989, 149}

\bibitem[\protect\citeauthoryear{{Schmidt}}{{Schmidt}}{1959}]{1959ApJ...129..243S}
{Schmidt} M.,  1959, \mn@doi [\apj] {10.1086/146614}, \href {https://ui.adsabs.harvard.edu/abs/1959ApJ...129..243S} {129, 243}

\bibitem[\protect\citeauthoryear{{Scudder}, {Ellison}, {Torrey}, {Patton}  \& {Mendel}}{{Scudder} et~al.}{2012}]{Scudder2012}
{Scudder} J.~M.,  {Ellison} S.~L.,  {Torrey} P.,  {Patton} D.~R.,   {Mendel} J.~T.,  2012, \mn@doi [\mnras] {10.1111/j.1365-2966.2012.21749.x}, \href {https://ui.adsabs.harvard.edu/abs/2012MNRAS.426..549S} {426, 549}

\bibitem[\protect\citeauthoryear{{Shah} et~al.,}{{Shah} et~al.}{2022}]{2022ApJ...940....4S}
{Shah} E.~A.,  et~al., 2022, \mn@doi [\apj] {10.3847/1538-4357/ac96eb}, \href {https://ui.adsabs.harvard.edu/abs/2022ApJ...940....4S} {940, 4}

\bibitem[\protect\citeauthoryear{{Sharda} et~al.,}{{Sharda} et~al.}{2026}]{2026arXiv260625995S}
{Sharda} P.,  et~al., 2026, \mn@doi [arXiv e-prints] {10.48550/arXiv.2606.25995}, \href {https://ui.adsabs.harvard.edu/abs/2026arXiv260625995S} {p. arXiv:2606.25995}

\bibitem[\protect\citeauthoryear{{Silva} et~al.,}{{Silva} et~al.}{2018}]{2018ApJ...868...46S}
{Silva} A.,  et~al., 2018, \mn@doi [\apj] {10.3847/1538-4357/aae847}, \href {https://ui.adsabs.harvard.edu/abs/2018ApJ...868...46S} {868, 46}

\bibitem[\protect\citeauthoryear{{Simard}, {Mendel}, {Patton}, {Ellison}  \& {McConnachie}}{{Simard} et~al.}{2011}]{Simard2011}
{Simard} L.,  {Mendel} J.~T.,  {Patton} D.~R.,  {Ellison} S.~L.,   {McConnachie} A.~W.,  2011, \mn@doi [\apjs] {10.1088/0067-0049/196/1/11}, \href {https://ui.adsabs.harvard.edu/abs/2011ApJS..196...11S} {196, 11}

\bibitem[\protect\citeauthoryear{{Springel}}{{Springel}}{2000}]{2000MNRAS.312..859S}
{Springel} V.,  2000, \mn@doi [\mnras] {10.1046/j.1365-8711.2000.03187.x}, \href {https://ui.adsabs.harvard.edu/abs/2000MNRAS.312..859S} {312, 859}

\bibitem[\protect\citeauthoryear{{Springel}, {White}, {Tormen}  \& {Kauffmann}}{{Springel} et~al.}{2001}]{Springel2001}
{Springel} V.,  {White} S. D.~M.,  {Tormen} G.,   {Kauffmann} G.,  2001, \mn@doi [\mnras] {10.1046/j.1365-8711.2001.04912.x}, \href {https://ui.adsabs.harvard.edu/abs/2001MNRAS.328..726S} {328, 726}

\bibitem[\protect\citeauthoryear{{Springel}, {Di Matteo}  \& {Hernquist}}{{Springel} et~al.}{2005}]{2005MNRAS.361..776S}
{Springel} V.,  {Di Matteo} T.,   {Hernquist} L.,  2005, \mn@doi [\mnras] {10.1111/j.1365-2966.2005.09238.x}, \href {https://ui.adsabs.harvard.edu/abs/2005MNRAS.361..776S} {361, 776}

\bibitem[\protect\citeauthoryear{{Strauss} et~al.,}{{Strauss} et~al.}{2002}]{Strauss2002}
{Strauss} M.~A.,  et~al., 2002, \mn@doi [\aj] {10.1086/342343}, \href {https://ui.adsabs.harvard.edu/abs/2002AJ....124.1810S} {124, 1810}

\bibitem[\protect\citeauthoryear{{Thorp}, {Ellison}, {Simard}, {S{\'a}nchez}  \& {Antonio}}{{Thorp} et~al.}{2019}]{2019MNRAS.482L..55T}
{Thorp} M.~D.,  {Ellison} S.~L.,  {Simard} L.,  {S{\'a}nchez} S.~F.,   {Antonio} B.,  2019, \mn@doi [\mnras] {10.1093/mnrasl/sly185}, \href {https://ui.adsabs.harvard.edu/abs/2019MNRAS.482L..55T} {482, L55}

\bibitem[\protect\citeauthoryear{{Toomre} \& {Toomre}}{{Toomre} \& {Toomre}}{1972}]{1972ApJ...178..623T}
{Toomre} A.,  {Toomre} J.,  1972, \mn@doi [\apj] {10.1086/151823}, \href {https://ui.adsabs.harvard.edu/abs/1972ApJ...178..623T} {178, 623}

\bibitem[\protect\citeauthoryear{{Trayford} et~al.,}{{Trayford} et~al.}{2026}]{2026MNRAS.545f2040T}
{Trayford} J.~W.,  et~al., 2026, \mn@doi [\mnras] {10.1093/mnras/staf2040}, \href {https://ui.adsabs.harvard.edu/abs/2026MNRAS.545f2040T} {545, staf2040}

\bibitem[\protect\citeauthoryear{{Wetzel}, {Tinker}  \& {Conroy}}{{Wetzel} et~al.}{2012}]{2012MNRAS.424..232W}
{Wetzel} A.~R.,  {Tinker} J.~L.,   {Conroy} C.,  2012, \mn@doi [\mnras] {10.1111/j.1365-2966.2012.21188.x}, \href {https://ui.adsabs.harvard.edu/abs/2012MNRAS.424..232W} {424, 232}

\bibitem[\protect\citeauthoryear{{Wilkinson}, {Ellison}, {Bottrell}, {Bickley}, {Gwyn}, {Cuillandre}  \& {Wild}}{{Wilkinson} et~al.}{2022}]{Wilkinson2022}
{Wilkinson} S.,  {Ellison} S.~L.,  {Bottrell} C.,  {Bickley} R.~W.,  {Gwyn} S.,  {Cuillandre} J.-C.,   {Wild} V.,  2022, \mn@doi [\mnras] {10.1093/mnras/stac1962}, \href {https://ui.adsabs.harvard.edu/abs/2022MNRAS.516.4354W} {516, 4354}

\bibitem[\protect\citeauthoryear{{Wilkinson}, {Ludlow}, {Lagos}, {Fall}, {Schaye}  \& {Obreschkow}}{{Wilkinson} et~al.}{2023}]{2023MNRAS.519.5942W}
{Wilkinson} M.~J.,  {Ludlow} A.~D.,  {Lagos} C. d.~P.,  {Fall} S.~M.,  {Schaye} J.,   {Obreschkow} D.,  2023, \mn@doi [\mnras] {10.1093/mnras/stad055}, \href {https://ui.adsabs.harvard.edu/abs/2023MNRAS.519.5942W} {519, 5942}

\bibitem[\protect\citeauthoryear{{Woods} \& {Geller}}{{Woods} \& {Geller}}{2007}]{2007AJ....134..527W}
{Woods} D.~F.,  {Geller} M.~J.,  2007, \mn@doi [\aj] {10.1086/519381}, \href {https://ui.adsabs.harvard.edu/abs/2007AJ....134..527W} {134, 527}

\bibitem[\protect\citeauthoryear{{Wright}, {Somerville}, {Lagos}, {Schaller}, {Dav{\'e}}, {Angl{\'e}s-Alc{\'a}zar}  \& {Genel}}{{Wright} et~al.}{2024}]{2024MNRAS.532.3417W}
{Wright} R.~J.,  {Somerville} R.~S.,  {Lagos} C. d.~P.,  {Schaller} M.,  {Dav{\'e}} R.,  {Angl{\'e}s-Alc{\'a}zar} D.,   {Genel} S.,  2024, \mn@doi [\mnras] {10.1093/mnras/stae1688}, \href {https://ui.adsabs.harvard.edu/abs/2024MNRAS.532.3417W} {532, 3417}

\bibitem[\protect\citeauthoryear{{Xu} et~al.,}{{Xu} et~al.}{2010}]{2010ApJ...713..330X}
{Xu} C.~K.,  et~al., 2010, \mn@doi [\apj] {10.1088/0004-637X/713/1/330}, \href {https://ui.adsabs.harvard.edu/abs/2010ApJ...713..330X} {713, 330}

\bibitem[\protect\citeauthoryear{{York} et~al.,}{{York} et~al.}{2000}]{York2000}
{York} D.~G.,  et~al., 2000, \mn@doi [\aj] {10.1086/301513}, \href {https://ui.adsabs.harvard.edu/abs/2000AJ....120.1579Y} {120, 1579}

\bibitem[\protect\citeauthoryear{{Yuan}, {Takeuchi}, {Matsuoka}, {Buat}, {Burgarella}  \& {Iglesias-P{\'a}ramo}}{{Yuan} et~al.}{2012}]{Yuan2012}
{Yuan} F.-T.,  {Takeuchi} T.~T.,  {Matsuoka} Y.,  {Buat} V.,  {Burgarella} D.,   {Iglesias-P{\'a}ramo} J.,  2012, \mn@doi [\aap] {10.1051/0004-6361/201220451}, \href {https://ui.adsabs.harvard.edu/abs/2012A&A...548A.117Y} {548, A117}

\bibitem[\protect\citeauthoryear{{Zepf} \& {Koo}}{{Zepf} \& {Koo}}{1989}]{Zepf1989}
{Zepf} S.~E.,  {Koo} D.~C.,  1989, \mn@doi [\apj] {10.1086/167085}, \href {https://ui.adsabs.harvard.edu/abs/1989ApJ...337...34Z} {337, 34}

\makeatother
\end{thebibliography}

\appendix

\section{Matching interacting galaxies to controls}
\label{app:matching}

Section~\ref{sec:methods_samples_3d} describes how interacting and control galaxies are matched based on stellar mass $M_*$, local density $N_2$, isolation $r_{\rm 3D}$ (and $r_{\mathrm{3D},2}$), and redshift $z$. Further details about the matching are discussed in this appendix.

Throughout most of this work (except for Section \ref{sec:results_sdss_sdss}, which compares \colibre{} to SDSS), we select one control galaxy for each interacting galaxy. A galaxy can be selected as a control if both galaxies have the same redshift, i.e. are taken from the same simulation snapshot, and if its stellar mass, local density, and isolation are within 0.05~dex, 10~per cent, and 10~per cent of those of the interacting galaxy, respectively. When multiple controls satisfy these criteria, we select the galaxy with the largest weight, following the weighting scheme of \citet{Patton2016}. The weights are defined such that a perfect match between an interacting galaxy and a control galaxy results in a weight of 1, while values at the edge of the tolerance range in each property correspond to 0. Specifically, the weight for each property is given by
\begin{equation}
w_{x_i} = \frac{|x - x_i|}{x_{\mathrm{tol}}},
\label{eq:weights}
\end{equation}
where $x$ is one of the three matching properties, $w_{x_i}$ is the weight of control galaxy $i$ relative to the interacting galaxy, and $x_{\mathrm{tol}}$ is the tolerance in property $x$. For the stellar mass, the weight is calculated with the log$_{10}$ of the stellar mass. The total weight of each control candidate is taken as the product of the individual weights,
\begin{equation}
w_i = w_{M^i_*} \, w_{N^i_2} \, w_{r^i_2}.
\end{equation}
Whenever no control candidates exist with matching properties within the tolerance range, the tolerance is increased by 50 per cent, leading to tolerances of 0.075~dex, 15 per cent, and 15 per cent for stellar mass, local density, and isolation, respectively. If still no match is found, this procedure is repeated once more, and if no controls are found, the interacting galaxy is removed from the interacting sample. We note, however, that approximately $99$~per cent of interacting galaxies are matched to a control within the original tolerance range. However, at large stellar masses ($M_*>10^{11}~\mathrm{M}_\odot$), $\approx21$~per~cent of interacting galaxies are rejected because no suitable match exists, even after expanding the tolerance window twice. This number drops below 1~per~cent for stellar masses $10^{10} < M_*/\mathrm{M}_\odot < 10^{11}$. An interacting galaxy can also serve as the control of another interacting galaxy, and a galaxy can be the control of multiple interacting galaxies.

\begin{figure}
    \centering
    \includegraphics[width=0.5\textwidth]{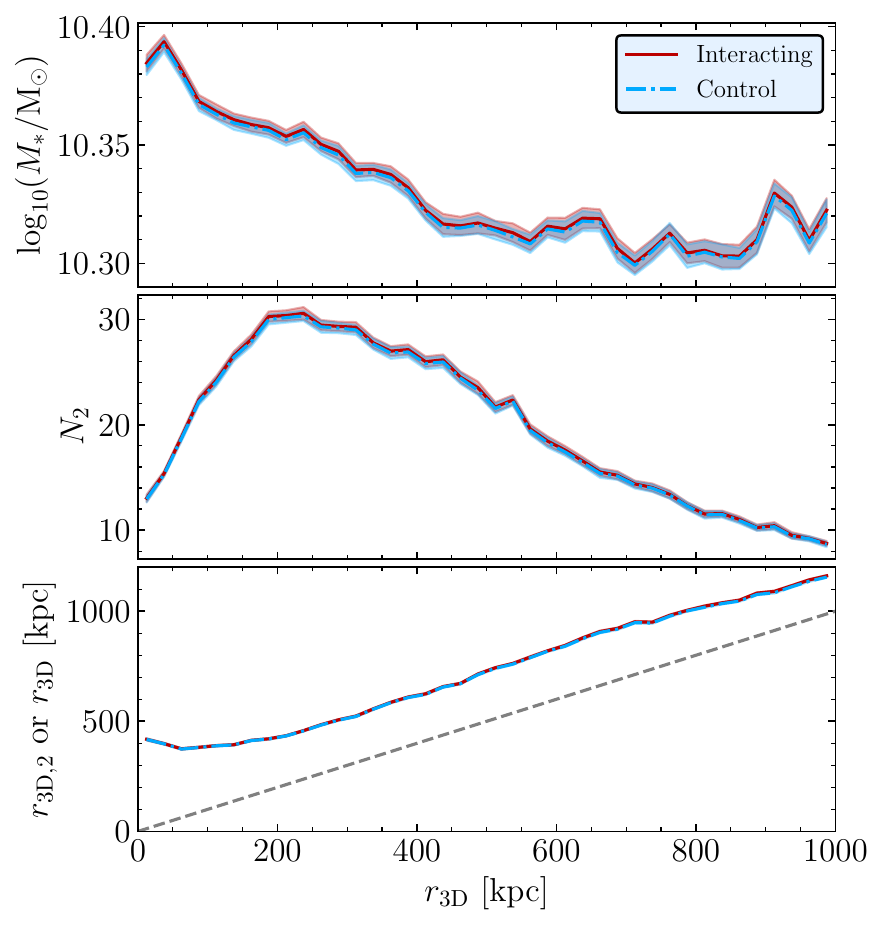}
    \caption{The outcome of matching the interacting and control galaxies. We show the mean stellar mass ($\log_{10}\, M_*/\mathrm{M}_\odot$; \textit{top panel}), local galaxy density ($N_2$; \textit{middle panel}), and isolation ($r_{\rm 3D}$ for controls and $r_{\rm 3D,2}$ for interacting galaxies; \textit{bottom panel}) of interacting galaxies (solid red lines) and their controls (dash-dotted blue lines), all as a function of separation between the interacting galaxies and their closest companions, $r_{\rm 3D}$. For reference, the bottom panel also shows the $x = y$ relation as a grey dashed line. The matching results in nearly identical masses, local densities, and isolations between interacting and control galaxies.}
    \label{fig:matching}
\end{figure}

Fig.~\ref{fig:matching} shows the outcome of this matching procedure. The mean $\log_{10}$ stellar mass, local density, and isolation are shown for interacting galaxies (solid red) and their controls (dash-dotted blue), binned by separation from the interacting galaxy’s closest companion, along with the $1\sigma$ error on the mean indicated by the shaded region. For all properties, no systematic differences are found between the two samples, thereby validating the matching procedure.

\begin{figure}
    \centering
    \includegraphics[width=0.5\textwidth]{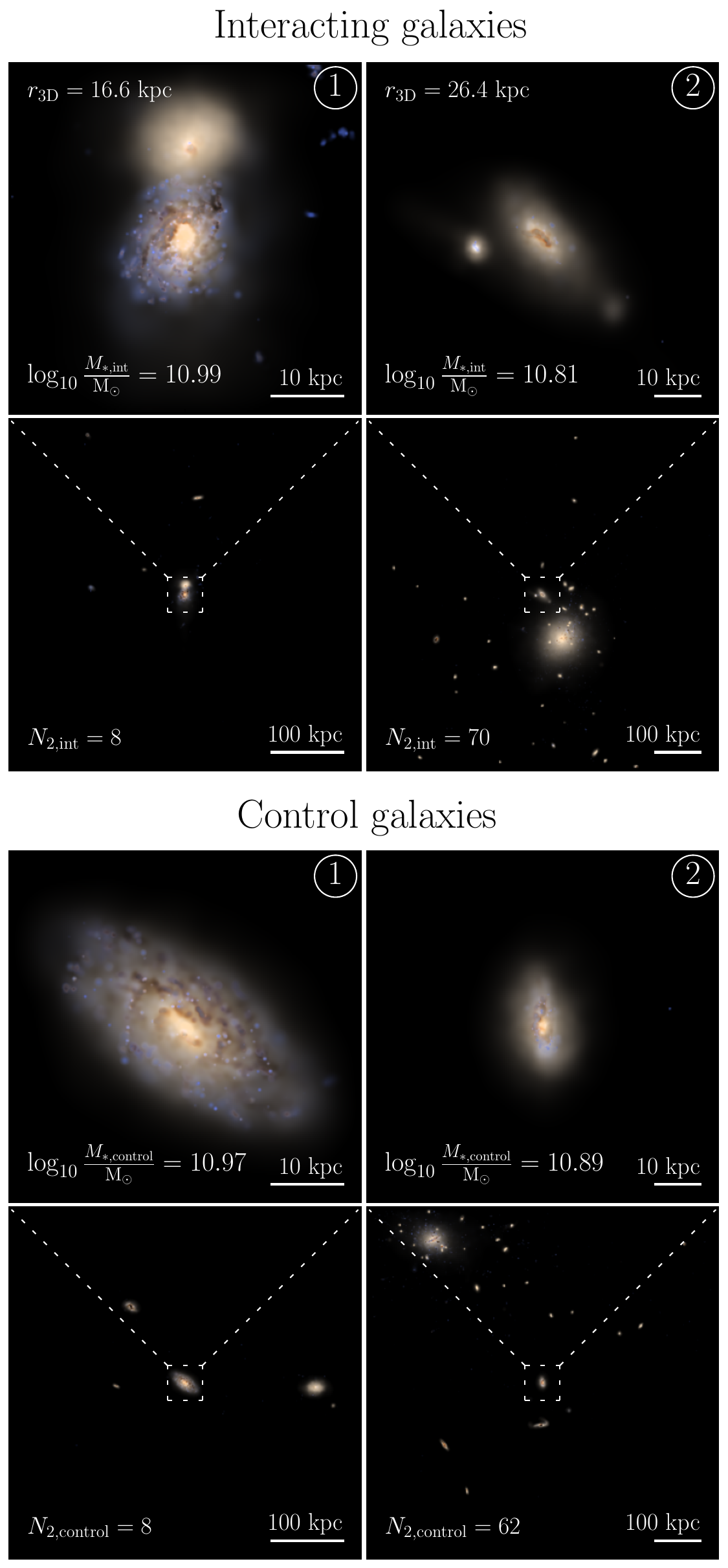}
       \vspace{-0.35cm}
    \caption{Visual impression of two example interacting galaxies (\textit{top row}) and their large-scale environments (\textit{second row}), compared to those of their matched control galaxies (\textit{third and fourth rows}), as indicated by the number in the top right corner of the odd rows. Each galaxy image shows the stellar light in \textit{HST} colours, including dust attenuation. The stellar mass of each galaxy in solar mass units, $M_*$, and its local galaxy density, $N_2$, are indicated in the bottom left corner of the zoom-in and large-scale panels, respectively. The galaxies are taken from the \colibre{} L200m6 simulation at $z = 0$ and are viewed along the simulation domain's native $x$-axis. The images are centred on the interacting (top two rows) and control (bottom two rows) galaxies. The size and depth of the small-scale (large-scale) images are equal to three (30) times the 3D separation between the corresponding interacting galaxy and its closest companion, which is indicated in the top left corner of each panel in the top row.}
    \label{fig:visualisation_environment}
\end{figure}

As an illustration of the matching procedure, in Fig.~\ref{fig:visualisation_environment} we show examples of two interacting galaxies and their large-scale environments (top two rows) taken from the \colibre{} L200m6 simulation at $z=0$, along with the small-scale view and large-scale environments of their matched control galaxies (bottom two rows). 

Finally, in the section comparing \colibre{} with SDSS (\S\ref{sec:results_sdss_sdss}), each interacting galaxy is matched to multiple controls (at least 10), from which a statistical control is constructed. The properties of the statistical control are computed as the weighted averages of the properties of the individual matched control galaxies, where the weights are determined by equation~(\ref{eq:weights}). We use a stellar mass tolerance of 0.1~dex instead of 0.05~dex to be consistent with the SDSS sample of \citet{Patton2013}. Similar to the selection of individual controls, the tolerance range can be widened at most two times to find at least 10 suitable control galaxies before an interacting galaxy is rejected.

\section{Residual environmental differences}
\label{app:environment}

Fig.~\ref{fig:matching} showed that the matching described in Section~\ref{sec:methods_samples_3d} yields samples of interacting and control galaxies with statistically indistinguishable distributions of stellar mass, local density, and isolation. However, while the local density constraint ensures that the number of galaxies (with stellar masses of more than 10 per cent of the interacting galaxy’s) is the same within a 3D sphere of 2 Mpc, it does not constrain their spatial distribution within that sphere. In Fig.~\ref{fig:q_r_mstar}, we found that the sSFR enhancement of interacting galaxies with $10^8 < M_{*, \rm int}/\mathrm{M}_\odot < 10^9$ does not fully converge to unity at large separations, which may indicate that small residual differences remain in the environments of interacting galaxies and their controls. In this Appendix, we investigate this potential discrepancy.

\begin{figure}
    \centering
    \includegraphics[width=0.5\textwidth]{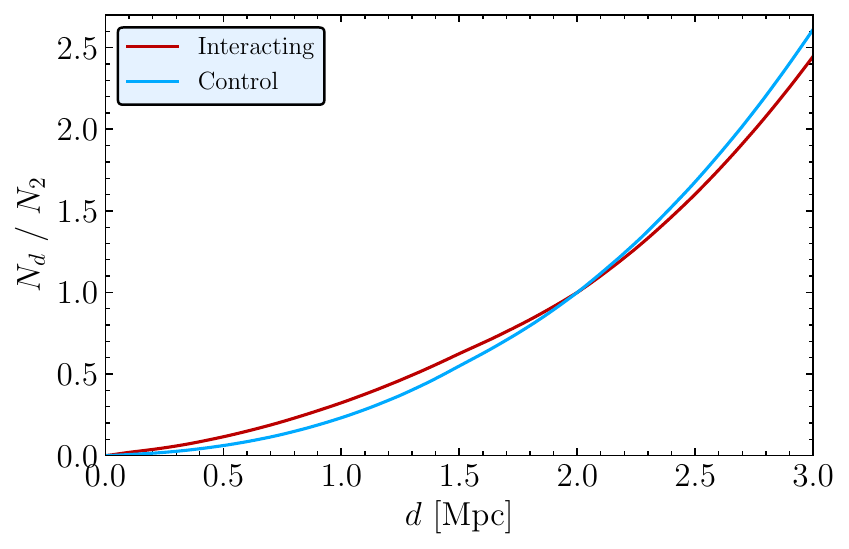}
    \caption{Mean ratio of the number of galaxies within a 3D spherical aperture of radius $d$, $N_d$, to that within an aperture of radius $2~\mathrm{Mpc}$, $N_2$ (referred to throughout this work as the local density), shown for interacting galaxies (red) and their matched controls (blue) at redshift $z=0$ for the L200m6 simulation. Only interacting galaxies with stellar masses $10^8 < M_{*, \rm int}/\mathrm{M}_\odot < 10^9$ are selected. The value of $N_d$ is systematically larger (smaller) for interacting galaxies than for controls at distances below (above) $2~\mathrm{Mpc}$, which corresponds to the radius within which the local densities of interacting and control galaxies are matched.}
    \label{fig:environment}
\end{figure}

Fig.~\ref{fig:environment} shows the mean of the ratio between the number of galaxies within a 3D aperture of radius $d$, $N_d$, and that within 2 Mpc, $N_2$, for interacting galaxies (red) and their controls (blue). We use the interacting galaxies with $10^8 < M_{*, \rm int}/\mathrm{M}_\odot < 10^9$ and their matched controls from the L200m6 simulation at redshift $z=0$. We plot the mean ratio as a function of aperture radius $d$, which is equal to 1 at $d = 2$~Mpc by construction, as the local density constraint in the matching procedure is defined using a 2~Mpc aperture. We find that for apertures smaller (larger) than the scale at which the interacting and control galaxies' local densities are matched ($d = 2~\mathrm{Mpc}$), the number density of interacting galaxies is slightly larger (smaller) than that of controls.

This discrepancy in the environments of interacting and control galaxies arises from the way in which we constrain them. By matching interacting galaxies with a closest companion at separation $r_{\rm 3D}$ and an isolation $r_{\rm 3D,2}$ to controls with an isolation $r_{\rm 3D}$, we bias interacting galaxies towards denser environments on small scales. By additionally matching on the local density $N_2$, we ensure that the density on larger scales is the same. This results in an overdensity of interacting galaxies relative to their controls on intermediate scales. These small differences may explain why the sSFR enhancement of galaxies with $10^8 < M_{*, \rm int}/\mathrm{M}_\odot < 10^9$ does not fully converge to unity at large separations (see Fig.~\ref{fig:q_r_mstar}).

\begin{figure}
    \centering
    \includegraphics[width=0.5\textwidth]{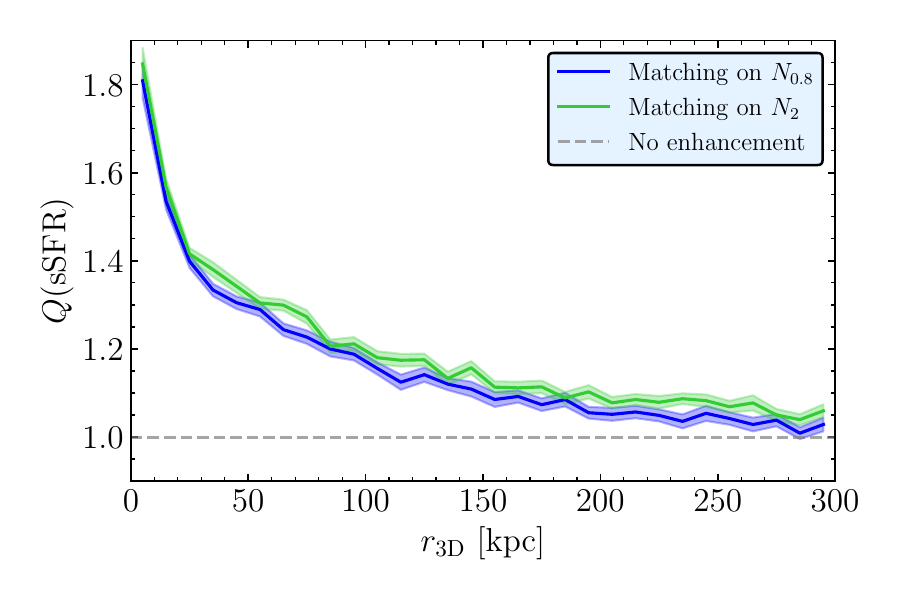}
    \caption{The mean sSFR enhancement of interacting galaxies with $10^8<M_{*, \rm int}/\mathrm{M}_\odot<10^9$ relative to their controls for the fiducial galaxy sample and a galaxy sample matched on the local density within a sphere of 0.8~Mpc, $N_{0.8}$, instead of the fiducial choice of 2~Mpc, $N_{2}$. The sample matched on $N_{0.8}$ converges significantly more closely to unity at large separations ($r_{\rm 3D}\gtrsim 200~\mathrm{kpc}$) than the fiducial sample, indicating that the lack of convergence in the latter case is driven by residual environmental differences between interacting and control galaxies.}
    \label{fig:q_environment}
\end{figure}

To verify this, Fig.~\ref{fig:q_environment} shows $Q$(sSFR) as a function of separation for our fiducial sample (green) and for a sample in which interacting and control galaxies are matched on the local density within an aperture of $0.8$~Mpc ($N_{0.8}$), instead of the fiducial choice of $2$~Mpc ($N_2$, blue). We choose a smaller aperture of $0.8$~Mpc because matching interacting and control galaxies on a number density within a smaller aperture should reduce differences in their environments on smaller scales, where we test the convergence of $Q$(sSFR) to unity. In both cases ($N_{0.8}$ and $N_2$), we select only interacting and control galaxies with $10^8 < M_{*, \rm int}/\mathrm{M}_\odot < 10^9$.

Fig.~\ref{fig:q_environment} shows that $Q$(sSFR) for the sample matched on $N_{0.8}$ is indeed closer to unity at large separations than for the sample matched on $N_2$, confirming that small residual environmental differences prevent $Q$(sSFR) from fully converging to unity at large separations. The smaller, but still non-negligible, deviations from unity for $N_{0.8}$ can also be expected, as the separations probed in Fig.~\ref{fig:q_environment} ($r_{\rm 3D} < 0.3$~Mpc) are smaller than the radius at which $N_{0.8}$ is defined ($d = 0.8$~Mpc). Finally, we note that although using $N_{0.8}$ results in better convergence to unity at large separations, we do not adopt it for the fiducial selection in this work, as $d=0.8$~Mpc is close enough to the largest separations we study ($r_{\rm 3D}=0.3$~Mpc) that it may introduce different biases in the analysis.

\section{The effect of very low-mass companions on SFR enhancement}
\label{app:mini_mergers}

\begin{figure}
    \centering
    \includegraphics[width=0.5\textwidth]{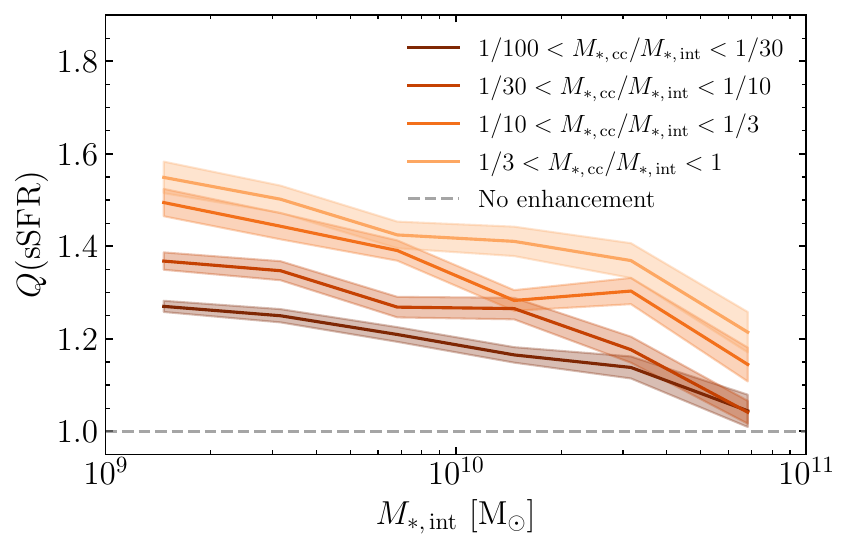}
    \caption{As Fig.~\ref{fig:mass_ratio}, but showing the sSFR enhancement in mass ratio bins extending down to $1/100 < M_{*, \rm cc}/M_{*, \rm int} < 1/30$. Unlike in the rest of this work, here the closest companions are galaxies with stellar masses of more than 1 per cent of their interacting counterpart (instead of the fiducial 10 per cent; see the main text for details). At fixed interacting galaxy stellar mass, the sSFR enhancement decreases monotonically with decreasing mass ratio.}
    \label{fig:mini_mergers}
\end{figure}

For completeness, Fig.~\ref{fig:mini_mergers} extends the results from Fig.~\ref{fig:mass_ratio} by showing the sSFR enhancement in two additional mass ratio bins: $1/30 < M_{*, \rm cc}/M_{*, \rm int} < 1/10$ and $1/100 < M_{*, \rm cc}/M_{*, \rm int} < 1/30$. As in Fig.~\ref{fig:mass_ratio}, we use the L200m6 simulation and consider only interacting galaxies with separations $r_{\rm 3D} < 50$~kpc. However, to probe these additional $M_{*, \rm cc}/M_{*, \rm int}$ bins, we relax the minimum stellar mass of the closest companion from the fiducial 10 per cent of the interacting galaxy stellar mass to 1~per cent, and re-run our interacting--control matching algorithm. Apart from this change, the sample construction follows Section~\ref{subsubsection:interacting_sample_construction}. Due to resolution limitations, in the main part of this work, closest companions are required to have stellar masses of at least $10^7~\mathrm{M}_\odot$ (which can be reached for an interacting galaxy with $M_{*,\rm int}=10^{8}~\mathrm{M_\odot}$ and a mass ratio of $1/10$). In Fig.~\ref{fig:mini_mergers}, we show the enhancement for interacting galaxies with stellar masses only down to $10^9~\mathrm{M_\odot}$, such that the closest companions in the lowest mass ratio bin ($1/100$) can also have stellar masses no lower than $10^7~\mathrm{M_\odot}$.

We find that, at fixed interacting galaxy stellar mass, the sSFR enhancement $Q$ decreases monotonically with decreasing $M_{*, \rm cc}/M_{*, \rm int}$. At $M_{*, \rm int} \sim 10^{9}~\mathrm{M}_\odot$, the enhancement in the $1/3 < M_{*, \rm cc}/M_{*, \rm int} < 1$ bin is $\approx 1.35$, decreasing to $\approx 1.25$ in the $1/100 < M_{*, \rm cc}/M_{*, \rm int} < 1/30$ bin. We note that at least part of the remaining enhancement in the lowest mass ratio bin may be due to small residual differences in the environments of interacting galaxies and their controls (see Appendix~\ref{app:environment}), rather than a genuine physical effect induced by the closest companion.

\section{Stellar mass stripping}
\label{app:numerical_stripping}

\begin{figure}
    \centering
    \includegraphics[width=0.5\textwidth]{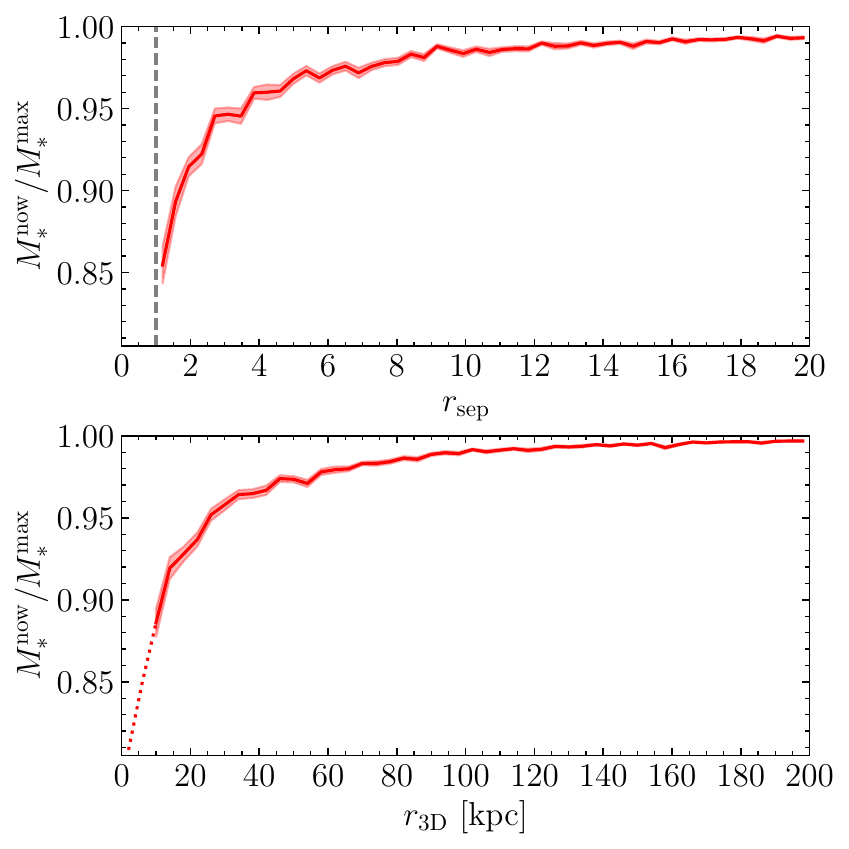}
    \caption{Mean ratio of stellar masses of interacting galaxies at redshift $z=0$ ($M_*^\text{now}$) and their maximum stellar masses within the last 500~Myr ($M_*^\text{max}$), as a function of dimensionless separation $r_{\rm sep}$ \textit{(top panel)} and 3D separation $r_\text{3D}$ \textit{(bottom panel)}. The mass ratios are shown in red, and the minimum allowed dimensionless separation of pairs, $r_\text{sep}=1$, is indicated by a grey dashed line. We include only interacting galaxies with stellar masses $M_{*, \rm int} > 10^{9}~\mathrm{M}_\odot$. The mass ratio decreases by up to 20~per cent at small separations due to the combined effects of numerical and physical stripping.}
    \label{fig:numerical_stripping}
\end{figure}

As a satellite subhalo orbits within a more massive central halo, it gradually loses its stellar (and dark matter) mass as a result of tidal effects. In numerical simulations, this mass loss can be further enhanced if, for example, the subhalo finders incorrectly assign some of the satellite's mass to the more massive central halo. This excess mass loss is termed `numerical stripping'. \citet{Patton2020} investigated this effect for galaxy pairs with small separations in \tng, finding that the stellar masses of galaxies in close encounters can decrease by up to $\approx 20$ per cent as a result of the combined effects of physical and numerical stripping (with mass from the less massive galaxy in the interacting pair transferred to the more massive one). By visually inspecting cases with small separations and severe mass loss to disentangle the physical stripping from the numerical component, they corrected for the numerical stripping effect by reassigning galaxy stellar masses to their values prior to the encounter (within the last 500~Myr). We note, however, that although important on a galaxy-by-galaxy basis, globally this process increased the stellar masses of only 1.2 per cent of the galaxies in their sample constructed from the TNG100-1 simulation.

While the \tng{} simulations used the SUBFIND halo finder \citep{Springel2001}, \colibre{} employs the halo finder HBT-HERONS \citep{Moreno2025}, which is expected to be less prone to numerical stripping due to its history-based approach to halo tracking. Following the method of \citet{Patton2020}, we investigate the magnitude of stripping in \colibre{} for interacting galaxies with $M_{*, \rm int} > 10^{9}~\mathrm{M}_\odot$ at redshift $z = 0$ by comparing the current stellar mass, $M_*^\text{now}$, of each galaxy with its maximal stellar mass over the last 500~Myr, $M_*^\text{max}$. When stripping occurs, the stellar mass can drop shortly before the merger. Recall that in our construction of the interacting galaxy sample, interacting galaxies can be more or less massive than their closest companion.

Fig.~\ref{fig:numerical_stripping} shows the ratio $M_*^\text{now} / M_*^\text{max}$ as a function of both the physical separation between the galaxies (bottom panel) and the dimensionless separation, $r_\text{sep}$ (top panel), defined as
\begin{equation}
r_\text{sep} = \frac{r_{\rm 3D}}{R_{1/2}^\text{int} + R_{1/2}^\text{cc}},
\end{equation}
where $R_{1/2}^\text{int}$ and $R_{1/2}^\text{cc}$ are the stellar half-mass radii of the interacting galaxy and its closest companion, respectively. The red solid lines indicate the mean values of the stellar mass ratios in each separation bin, with the shaded regions showing the bootstrap $1\sigma$ uncertainties on the mean. Bins with fewer than 50 pairs are indicated by dotted lines, and the grey dashed line marks the minimum separation required for a galaxy pair to be included in the sample. As pairs with overlapping stellar half-mass radii are excluded (see Section~\ref{sec:methods_samples_3d}), no galaxy pairs exist with $r_\text{sep} < 1$. The figure shows that at the smallest separation, $r_\text{sep} \approx 1$, the stellar mass decreases by up to $\approx 20$~per cent, consistent with the findings of \citet{Patton2020}. Since a significant fraction of this decrease is expected to arise from physical stripping rather than numerical effects, and because \citet{Patton2020} found that only the masses of $\approx 1.2$~per cent of galaxies required a correction (based on TNG100-1), we expect numerical stripping to have little to no impact on our results and therefore do not correct for it. We also note that \citet{2026arXiv260403105H} recently found that numerical disruption of satellites due to tidal effects in \colibre{} has only a marginal impact on the subhalo stellar mass function.

\bsp	
\label{lastpage}
\end{document}